\documentclass[3p,times]{elsarticle}

\usepackage{amssymb}
\usepackage{amsmath}
\usepackage{soul,xcolor}
\usepackage{caption}
\usepackage{subcaption}
\usepackage{graphicx}
\usepackage{tikz}
\usepackage{tkz-euclide}
\usepackage{tikz-dimline}
\usepackage{pgfplots}
\usetikzlibrary{patterns, pgfplots.fillbetween, bending, shapes}
\usepackage{color}
\usepackage{hyperref}
\usepackage[ruled,vlined]{algorithm2e}
\usepackage{comment}
\usepackage{listings}

\pgfdeclareverticalshading{distance}{100pt}{
  rgb(0pt)=(1,0.8,0.8); 
  rgb(5pt)=(1,1,1); 
  rgb(50pt)=(0.2,0.2,1) 
}

\definecolor{codebg}{rgb}{0.97, 0.97, 0.97}

\lstdefinestyle{kratos}{
  language=Python,
  basicstyle=\ttfamily\scriptsize,       
  backgroundcolor=\color{codebg},        
  keywordstyle=\color{blue}\bfseries,    
  commentstyle=\color{gray},             
  stringstyle=\color{red},               
  numbers=left,                          
  numberstyle=\tiny\color{gray},         
  stepnumber=1,
  numbersep=5pt,
  showstringspaces=false,
  breaklines=true,
  frame=single,
  captionpos=b,
  tabsize=4,
  morekeywords={Parameters, Model}
}

\usetikzlibrary{shapes.misc}
\tikzset{cross/.style={cross out, draw=black, minimum size=2*(#1-\pgflinewidth), inner sep=0pt, outer sep=0pt}, cross/.default={1pt}}

\newdefinition{definition}{Definition}
\newdefinition{rem}{Remark}

\newcommand{\ignore}[1]{}

\journal{Computational Mechanics}

\begin{document}

\begin{frontmatter}



\title{Consistent Mapping for Fluid–Structure Interaction with Geometrically Nonlinear Isogeometric Beams}

\author[tum]{Juan Ignacio Camarotti\fnref{equal}\corref{ca}}
\ead{juan.camarotti@tum.de}
\author[tum]{Ricky Aristio}
\ead{ricky.aristio@tum.de}
\author[tum]{Roland Wüchner}
\ead{wuechner@tum.de}

\address[tum]{Chair of Structural Analysis, Technical University of Munich, Arcisstr. 21, 80333 München, Germany}
\cortext[ca]{Corresponding author}

\begin{abstract}
Isogeometric beam formulations provide an efficient reduced-dimensional representation of slender structures, combining accurate geometry and high-order continuity with few structural degrees of freedom. Their application to fluid--structure interaction (FSI) requires transferring motion and loads between the one-dimensional beam centerline and the surrounding fluid interface. While mapping strategies exist for conventional finite element beams, geometrically nonlinear isogeometric formulations require operators consistent with their spline-based kinematics and distinct descriptions of cross-sectional rotation. This work develops consistent beam-to-surface mappings for geometrically nonlinear isogeometric Bernoulli and Timoshenko beams in partitioned FSI. Fluid-interface points are attached to the beam in the reference configuration and reconstructed using the nonlinear kinematics of each formulation. Consistent linearization yields a configuration-dependent mapping operator whose transpose ensures work-conjugate load transfer. The formulations are verified through analytical deformation tests, convergence studies, and virtual-work checks, and evaluated in beam-based adaptations of the Mok and Turek--Hron benchmarks and the Goland wing flutter benchmark. The results confirm accurate interface reconstruction and work-conjugate load transfer under large deformations, while demonstrating the applicability of both beam formulations to coupled FSI problems.
\end{abstract}




\end{frontmatter}


\section{Introduction}
\label{sec:introduction}

Fluid--structure interaction (FSI) plays an important role in many
engineering applications in which the deformation of a structure affects
the surrounding flow and, in turn, the resulting fluid loads influence the
structural response. Examples arise in aerospace, wind energy, civil
engineering, and other applications involving lightweight and flexible
structures. In aeroelastic applications, for instance, structural
deformations can significantly modify the aerodynamic loading, while the
resulting fluid forces determine the structural response. The numerical
prediction of this mutual interaction is therefore an important component
in the analysis and design of flexible structures.

At early design stages, repeatedly resolving the complete
three-dimensional structural geometry may be computationally
expensive. Reduced-dimensional structural models are therefore
particularly attractive for slender structures, as demonstrated
by their widespread application in the aeroelastic analysis of
aircraft wings~\cite{Patil2000,VarelloEtAl2010}, helicopter rotor
blades~\cite{ShangXia2024}, and wind-turbine
blades~\cite{HANSEN2006285,Streiner2011}. In such applications,
beam formulations can capture the dominant structural deformation
mechanisms with substantially fewer degrees of freedom than
shell or solid models.

Isogeometric analysis (IGA) provides an attractive framework for such
beam formulations by employing spline basis functions, most commonly
NURBS, to represent both the geometry and the unknown fields
\cite{Hughes2005}. In addition to accurately representing curved
geometries, spline-based approximations naturally provide higher
inter-element continuity than standard displacement-based finite element
discretizations. This property is particularly relevant for Bernoulli
beams, whose bending kinematics require at least $C^1$-continuous
centerline approximations. In the context of beam-based FSI, the
continuity of the underlying approximation is also relevant to the
reconstruction of the fluid interface, whose motion depends on both
the beam centerline displacement and the orientation of its
cross-sections.

These characteristics have motivated the development of various isogeometric beam and rod formulations for the geometrically nonlinear analysis of slender structures \cite{GrecoCuomo2013,Raknes2015,Marino2016,BAUER2016101,Vo2020}. In the present work, the spatial Bernoulli formulation of Bauer et al.~\cite{BAUER2016101} is adopted because it provides an explicit reconstruction of the cross-sectional orientation from the deformed centerline tangent and a scalar torsional rotation. This kinematic description, together with its consistent variational linearization, provides a suitable basis for deriving the beam-to-interface mapping operator. For the Timoshenko model, the total-Lagrangian Galerkin formulation of Vo et al.~\cite{Vo2020} is selected, providing independent cross-sectional rotations within an isogeometric finite-element framework. The two formulations allow the proposed mapping methodology to be developed for distinct representations of cross-sectional orientation while maintaining consistency with the underlying structural degrees of freedom.

Fluid--structure interaction (FSI) requires the satisfaction of
kinematic and dynamic coupling conditions at the fluid--structure
interface. Kinematic compatibility requires the fluid interface
to follow the structural motion, while dynamic equilibrium
requires the interface tractions exerted by the two fields to
be in balance. These conditions can be enforced using either
monolithic or partitioned solution strategies. In a monolithic
approach, the governing equations of the fluid and structure
are assembled and solved within a single coupled system
\cite{Heil2004,Hubner2004,HronTurek2006}. In contrast,
partitioned approaches solve the fluid and structural
subproblems separately, exchanging interface quantities to
satisfy the coupling conditions
\cite{Farhat1997,DEGROOTE2009793,DEGROOTE2010446}.
The latter strategy, adopted in this work, allows existing
specialized solvers to be retained and combined within a
modular framework \cite{DEBOER20071515}.

For transient partitioned FSI, coupling must be addressed in
both time and space. Temporal coupling determines how the
individual solvers advance and exchange information within
each time step, whereas spatial coupling concerns the transfer
of kinematic and load quantities between the fluid and
structural interface discretizations. When these discretizations
are non-matching, suitable mapping operators are required to
transfer interface quantities while preserving the relevant
coupling properties \cite{Farhat1998}. This becomes particularly
challenging when the structural model is represented by a
one-dimensional beam and the fluid interface by a
two-dimensional surface, as considered in this work.

Spatial coupling is already non-trivial when both fields provide surface
discretizations, since the fluid and structural interface meshes generally
do not coincide. Different resolution requirements and independent
discretizations of the two fields naturally lead to non-matching
interfaces. Consequently, considerable attention has been devoted to the
transfer of motion and loads between non-matching meshes
\cite{Farhat1998,DEBOER20071515}. Approaches include nearest-neighbor and
projection-based interpolation, radial basis function methods, and
mortar-type formulations. De Boer et al.~\cite{DEBOER20071515} provided a
comparative review of several such coupling methods, while Wang
et al.~\cite{Wang2016} subsequently assessed nearest-element,
standard mortar, and dual mortar mappings for non-matching meshes and
geometries in computational FSI.

Besides interpolation accuracy, spatial coupling must ensure an appropriate
transfer of motion and loads across the interface. Several approaches have
therefore addressed conservation properties for non-matching
discretizations, including consistent and conservative transfer strategies
\cite{Farhat1998,DeBoer2008}, conservative load transfer across curved
interfaces \cite{Jaiman2006}, and mortar-based coupling of isogeometric
B-Rep and fluid discretizations \cite{Apostolatos2021}. These methods
provide a general framework for non-matching interfaces, but typically
assume that the interface geometry is explicitly represented on both sides
of the coupling.

The use of beam models introduces an additional difficulty that cannot be
treated as a conventional non-matching interface problem.
Through dimensional reduction, the structural boundary is no longer
explicitly represented by the structural discretization: the
structure is described by a one-dimensional beam axis, whereas the
fluid loads act on an interface represented by a curve in two-dimensional
problems or a surface in three-dimensional problems. Consequently,
the motion of the fluid interface must be reconstructed from the beam
kinematics. Translational information alone is generally insufficient,
since rotations of the beam cross-sections produce additional
displacements of interface points located away from the beam axis.
Conversely, forces acting at an offset from the beam axis must
generate not only translational forces but also the corresponding
generalized moments in the beam model. A consistent coupling strategy
must therefore account for both translational and rotational effects
in the transfer of interface motion and loads.

Mixed-dimensional coupling between one-dimensional beam models and higher-dimensional structural discretizations has also been investigated. Steinbrecher et al.~\cite{Steinbrecher2022} developed a variationally consistent formulation for embedding geometrically exact Cosserat beams into three-dimensional solid volumes, enforcing positional and rotational compatibility. This approach was subsequently extended to the coupling of beams with two-dimensional deformable surfaces~\cite{Steinbrecher2025}, where the positional constraints account for the offset between the beam centerline and the surface, and rotational compatibility is established using the local surface orientation. Unlike these structural coupling formulations, beam-to-surface mapping in partitioned FSI involves transferring motion and fluid loads between independently discretized fluid and structural models.

The coupling of reduced-dimensional beam models to surface-based fluid
discretizations has been addressed through specialized mapping strategies.
Ahrem et al.~\cite{Ahrem2007} considered a conventional finite element
beam model and introduced a beam-spline-based coupling approach for the
transfer of structural motion to an aerodynamic surface. In addition to
the translational displacements of the beam, rotational information was
incorporated into the interpolation, allowing the motion of surface points
located away from the beam axis to be reconstructed. The formulation was
developed in the context of small structural rotations and demonstrated
the importance of accounting for rotational effects when transferring the
motion of a reduced-dimensional beam model to a surface discretization.

A further development was presented by Wang et al.~\cite{Wang2016}, who proposed a co-rotational mapping algorithm for geometrically nonlinear beam structures. Their approach is formulated for conventional finite element beams and determines the displacement of fluid-interface nodes from the rigid-body motion of the corresponding beam cross-sections. To accommodate large translations and rotations, the element motion is decomposed into a rigid-body displacement and reorientation, an axial rigid-body rotation, and a remaining deformation involving bending and torsion. After extracting the large rotations, the residual deformation is interpolated using conventional linear and cubic Hermite beam shape functions. This allows the surrounding fluid interface to follow large beam deformations while avoiding the direct interpolation of large nodal rotations. The formulation is, however, tied to the element-level co-rotational description and interpolation functions of conventional finite element beams, rather than the continuous kinematic fields of geometrically nonlinear isogeometric Bernoulli and Timoshenko formulations.

Geometrically nonlinear isogeometric beams introduce both a different discretization and, depending on the adopted beam theory, different cross-sectional kinematics. In contrast to nodal finite element beam formulations, the control points of an isogeometric beam do not generally correspond to material points on the beam axis and therefore cannot be used directly as attachment points for the surrounding interface. Surface points must instead be associated with the continuous beam geometry and reconstructed from the kinematic fields evaluated at their corresponding parametric locations. Moreover, the reconstruction depends on the adopted beam theory. In the Bernoulli formulation of Bauer et al. [5], the cross-section remains normal to the current beam axis and its orientation follows from the centerline geometry together with a scalar torsional rotation. In the Timoshenko formulation of Vo et al. [6], the cross-section orientation is instead an independent rotational field, allowing transverse shear deformation. Consequently, the mapping operators and their consistent linearization
must be derived from the respective nonlinear beam kinematics rather
than obtained through a direct extension of conventional nodal
interpolation schemes.

The objective of the present work is to develop a consistent
mapping framework for coupling geometrically nonlinear isogeometric
Bernoulli and Timoshenko beams with fluid discretizations.
The proposed mappings are derived from the nonlinear beam kinematics
of Bauer et al.~\cite{BAUER2016101} and Vo et al.~\cite{Vo2020},
respectively. Interface points are associated with the
beam in the reference configuration, and their cross-sectional coordinates
are retained throughout the deformation. Their current positions are then reconstructed according to the corresponding beam kinematics, accounting for both centerline motion and cross-sectional rotation. Linearization of this nonlinear reconstruction with respect to the admissible beam variations provides the configuration-dependent kinematic mapping operator. The corresponding load-transfer operator is obtained from its transpose through the principle of virtual work. In the present work, consistency therefore refers to deriving the kinematic and load-transfer operators from the same nonlinear interface reconstruction, such that the resulting transfer is work-conjugate at the discrete interface level. Forces acting at an offset from the beam axis thereby contribute naturally to both generalized forces and moments.

The main contribution of this work is therefore a consistent
mapping framework that connects one-dimensional isogeometric beam
models to the fluid--structure interface under large deformations,
accounting for both curve-based interfaces in two-dimensional problems
and surface-based interfaces in three-dimensional problems.
In particular, the framework (i) reconstructs the interface according to the nonlinear cross-sectional kinematics of the underlying Bernoulli or Timoshenko beam formulation, (ii) accounts for the different rotational descriptions of the two beam theories, and (iii) derives the corresponding load transfer from the linearization of the interface reconstruction through the principle of virtual work. The formulation is implemented in Kratos Multiphysics \cite{dadvand2010,dadvand2013,KratosMultiphysics2025} for partitioned FSI. It is first verified using prescribed analytical displacement and rotation fields and subsequently assessed in beam-based adaptations of the Mok and Turek–Hron benchmarks and in the Goland wing flutter benchmark.

The remainder of this paper is organized as follows.
Section~\ref{sec:beam_formulations} introduces the isogeometric beam
geometry and the geometrically nonlinear Bernoulli and Timoshenko
formulations underlying the proposed mappings.
Section~\ref{sec:kinematic_mapping} presents the reference attachment
procedure and the nonlinear reconstruction of the interface surface.
The corresponding work-conjugate transfer of interface loads is derived
in Section~\ref{sec:load_transfer}.
The proposed mappings are verified through analytical deformation
tests, convergence studies, and virtual-work consistency checks
in Section~\ref{sec:mapping_verification}.
Their application to fluid--structure interaction problems is
demonstrated in Section~\ref{sec:fsi_numerical_examples}.
Finally, Section~\ref{sec:conclusions} summarizes the main findings.

\section{Isogeometric beam geometry and structural formulations}
\label{sec:beam_formulations}

This section introduces the geometrically nonlinear isogeometric
Bernoulli and Timoshenko beam formulations employed in this work.
First, the spline-based representation of the beam centerline
and the definition of the cross-section directors are presented.
The governing kinematics and structural formulations of both
beam models are then summarized, providing the basis for the
consistent mapping operators developed in the following section.

\subsection{Isogeometric beam geometry and cross-section directors}
\label{sec:iga_beam_geometry}

In IGA, spline basis functions,
typically B-splines or NURBS, are employed to represent
the geometry and approximate the unknown fields
\cite{Hughes2005}. Their use is particularly attractive
for curved beam formulations, as they enable smooth
centerline representations and provide control over
the continuity of the approximation through the
polynomial degree and knot multiplicities.

Let
\begin{equation}
    \boldsymbol{\Xi}
    = \{\xi_1,\xi_2,\ldots,\xi_{n+p+1}\}
\end{equation}
be a non-decreasing knot vector, where $p$ is the polynomial degree and
$n$ is the number of basis functions. The univariate B-spline basis
functions are defined through the Cox--de Boor recursion
\cite{piegl1997nurbs},
\begin{equation}
    N_{i,0}(\xi)
    =
    \begin{cases}
        1, & \xi_i \leq \xi < \xi_{i+1},\\
        0, & \text{otherwise},
    \end{cases}
\end{equation}
and, for $p\geq 1$,
\begin{equation}
    N_{i,p}(\xi)
    =
    \frac{\xi-\xi_i}{\xi_{i+p}-\xi_i}N_{i,p-1}(\xi)
    +
    \frac{\xi_{i+p+1}-\xi}{\xi_{i+p+1}-\xi_{i+1}}
    N_{i+1,p-1}(\xi).
    \label{eq:bspline_recursion}
\end{equation}
Terms with a zero denominator are taken as zero. For positive weights
$w_i$, the corresponding non-uniform rational B-spline (NURBS) basis
functions are
\begin{equation}
    R_{i,p}(\xi)
    =
    \frac{N_{i,p}(\xi)w_i}
    {\displaystyle\sum_{j=1}^{n}N_{j,p}(\xi)w_j}.
    \label{eq:nurbs_basis}
\end{equation}
These functions are nonnegative, possess local support, and form a
partition of unity. At an interior knot of multiplicity $m$, the spline
space has continuity $C^{p-m}$. Non-empty knot spans define the
integration elements. In the following, an open knot vector is assumed,
with parametric domain
$\widehat{\Omega}=[\xi_{p+1},\xi_{n+1}]$.
At the right endpoint, the basis functions are evaluated using their
left-hand limits.

In the present beam setting, the NURBS curve defined by these basis
functions is identified with the beam centerline. The same parametric
description and basis functions are subsequently used to approximate
the kinematic fields governing its deformation. Hence, the beam geometry
and its unknown fields are described within a common NURBS framework,
in accordance with the isogeometric concept. 

\begin{figure}[h!]
    \centering
    \includegraphics[width=0.80\linewidth]
    {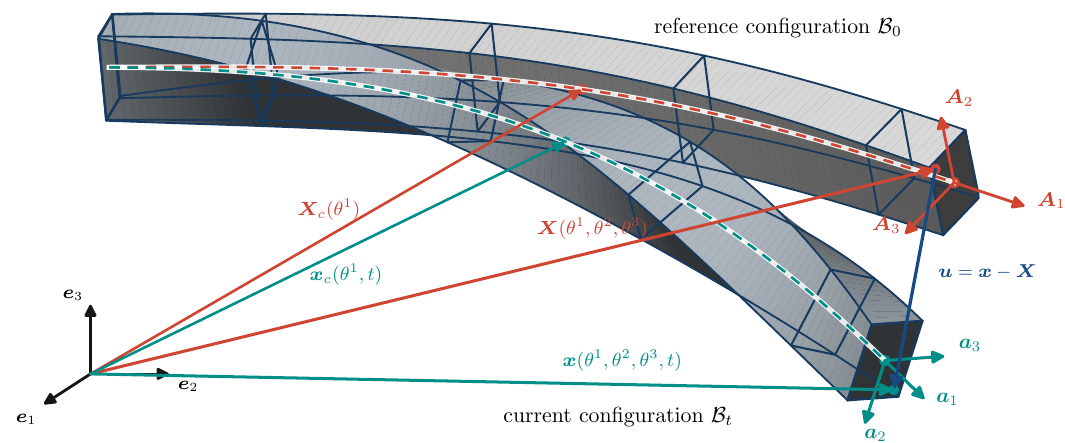}
    \caption{Reference and current beam configurations
    $\mathcal{B}_0$ and $\mathcal{B}_t$.
    The vectors $\boldsymbol{X}_c$ and $\boldsymbol{x}_c$ locate
    corresponding centerline points, whereas $\boldsymbol{X}$ and
    $\boldsymbol{x}$ locate the same off-centerline material point
    before and after deformation. Its displacement is
    $\boldsymbol{u}=\boldsymbol{x}-\boldsymbol{X}$.
    The vectors $\boldsymbol{A}_1$ and $\boldsymbol{a}_1$ denote the
    reference and current centerline tangent vectors, whereas
    $\boldsymbol{A}_2,\boldsymbol{A}_3$ and
    $\boldsymbol{a}_2,\boldsymbol{a}_3$ describe the reference and
    current cross-section orientations.}
    \label{fig:beam_reference_current_frames}
\end{figure}

With this isogeometric beam representation established, the geometry is
described in terms of a reference configuration $\mathcal{B}_0$ and a
current configuration $\mathcal{B}_t$, as illustrated in
Fig.~\ref{fig:beam_reference_current_frames}. Corresponding quantities in
the two configurations are denoted by uppercase and lowercase symbols,
respectively. In particular, $\boldsymbol{X}_c$ and
$\boldsymbol{x}_c$ denote the positions of corresponding points on the
reference and current centerlines with respect to a fixed Cartesian basis
$\{\boldsymbol{e}_1,\boldsymbol{e}_2,\boldsymbol{e}_3\}$, where the
subscript $c$ denotes a centerline quantity.

Accordingly, the reference beam centerline is represented by the NURBS
curve
\begin{equation}
    \boldsymbol{X}_c(\xi)
    =
    \sum_{i=1}^{n}R_{i,p}(\xi)\boldsymbol{P}_i,
    \qquad \xi\in\widehat{\Omega},
    \label{eq:reference_centerline}
\end{equation}
where $\boldsymbol{P}_i\in\mathbb{R}^3$ are the reference control points.
Control points are generally non-interpolatory and should therefore
not be identified with physical points on the centerline or on the
fluid--structure interface.

Using the same basis for the centerline displacement gives
\begin{subequations}
\label{eq:current_centerline}
\begin{align}
    \boldsymbol{u}_c(\xi,t)
    &=
    \sum_{i=1}^{n}R_{i,p}(\xi)\;\boldsymbol{u}_i(t),
    \label{eq:centerline_displacement}\\
    \boldsymbol{x}_c(\xi,t)
    &=
    \boldsymbol{X}_c(\xi)+\boldsymbol{u}_c(\xi,t)
    =
    \sum_{i=1}^{n}R_{i,p}(\xi)
    \bigl[\boldsymbol{P}_i+\boldsymbol{u}_i(t)\bigr].
\end{align}
\end{subequations}
Here, $\boldsymbol{u}_i$ are the displacement coefficients associated
with the control points. The interpolation and parametrization of the
rotational unknowns depend on the beam formulation and are introduced
in the following subsections.

To describe the beam continuum, material coordinates
$(\theta^1,\theta^2,\theta^3)$ are introduced. The longitudinal
coordinate is chosen as the reference arc length,
$\theta^1=s_0$, whereas $\theta^2$ and $\theta^3$ measure signed
distances along the reference cross-section directors. The relation
between the spline parameter $\xi$ and reference arc length is
\begin{equation}
    \mathrm{d}\theta^1
    =
    \mathrm{d}s_0
    =
    J_0(\xi)\,\mathrm{d}\xi,
    \qquad
    J_0(\xi)
    =
    \left\|\boldsymbol{X}_{c,\xi}(\xi)\right\|>0.
    \label{eq:reference_arclength}
\end{equation}
Consequently,
$\partial(\cdot)/\partial\theta^1
=J_0^{-1}\partial(\cdot)/\partial\xi$.
For brevity, the same symbols are used for fields parametrized by either
$\xi$ or $\theta^1$, with derivatives distinguished explicitly by the
corresponding subscript.

The reference and current covariant base vectors tangent to
the beam centerline are defined as
\begin{equation}
    \boldsymbol{A}_1
    =
    \boldsymbol{X}_{c,\theta^1},
    \qquad
    \boldsymbol{a}_1
    =
    \boldsymbol{x}_{c,\theta^1}.
    \label{eq:centerline_tangents}
\end{equation}
Because $\theta^1$ is reference arc length,
$\|\boldsymbol{A}_1\|=1$. In contrast, $\boldsymbol{a}_1$ is generally
not a unit vector. Its norm is the centerline stretch,
\begin{equation}
    \lambda_c
    =
    \|\boldsymbol{a}_1\|
    =
    \frac{\mathrm{d}s}{\mathrm{d}s_0},
    \qquad
    \boldsymbol{t}
    =
    \frac{\boldsymbol{a}_1}{\lambda_c},
    \label{eq:current_tangent_stretch}
\end{equation}
where $s$ is the current arc length and $\boldsymbol{t}$ is the current
unit tangent.

The centerline alone does not determine the orientation of the
cross-section. The reference cross-section is assumed to be
planar and perpendicular to $\mathbf A_1$. Its orientation is
defined by selecting a unit director $\mathbf A_2$ perpendicular
to $\mathbf A_1$, according to the prescribed reference
orientation of the cross-section. The third director
$\mathbf A_3$ is then determined to complete a right-handed
orthonormal basis,
\begin{equation}
    \boldsymbol{A}_i\cdot\boldsymbol{A}_j
    =
    \delta_{ij},
    \qquad
    \boldsymbol{A}_1
    =
    \boldsymbol{A}_2\times\boldsymbol{A}_3,
    \qquad i,j\in\{1,2,3\}.
    \label{eq:reference_material_basis}
\end{equation}
The corresponding current cross-section directors are denoted by
$\boldsymbol{a}_2$ and $\boldsymbol{a}_3$. Under the rigid-cross-section
assumption, they remain mutually orthogonal unit vectors,
\begin{equation}
    \boldsymbol{a}_\alpha\cdot\boldsymbol{a}_\beta
    =
    \delta_{\alpha\beta},
    \qquad \alpha,\beta\in\{2,3\}.
    \label{eq:current_directors}
\end{equation}
Cross-sectional deformation and warping are excluded from this
kinematic description.

\noindent An arbitrary material point, which need not lie on the centerline,
has reference and current position vectors
\begin{subequations}
\label{eq:beam_continuum_placements}
\begin{align}
    \boldsymbol{X}(\theta^1,\theta^2,\theta^3)
    &=
    \boldsymbol{X}_c(\theta^1)
    +\theta^2\boldsymbol{A}_2(\theta^1)
    +\theta^3\boldsymbol{A}_3(\theta^1),
    \label{eq:reference_beam_continuum}\\
    \boldsymbol{x}(\theta^1,\theta^2,\theta^3,t)
    &=
    \boldsymbol{x}_c(\theta^1,t)
    +\theta^2\boldsymbol{a}_2(\theta^1,t)
    +\theta^3\boldsymbol{a}_3(\theta^1,t).
    \label{eq:current_beam_continuum}
\end{align}
\end{subequations}
Thus, $\boldsymbol{X}_c$ and $\boldsymbol{x}_c$ locate the center of
the corresponding cross-section, whereas $\boldsymbol{X}$ and
$\boldsymbol{x}$ locate a material point within that section.
At $\theta^2=\theta^3=0$, the two descriptions coincide. Under the
rigid-cross-section assumption, $\theta^2$ and $\theta^3$ are material
coordinates and therefore remain fixed during deformation, identifying
the same material point in the reference and current configurations.

The displacement of this material point is
\begin{equation}
    \boldsymbol{u}
    =
    \boldsymbol{x}-\boldsymbol{X}
    =
    \boldsymbol{u}_c
    +\theta^2(\boldsymbol{a}_2-\boldsymbol{A}_2)
    +\theta^3(\boldsymbol{a}_3-\boldsymbol{A}_3).
    \label{eq:beam_material_displacement}
\end{equation}
This expression separates centerline translation from the displacement
induced by cross-section rotation. It provides the common kinematic
basis for the mapping operators developed in
Section~\ref{sec:kinematic_mapping}: an interface point is
identified by its centerline location and fixed reference
cross-sectional coordinates, while its current position follows from
the corresponding current centerline and cross-section directors.

Where the reference centerline has nonzero curvature, its Frenet basis
$\{\boldsymbol{A}_1,\boldsymbol{N},\boldsymbol{B}\}$ can be defined from
the centerline geometry. The reference curvature $\kappa_0$ measures the
variation of the unit tangent $\boldsymbol{A}_1$ along the reference arc
length, while $\boldsymbol{N}$ and $\boldsymbol{B}$ denote the corresponding
principal normal and binormal:
\begin{equation}
    \kappa_0
    =
    \left\|\boldsymbol{A}_{1,\theta^1}\right\|,
    \qquad
    \boldsymbol{N}
    =
    \frac{\boldsymbol{A}_{1,\theta^1}}{\kappa_0},
    \qquad
    \boldsymbol{B}
    =
    \boldsymbol{A}_1\times\boldsymbol{N}.
    \label{eq:reference_frenet_frame}
\end{equation}
The material directors describe the reference cross-section orientation
and may differ from the Frenet basis by a prescribed rotation
$\Psi(\theta^1)$ about $\boldsymbol{A}_1$,
\begin{subequations}
\label{eq:material_frame_from_frenet}
\begin{align}
    \boldsymbol{A}_2
    &=
    \cos\Psi\,\boldsymbol{N}
    +\sin\Psi\,\boldsymbol{B},\\
    \boldsymbol{A}_3
    &=
    -\sin\Psi\,\boldsymbol{N}
    +\cos\Psi\,\boldsymbol{B}.
\end{align}
\end{subequations}
Here, $\Psi$ is part of the prescribed reference geometry.

\begin{figure}[h!]
    \centering
    \includegraphics[width=0.50\linewidth]
    {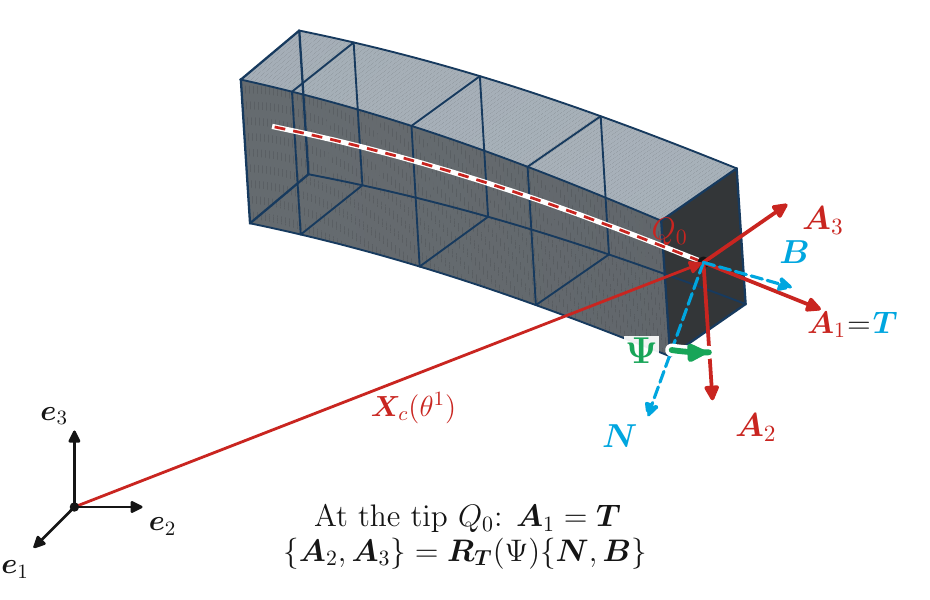}
    \caption{Frenet and material directors at the same reference centerline
    point. The transverse material directors
    $\boldsymbol{A}_2,\boldsymbol{A}_3$ differ from the Frenet directions
    $\boldsymbol{N},\boldsymbol{B}$ by a prescribed rotation $\Psi$
    about the common tangent $\boldsymbol{A}_1$.}
    \label{fig:frenet_material_basis}
\end{figure}

The Frenet frame is undefined where $\kappa_0=0$, including straight
segments. At such locations, the reference material directors may
instead be prescribed directly or obtained using a nonsingular
frame-transport procedure. The Frenet frame is therefore auxiliary,
as the structural model and the mapping only require a common reference
director field.

The preceding description applies to both beam theories considered
here. Their distinction concerns the relation between the current
centerline unit tangent $\boldsymbol{t}$ and the current cross-section
normal $\boldsymbol{a}_2\times\boldsymbol{a}_3$. For Bernoulli
kinematics,
\begin{equation}
    \boldsymbol{a}_2\times\boldsymbol{a}_3
    =
    \boldsymbol{t},
    \label{eq:bernoulli_section_normal}
\end{equation}
so the cross-section remains perpendicular to the current centerline.
For Timoshenko kinematics, the cross-section orientation is an
independent kinematic field and this constraint is relaxed to permit
transverse shear deformation. Consequently,
$\{\boldsymbol{a}_1,\boldsymbol{a}_2,\boldsymbol{a}_3\}$ is generally
not an orthonormal basis: $\boldsymbol{a}_1$ is the current centerline
tangent vector with $\|\boldsymbol{a}_1\|=\lambda_c$ and, in the
Timoshenko case, need not be perpendicular to the cross-section
directors.

\subsection{Geometrically nonlinear isogeometric Bernoulli beam}
\label{sec:bernoulli_beam}

The Bernoulli beam mapping operator developed in this work is derived
consistently from the geometrically nonlinear isogeometric spatial
beam formulation of Bauer et al.~\cite{BAUER2016101}. Therefore, only the
kinematic ingredients required for the subsequent mapping formulation are
summarized here. The complete derivation of the strain measures,
constitutive relations, internal force vector, and consistent tangent
stiffness is provided in the original reference.

The Bernoulli hypothesis requires the cross-sections to remain planar,
undeformed, and perpendicular to the current centerline. Torsional
deformation is admitted, whereas cross-sectional warping is neglected.
With the notation introduced in Section~\ref{sec:kinematic_mapping}, the
current cross-section directors therefore satisfy
\begin{equation}
    \boldsymbol{a}_\alpha\cdot\boldsymbol{t}=0,
    \qquad \alpha\in\{2,3\},
    \qquad
    \boldsymbol{a}_2\times\boldsymbol{a}_3=\boldsymbol{t},
    \label{eq:bernoulli_director_constraint}
\end{equation}
where $\boldsymbol{t}$ is the current unit tangent. Consequently, the
cross-section orientation is determined by the centerline deformation
together with one additional scalar rotation about its tangent.

The independent unknowns are the centerline displacement
$\boldsymbol{u}_c=[u\;v\;w]^{T}$ and the scalar twist field
$\phi$. Using the same NURBS basis functions as for the beam geometry,
they are approximated as
\begin{equation}
    \boldsymbol{u}_c(\xi,t)
    =
    \sum_{i=1}^{n}R_{i,p}(\xi)\;\boldsymbol{u}_{c,i}(t),
    \qquad
    \phi(\xi,t)
    =
    \sum_{i=1}^{n}R_{i,p}(\xi)\;\phi_i(t),
    \label{eq:bernoulli_kinematic_fields}
\end{equation}
such that each control point carries four degrees of freedom,
$\boldsymbol{q}_i = [u_i,\, v_i,\, w_i,\, \phi_i]^{T}$.

To describe the orientation of the cross-section, a material trihedral
$\{\boldsymbol{A}_1,\boldsymbol{A}_2,\boldsymbol{A}_3\}$ is introduced in
the reference configuration, with its first director coinciding with
the reference unit tangent, $\boldsymbol{A}_1=\boldsymbol{T}$.
The remaining directors need not coincide with the Frenet normal
and binormal. Instead, a prescribed initial cross-section orientation
$\Psi$ about the reference tangent is introduced through
\begin{equation}
    \boldsymbol{A}_2
    =
    \boldsymbol{R}_{\boldsymbol{T}}(\Psi)\boldsymbol{N},
    \qquad
    \boldsymbol{A}_3
    =
    \boldsymbol{R}_{\boldsymbol{T}}(\Psi)\boldsymbol{B}.
    \label{eq:bernoulli_initial_material_directors}
\end{equation}
Hence, $\Psi$ describes the initial orientation of the material
cross-section relative to the Frenet frame and is part of the reference
configuration.

Following Bauer et al.~\cite{BAUER2016101}, the current cross-section
directors are reconstructed by first aligning the reference tangent
$\boldsymbol{T}$ with the current tangent $\boldsymbol{t}$ and subsequently
applying the deformation-induced twist $\phi$,
\begin{equation}
    \boldsymbol{a}_\alpha
    =
    \boldsymbol{R}_{\boldsymbol{t}}(\phi)
    \boldsymbol{\Lambda}(\boldsymbol{T},\boldsymbol{t})
    \boldsymbol{A}_\alpha,
    \qquad \alpha\in\{2,3\}.
    \label{eq:bernoulli_director_transport}
\end{equation}
Here, $\boldsymbol{\Lambda}(\boldsymbol{T},\boldsymbol{t})$ denotes the
shortest rotation that maps $\boldsymbol{T}$ onto $\boldsymbol{t}$. With
$\boldsymbol{v}=\boldsymbol{T}\times\boldsymbol{t}$, it can be written as
\begin{equation}
    \boldsymbol{\Lambda}
    =
    \boldsymbol{I}
    +[\boldsymbol{v}]_\times
    +\frac{[\boldsymbol{v}]_\times^2}
    {1+\boldsymbol{T}\cdot\boldsymbol{t}},
    \label{eq:bernoulli_tangent_alignment}
\end{equation}
for $\boldsymbol{T}\cdot\boldsymbol{t}\neq-1$, where
$[\boldsymbol{v}]_\times\boldsymbol{w}
=\boldsymbol{v}\times\boldsymbol{w}$. The subsequent rotation about the
current tangent is given by Rodrigues' formula,
\begin{equation}
    \boldsymbol{R}_{\boldsymbol{t}}(\phi)
    =
    \cos\phi\,\boldsymbol{I}
    +(1-\cos\phi)\boldsymbol{t}\otimes\boldsymbol{t}
    +\sin\phi\,[\boldsymbol{t}]_\times.
    \label{eq:bernoulli_twist_rotation}
\end{equation}
The rotations are active and right-handed and act on vectors expressed in
the fixed Cartesian basis. Thus, $\Psi$ specifies the initial
cross-section orientation, whereas $\phi$ represents the additional
rotation arising during deformation. At the reference state,
$\phi=0$ and $\boldsymbol{t}=\boldsymbol{T}$, such that
$\boldsymbol{a}_\alpha=\boldsymbol{A}_\alpha$.

The centerline interpolation and director reconstruction in
Eqs.~\eqref{eq:bernoulli_kinematic_fields}--%
\eqref{eq:bernoulli_twist_rotation} provide the kinematic description
required by the Bernoulli beam mapping operator introduced in
Section~\ref{sec:kinematic_mapping}.

\subsection{Geometrically nonlinear isogeometric Timoshenko beam}
\label{sec:timoshenko_beam}

The Timoshenko beam mapping operator is based on the geometrically
nonlinear isogeometric spatial beam formulation of
Vo et al.~\cite{Vo2020}. As for the Bernoulli formulation, only
the kinematic quantities required for the subsequent mapping
are presented here.

In contrast to the Bernoulli hypothesis, the Timoshenko formulation
allows the cross-section to rotate independently of the centerline
tangent. Cross-sections remain planar and undeformed but are not
constrained to remain perpendicular to the current beam axis.
Consequently, transverse shear deformation is admitted, and the
cross-section orientation cannot be reconstructed from the
centerline deformation alone.

Accordingly, each control point has six degrees of freedom:
three translational and three rotational. In the geometrically
nonlinear formulation, finite rotations are updated incrementally,
so that the control-point increments are collected as
\begin{equation}
    \Delta\boldsymbol{q}_i
    =
    \begin{bmatrix}
        \Delta\boldsymbol{u}_i^{T} &
        \Delta\boldsymbol{\varphi}_i^{T}
    \end{bmatrix}^{T},
    \label{eq:timoshenko_cp_dofs}
\end{equation}
where $\Delta\boldsymbol{u}_i$ and
$\Delta\boldsymbol{\varphi}_i$ denote the translational
and rotational increments associated with control point $i$,
respectively.

The reference configuration is described by the centerline
$\boldsymbol{X}_c(\xi)$ and the orthonormal material directors
$\{\boldsymbol{A}_1,\boldsymbol{A}_2,\boldsymbol{A}_3\}$ introduced
in Section~\ref{sec:bernoulli_beam}. The current centerline follows
from the translational displacement field,
\begin{equation}
    \boldsymbol{x}_c(\xi,t)
    =
    \boldsymbol{X}_c(\xi)
    +
    \boldsymbol{u}_c(\xi,t),
    \qquad
    \boldsymbol{u}_c(\xi,t)
    =
    \sum_{i=1}^{n}
    R_{i,p}(\xi)\boldsymbol{u}_i(t).
    \label{eq:timoshenko_centerline}
\end{equation}
The corresponding reference and current unit tangents are
\begin{equation}
    \boldsymbol{T}
    =
    \frac{\boldsymbol{X}_{c,\xi}}
         {\|\boldsymbol{X}_{c,\xi}\|},
    \qquad
    \boldsymbol{t}
    =
    \frac{\boldsymbol{x}_{c,\xi}}
         {\|\boldsymbol{x}_{c,\xi}\|}.
    \label{eq:timoshenko_tangents}
\end{equation}

The orientation of the current cross-section is described by an
independent rotation tensor $\boldsymbol{R}(\xi,t)\in SO(3)$,
which maps the reference material directors onto their current
counterparts,
\begin{equation}
    \boldsymbol{a}_\alpha(\xi,t)
    =
    \boldsymbol{R}(\xi,t)\boldsymbol{A}_\alpha(\xi),
    \qquad
    \alpha\in\{2,3\}.
    \label{eq:timoshenko_director_update}
\end{equation}
The directors $\boldsymbol{a}_2$ and $\boldsymbol{a}_3$ remain
orthonormal, but their orthogonality to the current centerline
tangent is not enforced. Thus,
$\boldsymbol{a}_\alpha\cdot\boldsymbol{t}$ is not constrained
to vanish for $\alpha\in\{2,3\}$, allowing transverse shear
deformation.

Following Vo et al.~\cite{Vo2020}, finite cross-sectional
rotations are treated through incremental rotation vectors.
The incremental rotational field is interpolated using the
NURBS basis functions as
\begin{equation}
    \Delta\boldsymbol{\varphi}(\xi)
    =
    \sum_{i=1}^{n}
    R_{i,p}(\xi)\Delta\boldsymbol{\varphi}_i,
    \label{eq:timoshenko_rotation_interpolation}
\end{equation}
Since finite rotations cannot generally be updated additively,
the rotation tensor is updated multiplicatively,
\begin{equation}
    \boldsymbol{R}^{(k+1)}
    =
    \exp\!\left(
        [\Delta\boldsymbol{\varphi}^{(k)}]_\times
    \right)
    \boldsymbol{R}^{(k)},
    \label{eq:timoshenko_rotation_update}
\end{equation}
where $k$ denotes the iteration index of the nonlinear solution
procedure. The exponential map converts the incremental rotation
vector into a proper rotation tensor, preserving
$\boldsymbol{R}\in SO(3)$. It can be evaluated using Rodrigues'
formula,
\begin{equation}
    \exp([\boldsymbol{\varphi}]_\times)
    =
    \boldsymbol{I}
    +
    \frac{\sin\varphi}{\varphi}
    [\boldsymbol{\varphi}]_\times
    +
    \frac{1-\cos\varphi}{\varphi^2}
    [\boldsymbol{\varphi}]_\times^2,
    \qquad
    \varphi=\|\boldsymbol{\varphi}\|.
    \label{eq:timoshenko_rodrigues}
\end{equation}
with the limiting values understood for $\varphi=0$.
The rotations are active and right-handed and act on vectors
expressed in the fixed Cartesian basis.

For the Timoshenko beam mapping, an arbitrary material point
identified by the cross-sectional coordinates
$(\theta^2,\theta^3)$ is represented in the reference
configuration as
\begin{equation}
    \boldsymbol{X}
    =
    \boldsymbol{X}_c
    +
    \theta^2\boldsymbol{A}_2
    +
    \theta^3\boldsymbol{A}_3,
    \label{eq:timoshenko_reference_point}
\end{equation}
and in the current configuration as
\begin{equation}
    \boldsymbol{x}
    =
    \boldsymbol{x}_c
    +
    \theta^2\boldsymbol{a}_2
    +
    \theta^3\boldsymbol{a}_3.
    \label{eq:timoshenko_current_point}
\end{equation}
The position and displacement of any point attached to the
cross-section can therefore be reconstructed from the current
centerline and the independently rotated material directors,
providing the kinematic basis for the Timoshenko beam
mapping developed in Section~\ref{sec:kinematic_mapping}.

\section{Reference attachment and nonlinear interface reconstruction}
\label{sec:kinematic_mapping}

\begin{figure}[h!]
    \centering
    \includegraphics[width=0.80\textwidth]{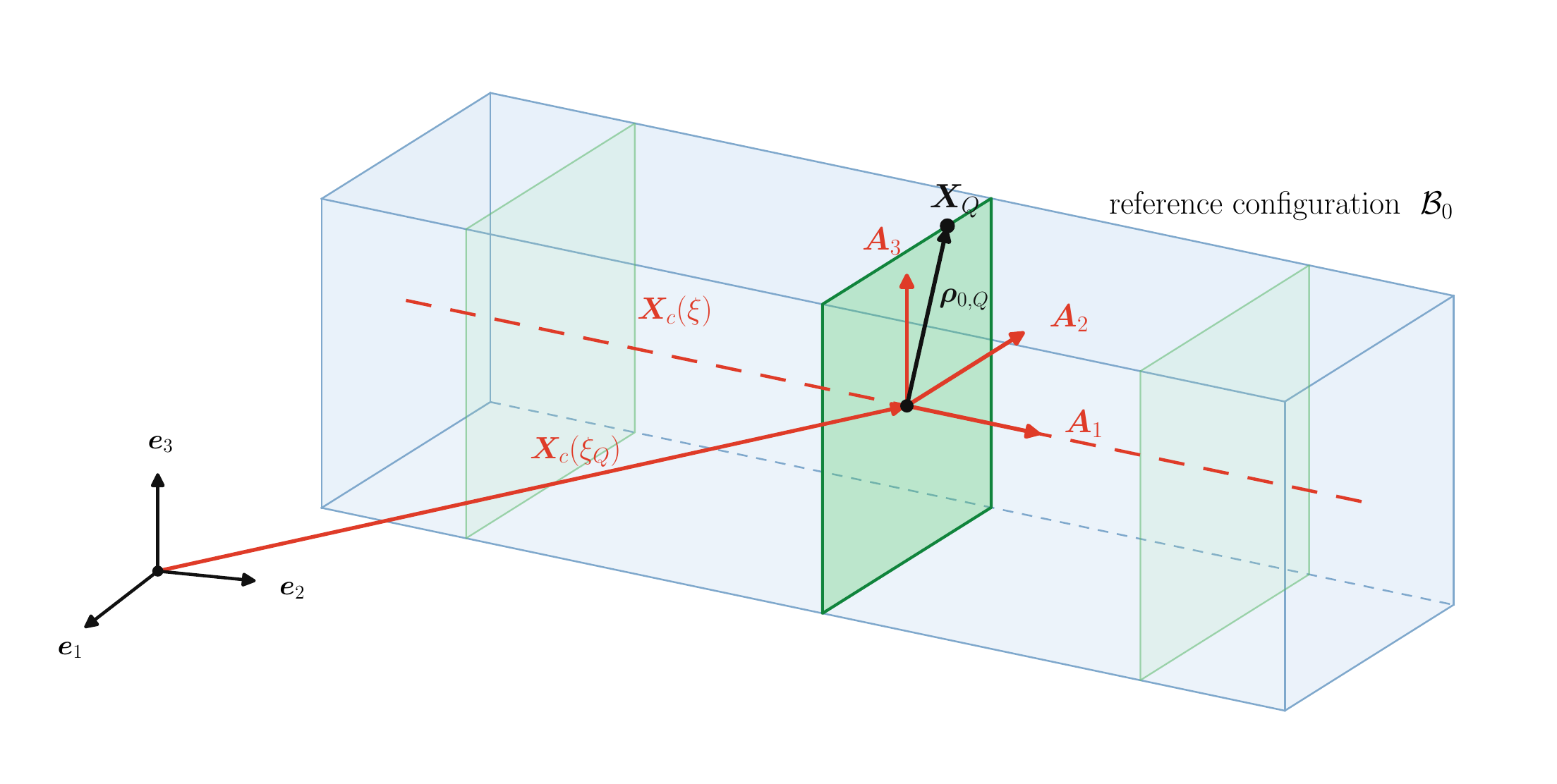}
    \caption{Reference attachment of a surface point $\boldsymbol{X}_Q$ to
    the beam. The projection onto the reference centerline
    $\boldsymbol{X}_c(\xi)$ determines the beam parameter $\xi_Q$ and the
    associated cross-section at $\boldsymbol{X}_c(\xi_Q)$. The material
    directors $\boldsymbol{A}_2$ and $\boldsymbol{A}_3$ span this
    cross-section, while $\boldsymbol{A}_1$ denotes the reference
    centerline tangent. The reference offset
    $\boldsymbol{\rho}_{0,Q}$ connects the projected centerline point to
    the attached surface point.}
    \label{fig:reference_attachment}
\end{figure}

The mapping is established by associating each fluid-side interface
point with the beam in the reference configuration, as illustrated in
Fig.~\ref{fig:reference_attachment}, and preserving this material
attachment throughout the nonlinear deformation. In particular,
each surface point $\boldsymbol{X}_Q$ is associated with a cross-section
through its projection onto the reference beam centerline
$\boldsymbol{X}_c(\xi)$. The attachment procedure is common to the
Bernoulli and Timoshenko formulations introduced above. The two
formulations differ only in the reconstruction of the current
cross-section orientation and, consequently, in the nonlinear map from the
beam configuration to the surface position.

\subsection{Minimum-distance projection onto the reference centerline}
\label{sec:reference_projection}

Let $\boldsymbol{X}_Q$ denote the reference position of a surface point
$Q$, and let $\boldsymbol{X}_c(\xi)$ denote the reference beam centerline.
The longitudinal coordinate of the material attachment is defined by the
closest-point projection onto the reference centerline,
\begin{equation}
    \xi_Q
    =
    \underset{\xi\;\in\;\Omega_b}{\operatorname{argmin}}\,
    d_Q^2(\xi),
    \qquad
    d_Q^2(\xi)
    =
    \left\|
        \boldsymbol{X}_Q-\boldsymbol{X}_c(\xi)
    \right\|^2,
    \label{eq:reference_projection}
\end{equation}
where $\Omega_b=[\xi_{\min},\xi_{\max}]$ denotes the admissible parametric
domain of the NURBS centerline. The corresponding projected point is
$\boldsymbol{X}_c(\xi_Q)$, as illustrated in
Fig.~\ref{fig:reference_attachment}.

For a minimum located in the interior of a smooth portion of $\Omega_b$,
differentiation of the squared distance gives
\begin{equation}
    \frac{\mathrm{d}d_Q^2}{\mathrm{d}\xi}
    =
    -2
    \left(
        \boldsymbol{X}_Q-\boldsymbol{X}_c(\xi)
    \right)
    \cdot
    \boldsymbol{X}_{c,\xi}(\xi).
    \label{eq:projection_distance_derivative}
\end{equation}
Hence, an interior stationary point $\xi^\star$ satisfies
\begin{equation}
    \left(
        \boldsymbol{X}_Q-\boldsymbol{X}_c(\xi^\star)
    \right)
    \cdot
    \boldsymbol{X}_{c,\xi}(\xi^\star)
    =
    0.
    \label{eq:projection_orthogonality}
\end{equation}
Geometrically, the vector connecting the centerline to the surface point
is orthogonal to the centerline tangent at an interior stationary point.
This condition alone does not identify the closest point, since a curved
centerline may possess several stationary points of the distance function.

The NURBS centerline is piecewise smooth over its non-zero knot spans.
\textcolor{black}{A multistart Newton-based search is therefore performed
using initial estimates distributed over these spans to identify
stationary points satisfying Eq.~\eqref{eq:projection_orthogonality}.
This strategy improves the robustness of the projection for curved
centerlines, although it does not guarantee that all stationary
points are identified.}

Stationary interior solutions are not the only possible minimizers of
Eq.~\eqref{eq:reference_projection}. In particular, the closest point
may occur at either end of the finite beam. Therefore, the global beam
endpoints
\begin{equation}
    \xi=\xi_{\min},
    \qquad
    \xi=\xi_{\max},
\end{equation}
\textcolor{black}{are also considered as projection candidates.
At these boundary locations, the orthogonality condition in
Eq.~\eqref{eq:projection_orthogonality} is not required.}

Let $\mathcal{C}_Q$ denote the resulting set of admissible projection
candidates, containing the identified interior stationary points
\textcolor{black}{and any admissible beam endpoints}. The distance is
evaluated for every candidate,
\begin{equation}
    d_{Q,j}
    =
    \left\|
        \boldsymbol{X}_Q
        -
        \boldsymbol{X}_c(\xi_{Q,j})
    \right\|,
    \qquad
    \xi_{Q,j}\in\mathcal{C}_Q,
    \label{eq:projection_candidate_distance}
\end{equation}
and the attachment parameter is selected as the minimum-distance
candidate,
\begin{equation}
    \xi_Q
    =
    \underset{\xi_{Q,j}\;\in\;\mathcal{C}_Q}
    {\operatorname{argmin}}\,
    d_{Q,j}.
    \label{eq:projection_candidate_selection}
\end{equation}
\textcolor{black}{The selected parameter therefore corresponds to the
minimum distance among the identified admissible candidates, without
a formal guarantee of global optimality.}

For strongly curved centerlines, multiple
parametric locations may have the same minimum distance to a
surface point, resulting in a non-unique closest-point projection.
In such cases, the minimum-distance criterion alone is insufficient
to establish a unique material attachment.

Once a \textcolor{black}{projection has been selected}, the parameter
$\xi_Q$ defines the candidate material attachment of surface point $Q$.
The admissibility of this attachment is subsequently verified based
on the position of the surface point relative to the associated
reference cross-section, as described in the following subsection.

\subsection{Material attachment coordinates and section offsets}
\label{sec:material_attachment}

Once $\xi_Q$ has been determined, the reference offset between the
projected centerline point and the surface point is defined as
\begin{equation}
    \boldsymbol{\rho}_{0,Q}
    =
    \boldsymbol{X}_Q
    -
    \boldsymbol{X}_c(\xi_Q).
    \label{eq:reference_offset}
\end{equation}
As illustrated in Fig.~\ref{fig:reference_attachment},
$\boldsymbol{\rho}_{0,Q}$ connects the centerline point
$\boldsymbol{X}_c(\xi_Q)$ with the surface point $\boldsymbol{X}_Q$ in the
reference configuration. Using the reference material directors introduced
in the previous section, the offset is expressed in cross-section
coordinates as
\begin{equation}
    \rho_{2,Q}
    =
    \boldsymbol{\rho}_{0,Q}\cdot
    \boldsymbol{A}_2(\xi_Q),
    \qquad
    \rho_{3,Q}
    =
    \boldsymbol{\rho}_{0,Q}\cdot
    \boldsymbol{A}_3(\xi_Q).
    \label{eq:section_coordinates}
\end{equation}
The coordinates $\rho_{2,Q}$ and $\rho_{3,Q}$ correspond to
the cross-sectional material coordinates $\theta^2$ and
$\theta^3$ introduced in Section~\ref{sec:iga_beam_geometry},
evaluated at the attachment point $Q$.

For an interior minimum-distance projection, the orthogonality
condition in Eq.~\eqref{eq:projection_orthogonality} ensures that
the reference offset is perpendicular to the beam tangent.
At beam endpoints, however, this condition is not guaranteed.
The longitudinal component of the reference offset is therefore
evaluated as
\begin{equation}
    \rho_{1,Q}
    =
    \boldsymbol{\rho}_{0,Q}\cdot\boldsymbol{T}(\xi_Q),
    \label{eq:longitudinal_attachment_offset}
\end{equation}
and the attachment is accepted only if
$|\rho_{1,Q}|\leq\varepsilon_{\mathrm{att}}$,
where $\varepsilon_{\mathrm{att}}$ is a prescribed geometric
tolerance. For accepted attachments, any residual longitudinal
component is discarded.

The reference offset is then expressed in the transverse
cross-section plane spanned by $\boldsymbol{A}_2(\xi_Q)$ and
$\boldsymbol{A}_3(\xi_Q)$, as illustrated in
Fig.~\ref{fig:reference_attachment},
\begin{equation}
    \boldsymbol{\rho}_{0,Q}
    =
    \rho_{2,Q}\boldsymbol{A}_2(\xi_Q)
    +
    \rho_{3,Q}\boldsymbol{A}_3(\xi_Q).
    \label{eq:reference_offset_decomposition}
\end{equation}
Accordingly, the reference position of the surface point is
reconstructed as
\begin{equation}
    \boldsymbol{X}_Q
    =
    \boldsymbol{X}_c(\xi_Q)
    +
    \rho_{2,Q}\boldsymbol{A}_2(\xi_Q)
    +
    \rho_{3,Q}\boldsymbol{A}_3(\xi_Q).
    \label{eq:reference_surface_reconstruction}
\end{equation}

The tuple
\begin{equation}
    \left(\xi_Q,\rho_{2,Q},\rho_{3,Q}\right)
    \label{eq:material_attachment_tuple}
\end{equation}
therefore completely defines the material attachment of point $Q$ to the
beam: $\xi_Q$ identifies the associated beam cross-section, while
$\rho_{2,Q}$ and $\rho_{3,Q}$ locate the point within that cross-section.
These quantities are computed only once in the reference configuration and
remain fixed throughout the subsequent deformation. In particular, the
surface point is not re-projected onto the current centerline, since doing
so would modify its material attachment and introduce artificial sliding
along the beam.

The construction assumes that the coupled surface can be represented
locally by points attached to the transverse cross-sections of the
beam centerline. Accordingly, a unique and sufficiently regular
association between surface points and beam sections must exist.
Geometries for which a surface point has several equally valid
centerline projections, a nonzero longitudinal attachment offset,
or an ill-defined reference material frame fall outside the
present representation unless additional attachment information
is supplied.

\subsection{Reconstruction of the fluid–structure interface for isogeometric Bernoulli beams}
\label{sec:bernoulli_surface_reconstruction}

For the Bernoulli beam, the current cross-section orientation is determined
by the deformation of the centerline together with the scalar twist field
$\phi$. At the stored coordinate $\xi_Q$, the reference tangent
$\boldsymbol{T}$ is first transported onto the current tangent
$\boldsymbol{t}$ by the shortest rotation
$\boldsymbol{\Lambda}(\boldsymbol{T},\boldsymbol{t})$ introduced in
Section~\ref{sec:bernoulli_beam}. The transported cross-section is then
rotated through the finite twist $\phi$ about the current tangent.
Accordingly, the current material directors are
\begin{equation}
    \boldsymbol{a}_\alpha
    =
    \boldsymbol{R}_{\boldsymbol{t}}(\phi)
    \boldsymbol{\Lambda}(\boldsymbol{T},\boldsymbol{t})
    \boldsymbol{A}_\alpha,
    \qquad
    \alpha\in\{2,3\}.
    \label{eq:bernoulli_mapper_directors}
\end{equation}

The material coordinates determined in the reference configuration are
then used to reconstruct the current surface position,
\begin{equation}
    \boldsymbol{x}_Q
    =
    \boldsymbol{x}_c(\xi_Q)
    +
    \rho_{2,Q}\boldsymbol{a}_2(\xi_Q)
    +
    \rho_{3,Q}\boldsymbol{a}_3(\xi_Q)
    \label{eq:bernoulli_surface_position}
\end{equation}
The displacement transferred to the surface is therefore
\begin{equation}
    \boldsymbol{u}_Q
    =
    \boldsymbol{x}_Q-\boldsymbol{X}_Q.
    \label{eq:bernoulli_surface_displacement}
\end{equation}

Equation~\eqref{eq:bernoulli_surface_position} is a geometrically nonlinear
mapping. Bending modifies the current tangent and hence the transported
cross-section, while the interpolated twist field produces an additional
finite rotation about the current tangent. The construction therefore
naturally accounts for curved reference centerlines and spatially varying
twist without introducing independent bending rotations.

The shortest-rotation operator becomes singular when
$\boldsymbol{T}\cdot\boldsymbol{t}=-1$, corresponding to antiparallel
reference and current tangents. At this configuration, the rotation
axis is not uniquely determined by the two tangents. This limitation
could be addressed by incrementally transporting the cross-sectional
frame between consecutive configurations, thereby avoiding the direct
rotation between antiparallel tangents. Such an extension would require
a consistent treatment of the twist field and the corresponding mapping
linearization and is not considered in the present work. Furthermore,
the reference material directors must vary consistently along the beam
to prevent artificial discontinuities in the reconstructed interface.

\subsection{Reconstruction of the fluid–structure interface for isogeometric Timoshenko beams}
\label{sec:timoshenko_surface_reconstruction}

For the Timoshenko beam, the material attachment defined in
Section~\ref{sec:material_attachment} remains unchanged, but the
cross-section orientation is reconstructed differently. In contrast to the
Bernoulli case, the current directors are obtained directly from the
independent cross-sectional rotation tensor $\boldsymbol{R}$,
\begin{equation}
    \boldsymbol{a}_\alpha
    =
    \boldsymbol{R}\boldsymbol{A}_\alpha,
    \qquad
    \alpha\in\{2,3\}.
    \label{eq:timoshenko_mapper_directors}
\end{equation}
The current offset associated with point $Q$ is therefore
\begin{equation}
    \boldsymbol{\rho}_Q
    =
    \rho_{2,Q}\boldsymbol{a}_2(\xi_Q)
    +
    \rho_{3,Q}\boldsymbol{a}_3(\xi_Q)
    =
    \boldsymbol{R}(\xi_Q)\boldsymbol{\rho}_{0,Q}.
    \label{eq:timoshenko_current_offset}
\end{equation}
The current surface position follows as
\begin{equation}
    \boldsymbol{x}_Q
    =
    \boldsymbol{x}_c(\xi_Q)
    +
    \boldsymbol{\rho}_Q
    =
    \boldsymbol{x}_c(\xi_Q)
    +
    \boldsymbol{R}(\xi_Q)\boldsymbol{\rho}_{0,Q}
    \label{eq:timoshenko_surface_position}
\end{equation}
Using the corresponding reference relation, the transferred displacement
can equivalently be written as
\begin{equation}
    \boldsymbol{u}_Q
    =
    \boldsymbol{u}_c(\xi_Q)
    +
    \left[
        \boldsymbol{R}(\xi_Q)-\boldsymbol{I}
    \right]\boldsymbol{\rho}_{0,Q}.
    \label{eq:timoshenko_surface_displacement}
\end{equation}

The essential difference with respect to
Eq.~\eqref{eq:bernoulli_surface_position} is that the Timoshenko
cross-section orientation is not reconstructed from the current centerline
tangent. In general,
\begin{equation}
    \boldsymbol{a}_\alpha\cdot\boldsymbol{t}\neq0,
    \qquad
    \alpha\in\{2,3\},
    \label{eq:timoshenko_mapper_shear}
\end{equation}
and the cross-section may therefore remain inclined with respect to the
current beam axis. This relative orientation represents the transverse
shear kinematics and must be retained by the mapper.

\subsection{Displacement variations and mapping linearization}
\label{sec:mapping_linearization}

Since the material attachment coordinates
$(\xi_Q,\rho_{2,Q},\rho_{3,Q})$ are fixed in the reference configuration,
the variation of the reconstructed interface position results exclusively
from variations of the current beam centerline and cross-sectional
orientation. Moreover, since the reference position
$\boldsymbol{X}_Q$ is fixed,
\begin{equation}
    \delta\boldsymbol{u}_Q
    =
    \delta\boldsymbol{x}_Q
    =
    \delta\boldsymbol{x}_c(\xi_Q)
    +
    \rho_{2,Q}\,\delta\boldsymbol{a}_2(\xi_Q)
    +
    \rho_{3,Q}\,\delta\boldsymbol{a}_3(\xi_Q).
    \label{eq:surface_displacement_variation}
\end{equation}
This expression holds for both the Bernoulli and Timoshenko
formulations. However, the variations of the cross-sectional
directors depend on the underlying beam kinematics, leading
to different mapping operators.

For the Bernoulli formulation, the director variations depend
on both the current centerline tangent and the scalar twist field.
Consequently, the mapping linearization must account for the
contributions of the translational and torsional degrees of freedom
to the motion of the attached interface points.

Let $\boldsymbol{q}_b^{\mathrm{Ber}}$ collect the Bernoulli beam degrees of
freedom,
\begin{equation}
    \boldsymbol{q}_b^{\mathrm{Ber}}
    =
    \begin{bmatrix}
        (\boldsymbol{q}_1^{\mathrm{Ber}})^{\mathsf T} &
        \cdots &
        (\boldsymbol{q}_n^{\mathrm{Ber}})^{\mathsf T}
    \end{bmatrix}^{\mathsf T},
    \qquad
    \boldsymbol{q}_i^{\mathrm{Ber}}
    =
    \begin{bmatrix}
        u_i & v_i & w_i & \phi_i
    \end{bmatrix}^{\mathsf T}.
    \label{eq:bernoulli_global_dofs}
\end{equation}
The nonlinear reconstruction of point $Q$ may then be written as
\begin{equation}
    \boldsymbol{x}_Q
    =
    \mathcal{M}^{\mathrm{Ber}}_Q
    (\boldsymbol{q}_b^{\mathrm{Ber}}),
    \label{eq:bernoulli_nonlinear_point_map}
\end{equation}
and its linearization about the current configuration as
\begin{equation}
    \delta\boldsymbol{x}_Q
    =
    \boldsymbol{H}^{\mathrm{Ber}}_Q
    \,\delta\boldsymbol{q}_b^{\mathrm{Ber}},
    \label{eq:bernoulli_point_linearization}
\end{equation}
where the configuration-dependent tangent operator is formally given by
\begin{equation}
    \boldsymbol{H}^{\mathrm{Ber}}_Q
    =
    \frac{\partial\boldsymbol{x}_c}
         {\partial\boldsymbol{q}_b^{\mathrm{Ber}}}
    +
    \rho_{2,Q}
    \frac{\partial\boldsymbol{a}_2}
         {\partial\boldsymbol{q}_b^{\mathrm{Ber}}}
    +
    \rho_{3,Q}
    \frac{\partial\boldsymbol{a}_3}
         {\partial\boldsymbol{q}_b^{\mathrm{Ber}}}.
    \label{eq:bernoulli_mapping_tangent}
\end{equation}
Unlike the Timoshenko formulation, the Bernoulli beam does not
possess independent three-dimensional rotational degrees of freedom.
Instead, cross-sectional rotations induced by bending are determined
by variations of the centerline tangent, while the scalar twist
degree of freedom accounts for rotation about this tangent.
Both contributions are included in the director derivatives
of $\boldsymbol{H}^{\mathrm{Ber}}_Q$. The detailed linearization
is provided in ~\ref{app:derivatives}.

For the Timoshenko formulation, the cross-sectional orientation is described by the independent rotation tensor $\boldsymbol{R}$ introduced in Section~\ref{sec:timoshenko_beam}. Consistently with the left-multiplicative rotation update in Eq.~\eqref{eq:timoshenko_rotation_update}, the variation of the current directors is expressed in terms of the spatial virtual rotation vector $\delta\boldsymbol{\varphi}$,
\begin{equation}
    \delta\boldsymbol{a}_\alpha
    =
    \delta\boldsymbol{\varphi}
    \times\boldsymbol{a}_\alpha,
    \qquad
    \alpha\in\{2,3\}.
    \label{eq:timoshenko_director_variation}
\end{equation}
Using the current offset vector $\boldsymbol{\rho}_Q$ defined in Eq.~\eqref{eq:timoshenko_current_offset}, the variation of the reconstructed interface position becomes
\begin{equation}
    \delta\boldsymbol{x}_Q
    =
    \delta\boldsymbol{x}_c
    -
    [\boldsymbol{\rho}_Q]_\times
    \delta\boldsymbol{\varphi},
    \label{eq:timoshenko_surface_variation}
\end{equation}
where $[\boldsymbol{\rho}_Q]_\times$ denotes the skew-symmetric matrix associated with the cross product.

The translational and virtual rotational fields are interpolated using the NURBS basis functions,
\begin{equation}
    \delta\boldsymbol{x}_c(\xi_Q)
    =
    \sum_{i=1}^{n}
    R_{i,p}(\xi_Q)\delta\boldsymbol{u}_i,
    \qquad
    \delta\boldsymbol{\varphi}(\xi_Q)
    =
    \sum_{i=1}^{n}
    R_{i,p}(\xi_Q)\delta\boldsymbol{\varphi}_i.
    \label{eq:timoshenko_virtual_interpolation}
\end{equation}
Consequently, the contribution of control point $i$ to the Timoshenko mapping tangent is
\begin{equation}
    \boldsymbol{H}^{\mathrm{Tim}}_{Qi}
    =
    R_{i,p}(\xi_Q)
    \begin{bmatrix}
        \boldsymbol{I} &
        -[\boldsymbol{\rho}_Q]_\times
    \end{bmatrix}.
    \label{eq:timoshenko_mapping_tangent}
\end{equation}
The complete pointwise tangent $\boldsymbol{H}^{\mathrm{Tim}}_Q$ follows by assembling the contributions of all control points whose basis functions are non-zero at $\xi_Q$. The detailed derivation is provided in \ref{app:derivatives}.

The physical virtual rotation $\delta\boldsymbol{\phi}$ is
expressed in the spatial Cartesian frame, consistently with
the left-multiplicative rotation update of the Timoshenko beam.
It corresponds to the infinitesimal variation of the solver's
incremental rotation evaluated at zero increment. Consequently,
the rotational contribution to the mapping tangent acts directly
on the rotational increments of the structural solver, without
requiring an additional transformation.

For both beam formulations, the tangent operator is evaluated
at the current configuration and represents the linearization
of the nonlinear interface reconstruction. It maps admissible
virtual beam variations onto the corresponding virtual interface
motion. The tangent operator does not replace the finite position
maps in Eqs.~\eqref{eq:bernoulli_surface_position} and
\eqref{eq:timoshenko_surface_position}. As shown in the following
section, its transpose provides the corresponding work-conjugate
interface-to-beam force mapping.

\section{Work-conjugate transfer of interface loads}
\label{sec:load_transfer}

The nonlinear interface reconstruction introduced in the previous
section determines the kinematic transfer from the beam to the coupled
FSI interface. For the transfer of interface loads in the opposite
direction, the corresponding mapping operator is obtained from the
principle of virtual work. This ensures that the fluid-interface forces
and the generalized beam forces are work-conjugate, preserving the
discrete virtual work associated with admissible variations of the
beam configuration.

The fluid traction acting on the fluid--structure interface
$\Gamma_{\mathrm{FSI}}$ is obtained from the fluid Cauchy stress tensor as
\begin{equation}
    \boldsymbol{t}_f
    =
    \boldsymbol{\sigma}_f \boldsymbol{n}_f,
    \qquad
    \boldsymbol{\sigma}_f
    =
    -p\boldsymbol{I}
    +
    2\mu_f\boldsymbol{\varepsilon}(\boldsymbol{v}_f),
    \label{eq:fluid_traction}
\end{equation}
where $p$ is the fluid pressure, $\mu_f$ is the dynamic viscosity,
$\boldsymbol{\varepsilon}(\boldsymbol{v}_f)$ is the fluid strain-rate
tensor, and $\boldsymbol{n}_f$ denotes the outward unit normal to
the fluid domain. The corresponding discrete force acting on an
interface node $Q$ is obtained by integrating the traction over
the fluid--structure interface,
\begin{equation}
    \boldsymbol{f}_Q
    =
    -\int_{\Gamma_{\mathrm{FSI}}}
    N_Q^f\,\boldsymbol{t}_f
    \,\mathrm{d}\Gamma,
    \label{eq:fluid_nodal_force}
\end{equation}
where $N_Q^f$ is the fluid-interface shape function associated
with node $Q$. The negative sign accounts for the force exerted
by the fluid on the structure.

Let $\delta\boldsymbol{x}_Q$ be an admissible variation of the
current position of interface node $Q$. The virtual work
associated with the corresponding nodal force is
\begin{equation}
    \delta W_Q
    =
    \delta\boldsymbol{x}_Q^{\mathrm{T}}
    \boldsymbol{f}_Q.
    \label{eq:surface_virtual_work}
\end{equation}
The linearization of the nonlinear mapping operator can be
written in compact form as
\begin{equation}
    \delta\boldsymbol{x}_Q
    =
    \boldsymbol{H}_Q\,\delta\boldsymbol{q},
    \label{eq:generic_mapping_tangent}
\end{equation}
where $\delta\boldsymbol{q}$ collects the admissible variations of the
beam degrees of freedom and $\boldsymbol{H}_Q$ denotes the pointwise
mapping tangent evaluated at the current beam configuration. Depending on
the adopted beam kinematics, $\boldsymbol{H}_Q$ corresponds to the
Bernoulli or Timoshenko linearization derived in the preceding section.

Substitution of Eq.~\eqref{eq:generic_mapping_tangent} into
Eq.~\eqref{eq:surface_virtual_work} yields
\begin{equation}
    \delta W_Q
    =
    \delta\boldsymbol{q}^{\mathrm{T}}
    \boldsymbol{H}_Q^{\mathrm{T}}
    \boldsymbol{f}_Q.
    \label{eq:mapped_virtual_work}
\end{equation}
Requiring the same virtual work on the beam side,
\begin{equation}
    \delta W_Q
    =
    \delta\boldsymbol{q}^{\mathrm{T}}
    \boldsymbol{f}_{b,Q},
    \label{eq:beam_virtual_work}
\end{equation}
gives the work-conjugate load transfer
\begin{equation}
    \boldsymbol{f}_{b,Q}
    =
    \boldsymbol{H}_Q^{\mathrm{T}}
    \boldsymbol{f}_Q
    \label{eq:work_conjugate_load_transfer}
\end{equation}
where $\boldsymbol{f}_{b,Q}$ denotes the contribution of surface point
$Q$ to the generalized beam force vector. Thus, the transpose of the
current kinematic mapping tangent provides the consistent transfer of
surface forces to the beam degrees of freedom.

For the Timoshenko formulation, the pointwise variation derived in
Section~\ref{sec:kinematic_mapping} is
\begin{equation}
    \delta\boldsymbol{x}_Q
    =
    \sum_{i=1}^{n}
    R_{i,p}(\xi_Q)
    \begin{bmatrix}
        \boldsymbol{I} &
        -[\boldsymbol{\rho}_Q]_\times
    \end{bmatrix}
    \begin{bmatrix}
        \delta\boldsymbol{u}_i\\
        \delta\boldsymbol{\varphi}_i
    \end{bmatrix}.
    \label{eq:timoshenko_variation_load_transfer}
\end{equation}
Consequently, the force contribution associated with control point $i$
follows from the transpose of the corresponding Timoshenko mapping
tangent,
\begin{equation}
    \begin{bmatrix}
        \boldsymbol{f}_{i,Q}\\
        \boldsymbol{m}_{i,Q}
    \end{bmatrix}
    =
    R_{i,p}(\xi_Q)
    \begin{bmatrix}
        \boldsymbol{I}\\
        [\boldsymbol{\rho}_Q]_\times
    \end{bmatrix}
    \boldsymbol{f}_Q,
    \label{eq:timoshenko_force_transfer}
\end{equation}
where $\boldsymbol{f}_{i,Q}$ and $\boldsymbol{m}_{i,Q}$ are the
translational-force and rotational-moment contributions, respectively.
Using
$[\boldsymbol{\rho}_Q]_\times^{\mathrm{T}}
=-[\boldsymbol{\rho}_Q]_\times$, the individual contributions can
equivalently be written as
\begin{equation}
    \boldsymbol{f}_{i,Q}
    =
    R_{i,p}(\xi_Q)\boldsymbol{f}_Q,
    \qquad
    \boldsymbol{m}_{i,Q}
    =
    R_{i,p}(\xi_Q)
    \left(
        \boldsymbol{\rho}_Q
        \times
        \boldsymbol{f}_Q
    \right).
    \label{eq:timoshenko_force_moment_transfer}
\end{equation}
Hence, a force acting away from the beam centerline generates
both a resultant force and the corresponding moment about the
centerline.
These moments are expressed in the spatial
Cartesian frame and are work-conjugate to the rotational
variations of the Timoshenko beam.

The same work-conjugacy argument applies to the Bernoulli formulation.
In this case, however, the rotational variation of the cross-section is
not an independent three-component field, since the Bernoulli constraint
relates the cross-section orientation to the centerline tangent together
with the scalar twist field. Therefore, the corresponding generalized
force and moment contributions are obtained directly by applying the
transpose of the Bernoulli mapping tangent derived in the previous
section,
\begin{equation}
    \boldsymbol{f}_{b,Q}^{\mathrm{Ber}}
    =
    \left(
        \boldsymbol{H}_Q^{\mathrm{Ber}}
    \right)^{\mathrm{T}}
    \boldsymbol{f}_Q.
    \label{eq:bernoulli_force_transfer}
\end{equation}
This automatically accounts for the dependence of the Bernoulli
cross-section rotation on the centerline degrees of freedom. The resulting rotational generalized forces
are scalar quantities conjugate to the twist variations,
while bending-related contributions are transferred through
the translational degrees of freedom.

For an interface containing several surface points, the total generalized
beam load is obtained by assembling the contributions from all points,
\begin{equation}
    \boldsymbol{f}_b
    =
    \sum_{Q=1}^{n_Q}
    \boldsymbol{H}_Q^{\mathrm{T}}
    \boldsymbol{f}_Q.
    \label{eq:assembled_load_transfer}
\end{equation}
Equivalently, after assembling the pointwise mapping tangents into a
global operator $\boldsymbol{H}$ and collecting the surface forces in
$\boldsymbol{f}_s$, the kinematic and load transfers take the compact
forms
\begin{equation}
    \delta\boldsymbol{x}_s
    =
    \boldsymbol{H}\,\delta\boldsymbol{q},
    \qquad
    \boldsymbol{f}_b
    =
    \boldsymbol{H}^{\mathrm{T}}\boldsymbol{f}_s.
    \label{eq:global_kinematic_load_transfer}
\end{equation}
It follows immediately that
\begin{equation}
    \delta\boldsymbol{x}_s^{\mathrm{T}}\boldsymbol{f}_s
    =
    \delta\boldsymbol{q}^{\mathrm{T}}\boldsymbol{f}_b,
    \label{eq:global_work_conservation}
\end{equation}
which establishes discrete virtual-work conjugacy between the
surface and beam representations. This identity holds for
admissible variations at a fixed configuration and does not,
by itself, imply energy conservation over finite time steps,
which also depends on the temporal discretization and the
fluid--structure coupling algorithm.

An important consequence of the geometrically nonlinear formulation is
that $\boldsymbol{H}$ is configuration-dependent. Although the material
attachment coordinates
$(\xi_Q,\rho_{2,Q},\rho_{3,Q})$ remain fixed, the current offset and
cross-section orientation change during the deformation. The mapping
tangent must therefore be evaluated consistently at the current beam
configuration when transferring interface loads. Using the transpose of
this current tangent ensures that the load transfer remains
work-conjugate to the nonlinear kinematic reconstruction throughout the
deformation.

\section{Verification of the mapping operators}
\label{sec:mapping_verification}

\begin{figure}[h!]
    \centering
    \includegraphics[width=0.95\linewidth]{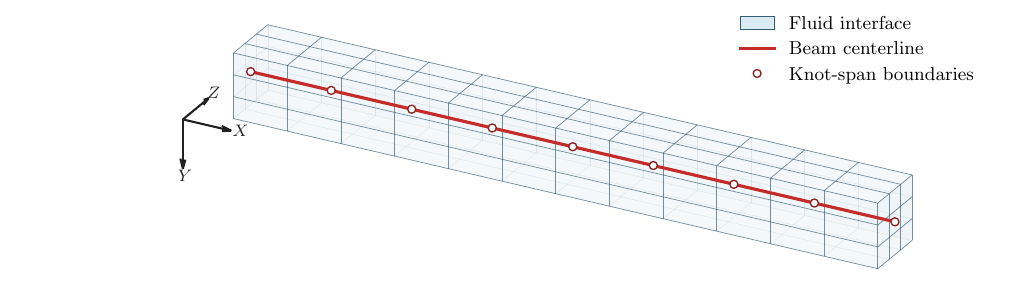}
    \caption{Schematic representation of the isogeometric beam
    discretization embedded within the surrounding fluid--structure
    interface mesh. Markers along the beam centerline indicate the
    knot-span boundaries. A coarsened surface discretization is shown
    for visualization purposes.}
    \label{fig:mapping_verification_discretizations}
\end{figure}

The proposed Bernoulli and Timoshenko beam mapping operators are
verified using prescribed analytical displacement and rotation fields. All
verification examples employ the same reference geometry, consisting of a
straight beam of length $L=10$ aligned with the global $x$-axis and a
surrounding prismatic surface with a square cross-section of unit width and
height,
\begin{equation}
    0\leq X\leq L,
    \qquad
    -0.5\leq Y\leq0.5,
    \qquad
    -0.5\leq Z\leq0.5.
    \label{eq:verification_geometry}
\end{equation}
Figure~\ref{fig:mapping_verification_discretizations} illustrates the
one-dimensional beam discretization embedded within the surrounding
fluid--structure interface mesh. The surface discretization used in the
verification studies is kept fixed and consists of $101$ sections along
the beam axis and $11$ points along each cross-section direction,
retaining only the points on the boundary of the prism. The beam
discretization, in contrast, is successively refined to assess the
convergence of the mapping operators.

For each verification test, the prescribed analytical displacement field
is represented on the isogeometric beam through interpolation at the
Greville abscissae. This procedure determines the control-point
coefficients of the NURBS approximation from the analytical solution,
without requiring the solution of the structural equilibrium equations.
The resulting discrete beam fields are subsequently transferred to the
surface using the proposed mapping operators.

The mapped displacement $\boldsymbol{u}^{\mathrm{map}}_Q$ at each surface
point $Q$ is compared with the analytical displacement
$\boldsymbol{u}^{\mathrm{ex}}(\boldsymbol{X}_Q)$ evaluated directly at
the same reference position. The pointwise error is defined as
$\boldsymbol{u}^{\mathrm{ex}}(\boldsymbol{X}_Q)$ evaluated directly at the
same reference position. The pointwise error is defined as
\begin{equation}
    \boldsymbol{e}_Q
    =
    \boldsymbol{u}^{\mathrm{map}}_Q
    -
    \boldsymbol{u}^{\mathrm{ex}}(\boldsymbol{X}_Q).
    \label{eq:mapping_pointwise_error}
\end{equation}
The overall accuracy is quantified by the root-mean-square error over all
$n_Q$ surface points and the three spatial components,
\begin{equation}
    e_{\mathrm{RMS}}
    =
    \sqrt{
        \frac{1}{3n_Q}
        \sum_{Q=1}^{n_Q}
        \left\|
            \boldsymbol{e}_Q
        \right\|^2
    },
    \label{eq:mapping_rms_error}
\end{equation}
Theis measure is used in the following examples to assess the accuracy
and convergence of both mapping formulations under spline refinement.

\subsection{Pure bending}
\label{sec:verification_bending}

The first verification case considers the pure bending of the initially
straight beam into a circular arc in the $x$--$y$ plane, following the
bending test of Wang et al.~\cite{Wang2016}. The deformation preserves the
centerline length $L$, and the total rotation of the cross-section at the
right end is prescribed by the bending angle $\alpha$. Consequently, the
curvature is constant and the radius of the deformed centerline is
\begin{equation}
    r=\frac{L}{\alpha},
    \label{eq:bending_radius}
\end{equation}
where $\alpha$ is expressed in radians.

The cross-section rotation varies linearly along the initially straight
centerline according to
\begin{equation}
    \theta(X)=\frac{\alpha X}{L}.
    \label{eq:bending_rotation}
\end{equation}
The exact displacement of the beam centerline is therefore
\begin{equation}
    \boldsymbol{u}^{\mathrm{ex}}_c(X)
    =
    \begin{bmatrix}
        r\sin\theta(X)-X\\[1mm]
        r\left[\cos\theta(X)-1\right]\\[1mm]
        0
    \end{bmatrix}.
    \label{eq:bending_centerline_displacement}
\end{equation}
Thus, the left end remains fixed, whereas the initially straight
centerline is transformed into a circular arc with constant curvature
$\alpha/L$.

To evaluate the proposed mapping, the analytical solution is
extended from the beam centerline to the entire prismatic surface. A surface
point with reference coordinates
$\boldsymbol{X}=(X,Y,Z)^{\mathrm{T}}$ follows the translation of the
centerline and the rotation of its associated cross-section. Its exact
current position is therefore
\begin{equation}
    \boldsymbol{x}^{\mathrm{ex}}(\boldsymbol{X})
    =
    \begin{bmatrix}
        (r+Y)\sin\theta(X)\\[1mm]
        r\left[\cos\theta(X)-1\right]
        +Y\cos\theta(X)\\[1mm]
        Z
    \end{bmatrix},
    \label{eq:bending_exact_position}
\end{equation}
and the corresponding analytical displacement is
\begin{equation}
    \boldsymbol{u}^{\mathrm{ex}}(\boldsymbol{X})
    =
    \boldsymbol{x}^{\mathrm{ex}}(\boldsymbol{X})
    -
    \boldsymbol{X}.
    \label{eq:bending_exact_displacement}
\end{equation}
For $Y=0$, Eq.~\eqref{eq:bending_exact_position} reduces to the circular
centerline deformation, whereas for $Y\neq0$ it additionally accounts for
the displacement induced by the rotation of the cross-section.

Three bending angles,
$\alpha=20^\circ$, $40^\circ$, and $60^\circ$, are considered. Uniform
beam discretizations ranging from $2$ to $64$ knot spans are employed,
while the surface discretization is kept fixed. The characteristic beam
discretization size is defined as
\begin{equation}
    h=\frac{L}{n_{\mathrm{span}}},
    \label{eq:bending_mesh_size}
\end{equation}
and the mapped surface displacements are compared with
Eq.~\eqref{eq:bending_exact_displacement} using the error measure
introduced in Eqs.~\eqref{eq:mapping_rms_error}.

The Bernoulli mapper is considered first. The analytical centerline
displacement is prescribed on the beam, while the independent twist field
is set to zero. The cross-section orientation required for the surface
reconstruction is therefore obtained from the tangent of the deformed
centerline, consistently with the Bernoulli kinematic constraint. The test
thus assesses both the transfer of the centerline motion and the
derivative-based reconstruction of the cross-section orientation.

\begin{figure}[h!]
    \centering
    \begin{subfigure}[t]{0.32\textwidth}
        \centering
        \includegraphics[width=\textwidth]{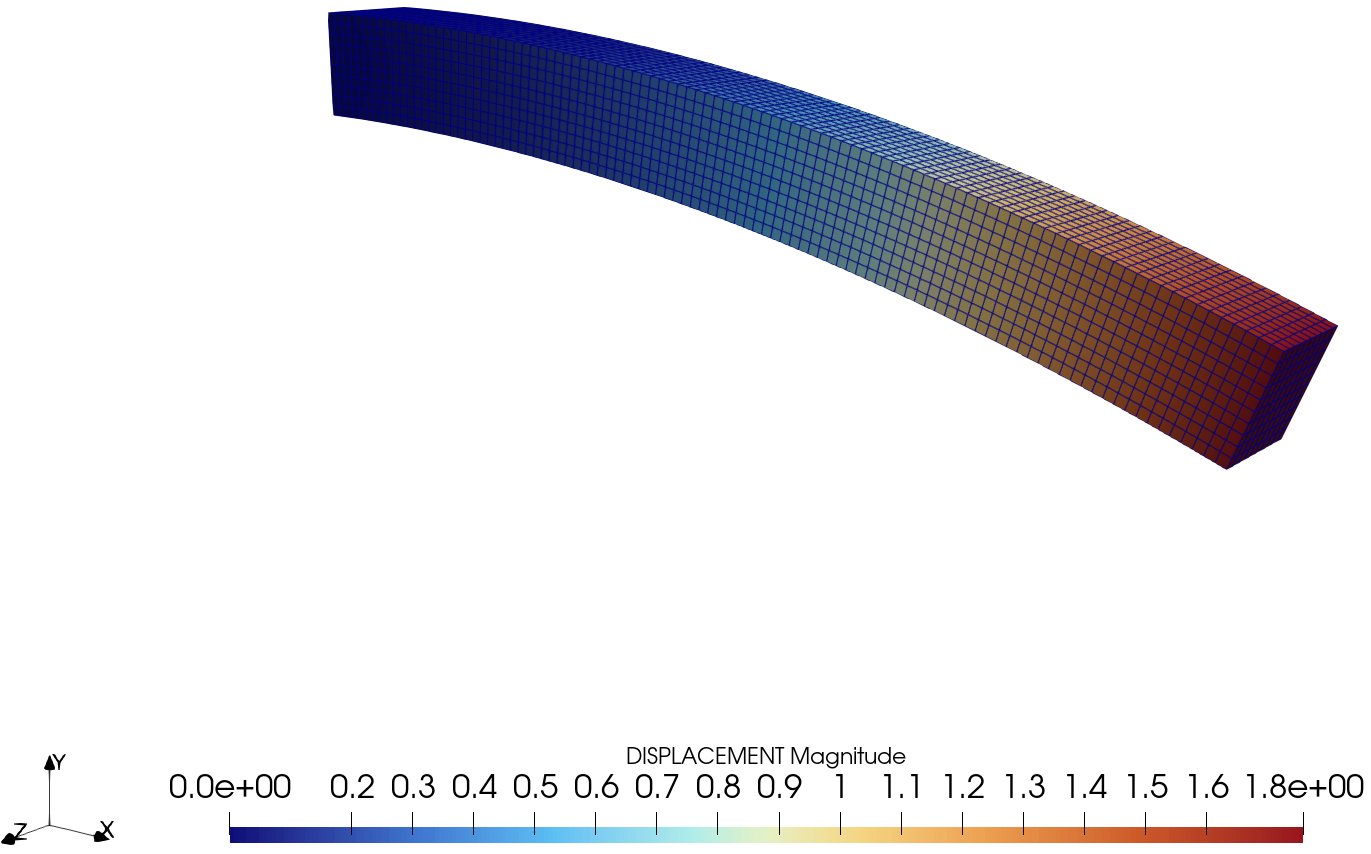}
        \caption{$\alpha=20^\circ$.}
    \end{subfigure}
    \hfill
    \begin{subfigure}[t]{0.32\textwidth}
        \centering
        \includegraphics[width=\textwidth]{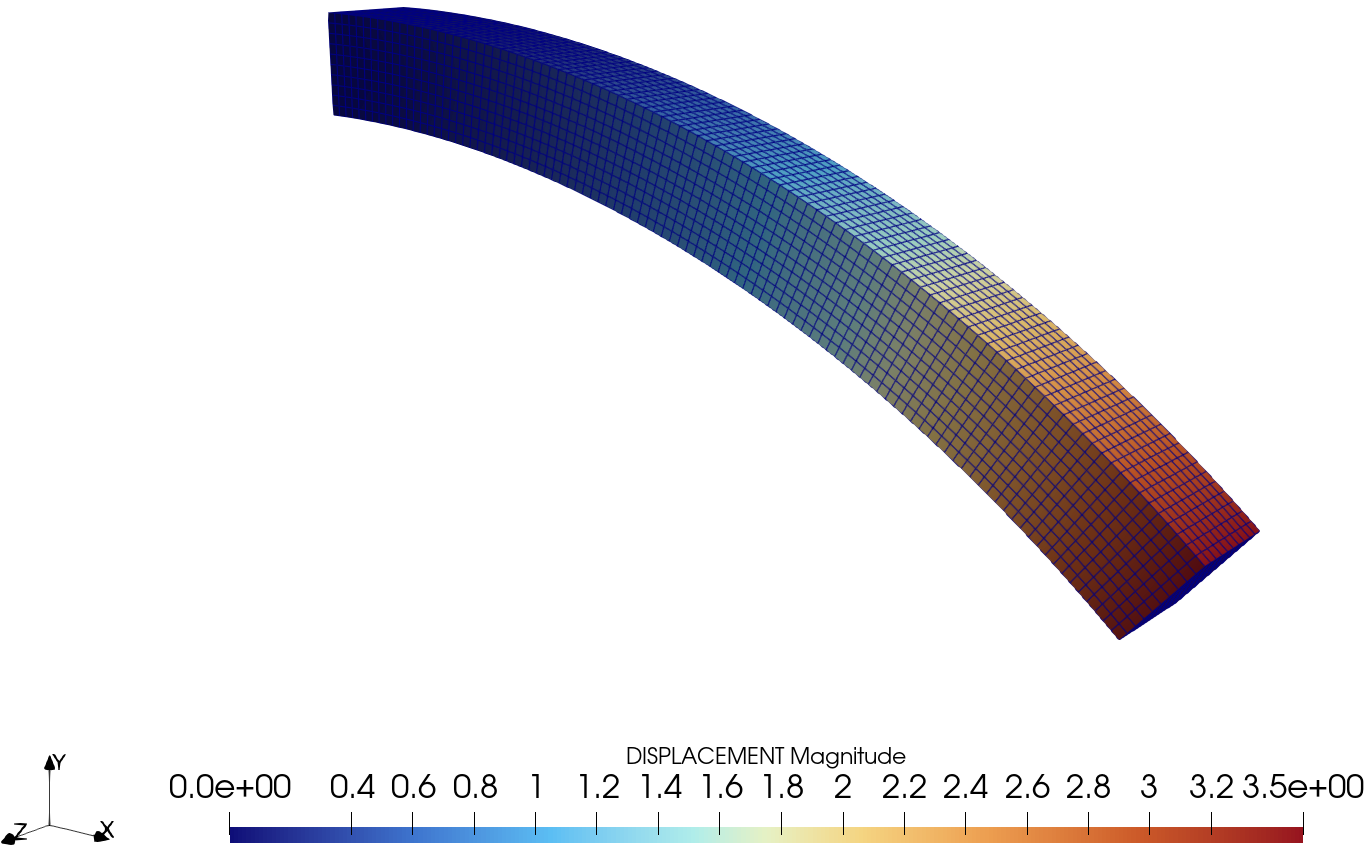}
        \caption{$\alpha=40^\circ$.}
    \end{subfigure}
    \hfill
    \begin{subfigure}[t]{0.32\textwidth}
        \centering
        \includegraphics[width=\textwidth]{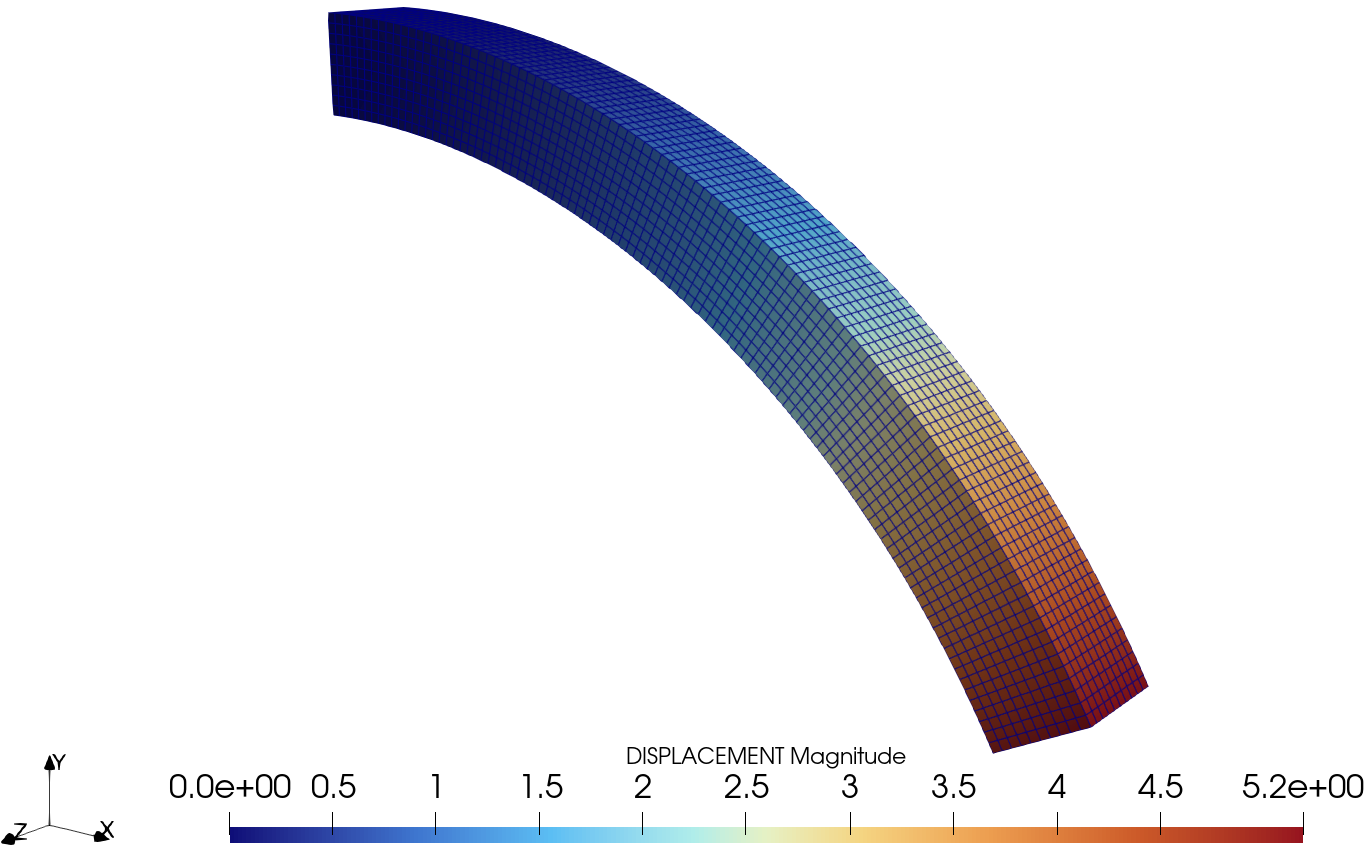}
        \caption{$\alpha=60^\circ$.}
    \end{subfigure}
    \caption{Surface configurations obtained with the Bernoulli
    beam mapper for the pure-bending test using the finest beam
    discretization with $64$ knot spans: (a) $\alpha=20^\circ$,
    (b) $\alpha=40^\circ$, and (c) $\alpha=60^\circ$. The surface is
    colored by the magnitude of the mapped displacement.}
    \label{fig:bernoulli_bending_deformed_configurations}
\end{figure}

Figure~\ref{fig:bernoulli_bending_deformed_configurations} shows the
surface configurations obtained with the Bernoulli mapper for the three
bending angles using the finest beam discretization with $64$ knot spans and $p=2$.
The surface is colored by the magnitude of the mapped displacement. With
increasing $\alpha$, the surface follows the prescribed circular
centerline deformation while the attached cross-sections rotate according
to the orientation reconstructed by the mapper.

\begin{figure}[h!]
    \centering
    \begin{subfigure}[t]{0.32\textwidth}
        \centering
        \includegraphics[width=\textwidth]
        {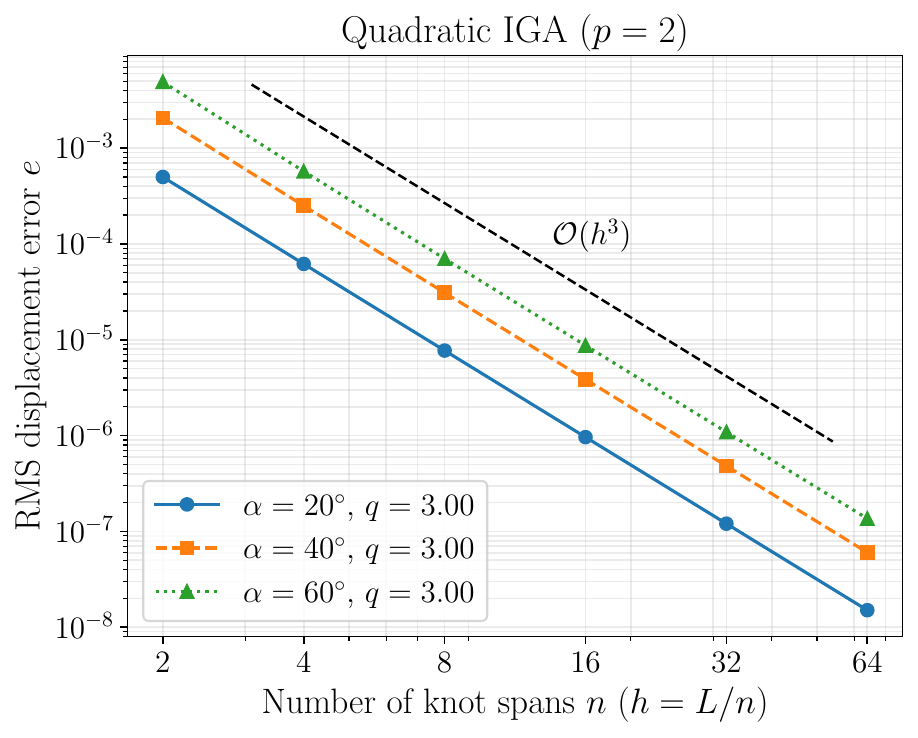}
        \caption{$p=2$.}
    \end{subfigure}
    \hfill
    \begin{subfigure}[t]{0.32\textwidth}
        \centering
        \includegraphics[width=\textwidth]
        {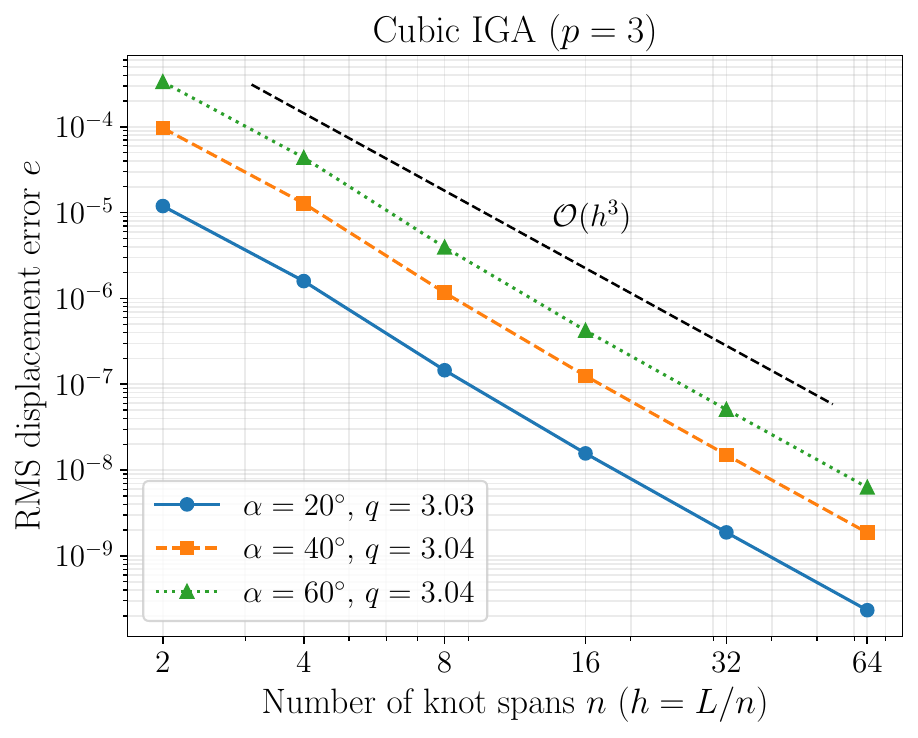}
        \caption{$p=3$.}
    \end{subfigure}
    \hfill
    \begin{subfigure}[t]{0.32\textwidth}
        \centering
        \includegraphics[width=\textwidth]
        {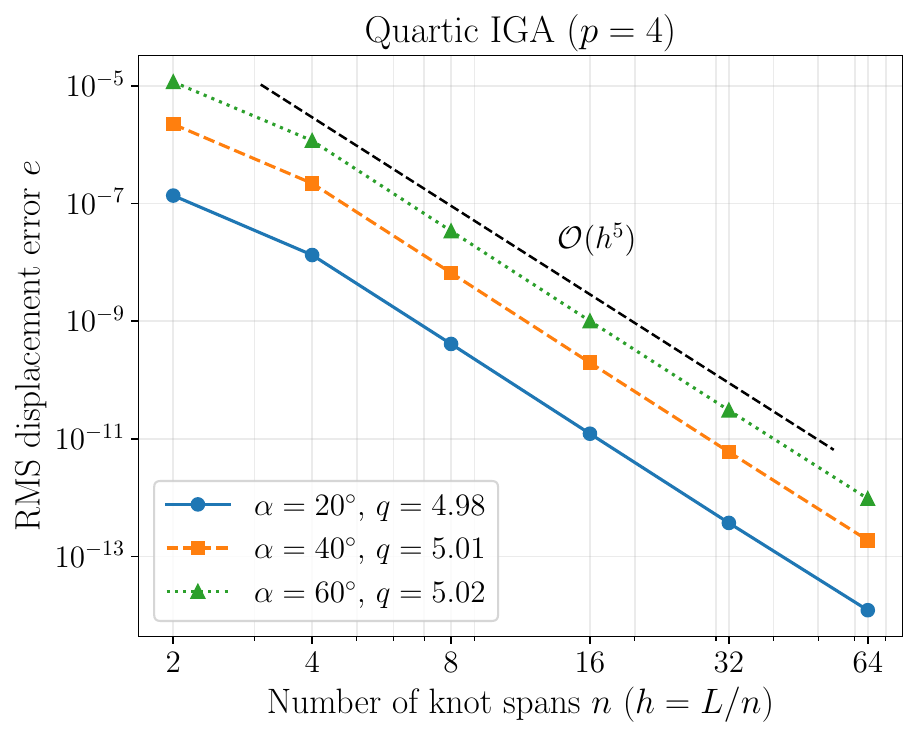}
        \caption{$p=4$.}
    \end{subfigure}
    \caption{Convergence of the RMS displacement error for the Bernoulli
    beam mapper under pure bending for
    $\alpha=20^\circ$, $40^\circ$, and $60^\circ$:
    (a) $p=2$, (b) $p=3$, and (c) $p=4$.}
    \label{fig:bernoulli_bending_convergence}
\end{figure}

The quantitative convergence of the Bernoulli mapper is shown in
Fig.~\ref{fig:bernoulli_bending_convergence} for spline degrees $p=2$,
$3$, and $4$. For $p=2$, an order of $q=3.00$ is obtained for all three
bending angles. For $p=3$, the corresponding orders are $q=3.03$, $3.04$,
and $3.04$ for $\alpha=20^\circ$, $40^\circ$, and $60^\circ$,
respectively, whereas for $p=4$ they increase to $q=4.98$, $5.01$, and
$5.02$. The convergence rates are therefore essentially independent of
the considered bending angle, although the magnitude of the mapping error
increases with $\alpha$.

The results exhibit an even--odd pairing of the convergence rates,
with approximately $\mathcal{O}(h^3)$ for both $p=2$ and $p=3$,
and $\mathcal{O}(h^5)$ for $p=4$. In general, the convergence
of the Bernoulli mapping is governed by the derivative-based
reconstruction of the cross-section orientation, leading to an
expected order of $\mathcal{O}(h^p)$ for smooth deformations.

The enhanced convergence for even polynomial degrees is associated
with the constant-curvature bending deformation considered here.
For a circular centerline, the unit tangent rotates at a constant
rate $\kappa$, such that its derivatives satisfy
$\boldsymbol{t}''=-\kappa^2\boldsymbol{t}$ and, more generally,
$\boldsymbol{t}^{(2m)}=(-1)^m\kappa^{2m}\boldsymbol{t}$.
Thus, even-order derivatives are parallel to the tangent itself.
When the leading approximation error is tangent-parallel,
normalization eliminates this contribution to first order, since
$(\boldsymbol{I}-\boldsymbol{t}\otimes\boldsymbol{t})
\boldsymbol{t}=\boldsymbol{0}$.
This provides an explanation for the additional order of
convergence observed for even $p$, which is associated with
the particular bending deformation rather than a general
property of the Bernoulli mapping.

The Timoshenko mapper is considered next using the same analytical bending
field and discretizations. In contrast to the Bernoulli formulation, the
cross-section rotation constitutes an independent beam field and is
therefore prescribed together with the centerline displacement. For the
present constant-curvature deformation, the rotation field
$\theta(X)=\alpha X/L$ is linear along the beam and can consequently be
represented exactly by the employed spline spaces. The surface
reconstruction is then obtained from the independently interpolated
centerline displacement and cross-section rotation according to the
Timoshenko kinematics introduced in
Section~\ref{sec:timoshenko_surface_reconstruction}.

\begin{figure}[h]
    \centering
    \begin{subfigure}[t]{0.32\textwidth}
        \centering
        \includegraphics[width=\textwidth]
        {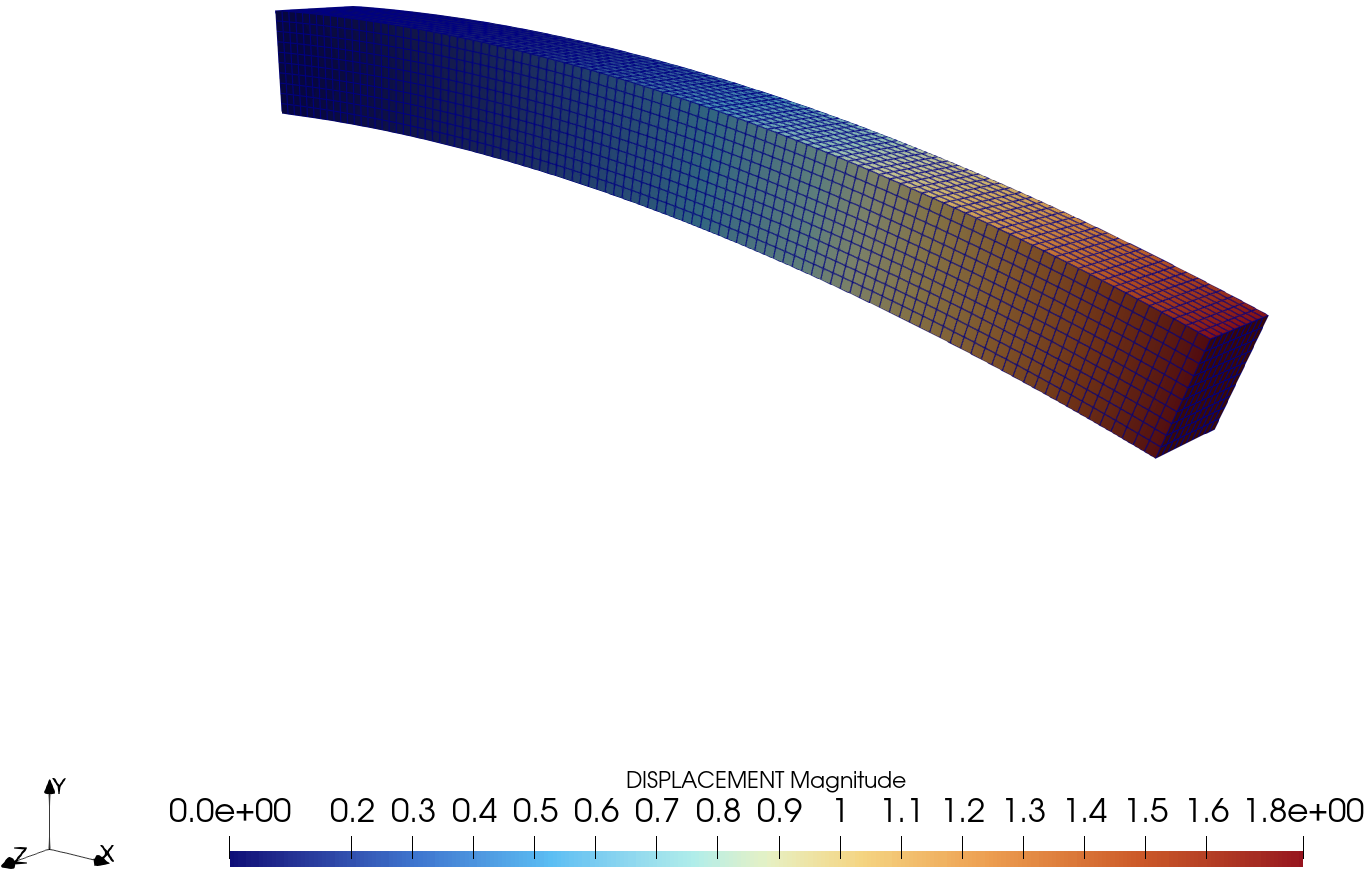}
        \caption{$\alpha=20^\circ$.}
    \end{subfigure}
    \hfill
    \begin{subfigure}[t]{0.32\textwidth}
        \centering
        \includegraphics[width=\textwidth]
        {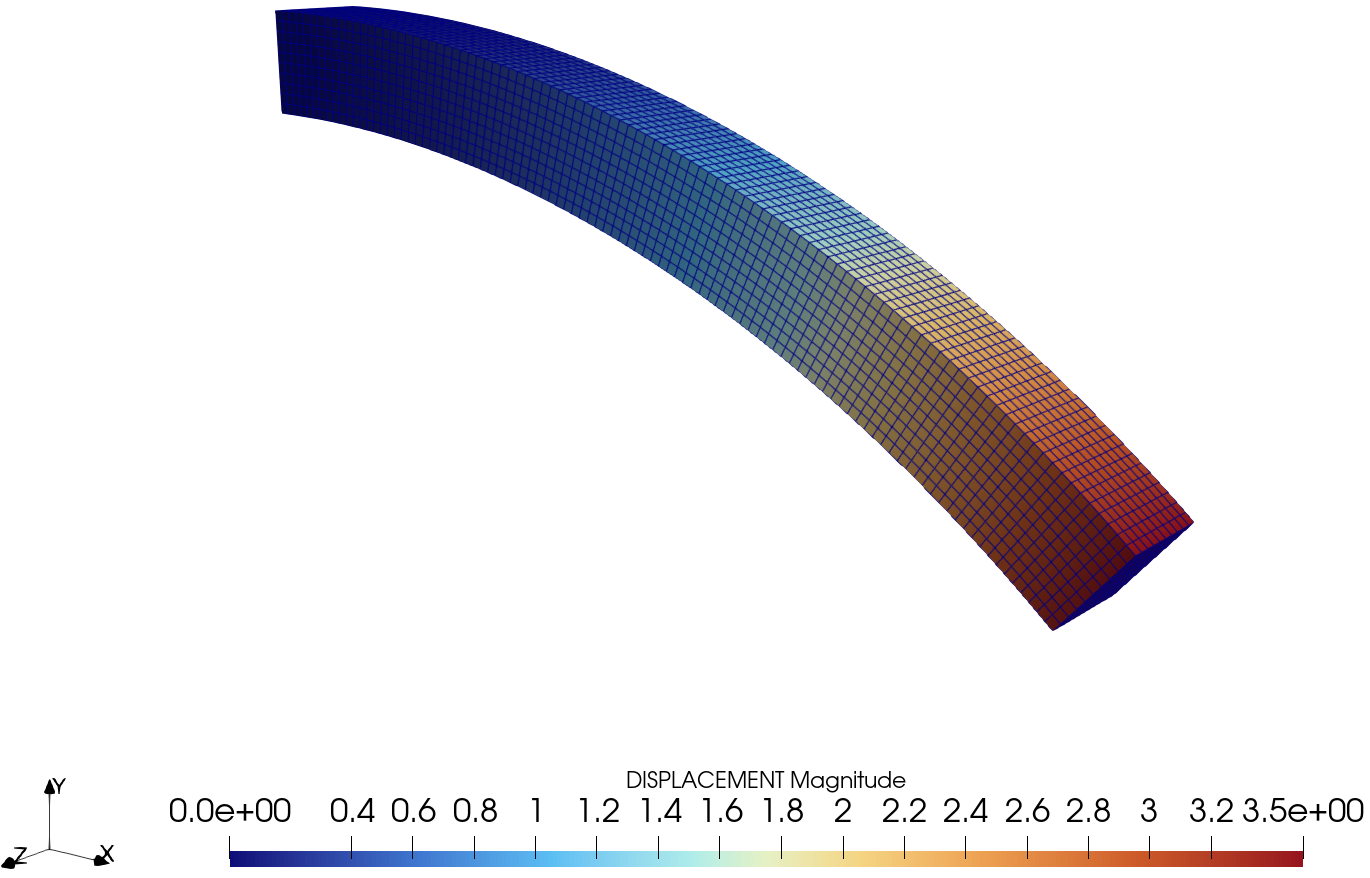}
        \caption{$\alpha=40^\circ$.}
    \end{subfigure}
    \hfill
    \begin{subfigure}[t]{0.32\textwidth}
        \centering
        \includegraphics[width=\textwidth]
        {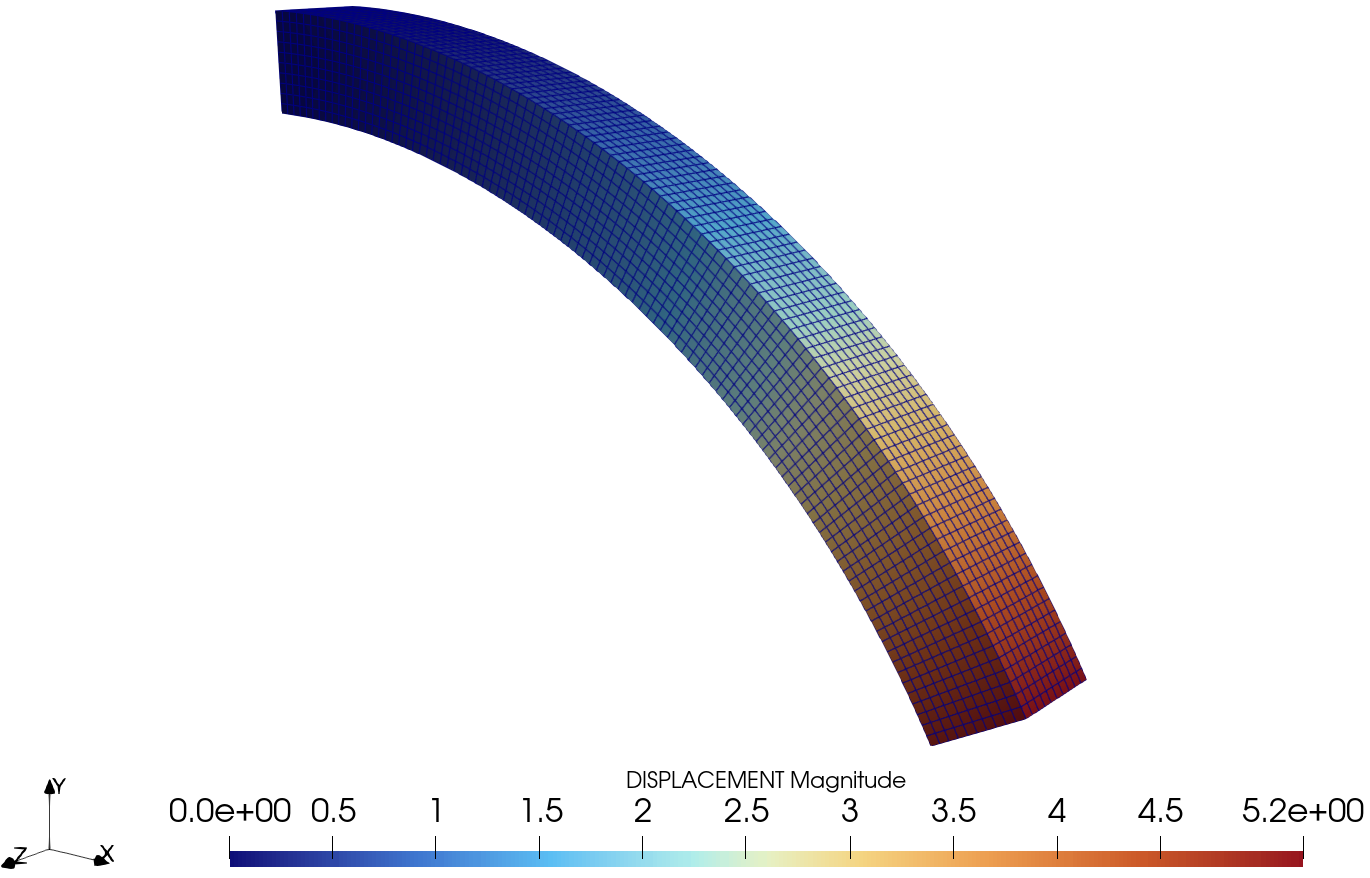}
        \caption{$\alpha=60^\circ$.}
    \end{subfigure}
    \caption{Surface configurations obtained with the Timoshenko
    beam mapper for the pure-bending test using the finest beam
    discretization with $64$ knot spans: (a) $\alpha=20^\circ$,
    (b) $\alpha=40^\circ$, and (c) $\alpha=60^\circ$. The surface is
    colored by the magnitude of the mapped displacement.}
    \label{fig:timoshenko_bending_deformed_configurations}
\end{figure}

Figure~\ref{fig:timoshenko_bending_deformed_configurations} shows the
surface configurations obtained with the Timoshenko mapper for the three
bending angles using the finest beam discretization with $64$ knot spans and $p=2$.
As for the Bernoulli case, the surface follows the prescribed circular
deformation, while the cross-sections are oriented according to the
independently prescribed rotational field.

\begin{figure}[h!]
    \centering
    \begin{subfigure}[t]{0.32\textwidth}
        \centering
        \includegraphics[width=\textwidth]
        {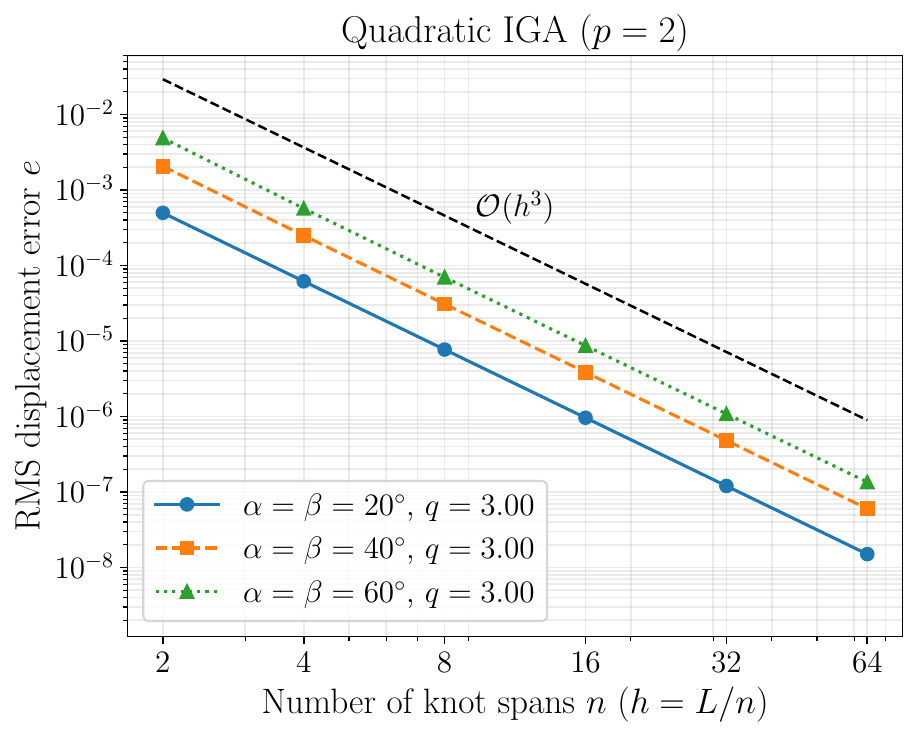}
        \caption{$p=2$.}
    \end{subfigure}
    \hfill
    \begin{subfigure}[t]{0.32\textwidth}
        \centering
        \includegraphics[width=\textwidth]
        {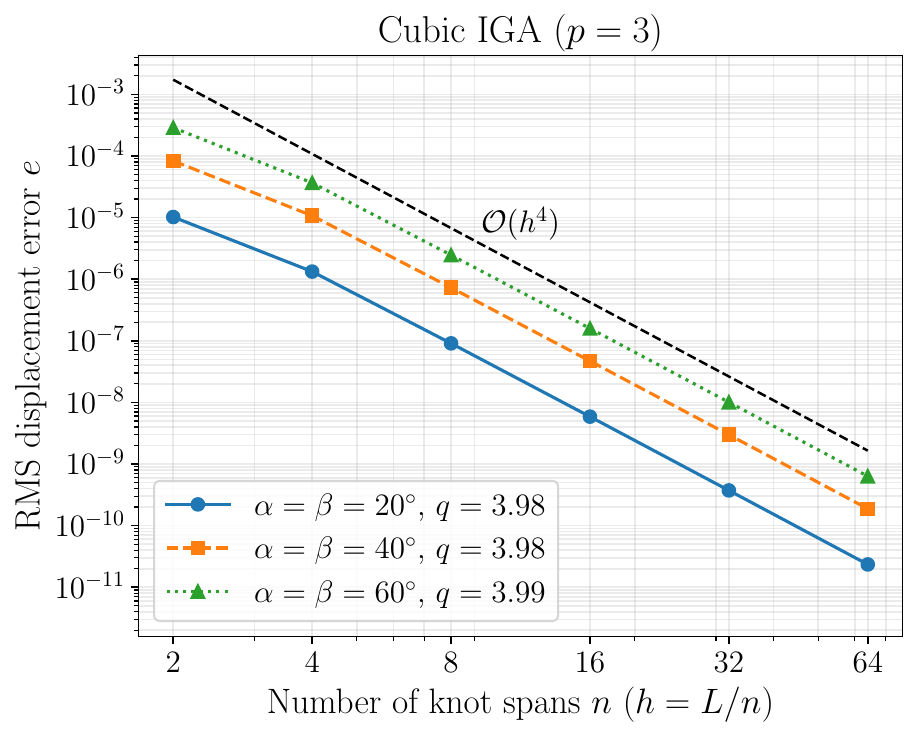}
        \caption{$p=3$.}
    \end{subfigure}
    \hfill
    \begin{subfigure}[t]{0.32\textwidth}
        \centering
        \includegraphics[width=\textwidth]
        {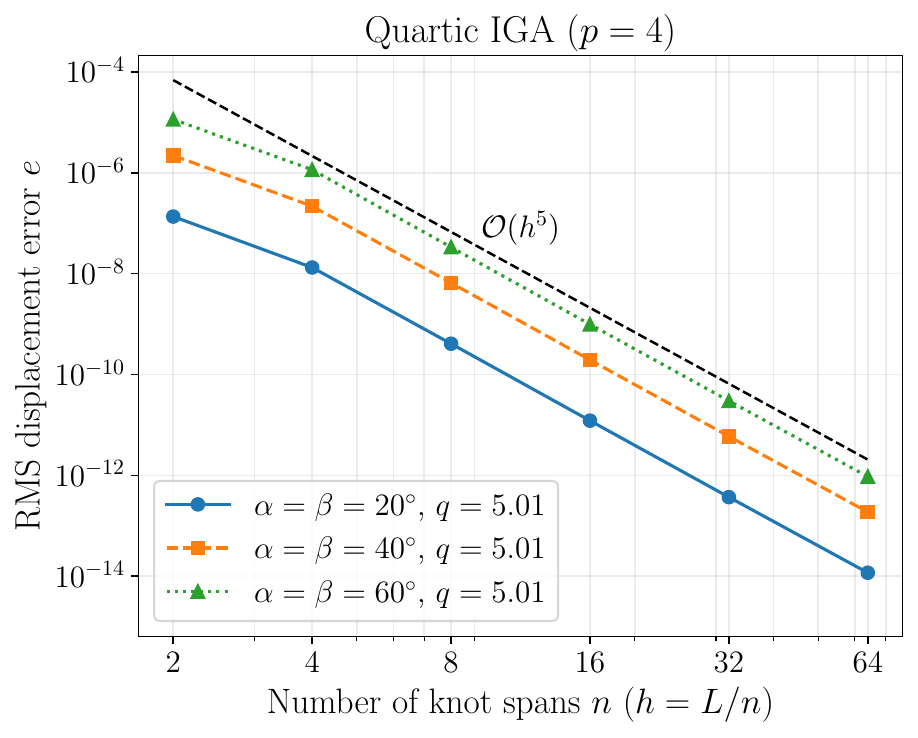}
        \caption{$p=4$.}
    \end{subfigure}
    \caption{Convergence of the RMS displacement error for the Timoshenko
    beam mapper under pure bending for
    $\alpha=20^\circ$, $40^\circ$, and $60^\circ$:
    (a) $p=2$, (b) $p=3$, and (c) $p=4$.}
    \label{fig:timoshenko_bending_convergence}
\end{figure}

The corresponding convergence results are shown in
Fig.~\ref{fig:timoshenko_bending_convergence}. For $p=2$, third-order
convergence is obtained for all three bending angles, with $q=3.00$.
For $p=3$, the observed orders are $q=3.98$, $3.98$, and $3.99$ for
$\alpha=20^\circ$, $40^\circ$, and $60^\circ$, respectively. For $p=4$,
an order of $q=5.01$ is obtained for all three cases. Thus, the
Timoshenko mapper exhibits $\mathcal{O}(h^{p+1})$
convergence for the investigated spline degrees.

The different convergence behavior of the two mapping formulations
follows from their treatment of the cross-section orientation.
In the Timoshenko case, the prescribed rotation field is linear
in the beam parameter and is therefore reproduced exactly by
the spline interpolation. The remaining error is governed by the
approximation of the smooth, non-polynomial centerline geometry,
whose trigonometric dependence on $\theta(X)$ cannot be represented
exactly by a finite-degree polynomial spline. For a smooth target
field, this approximation yields the optimal
$\mathcal{O}(h^{p+1})$ rate observed in
Fig.~\ref{fig:timoshenko_bending_convergence}.
In contrast, the Bernoulli mapper reconstructs the cross-section
orientation from the derivative of the approximated centerline,
generally limiting the expected convergence rate to
$\mathcal{O}(h^p)$. The even--odd pairing observed in this example
results from the particular constant-curvature bending deformation,
as discussed above, and is not a general characteristic of the
Bernoulli formulation.

\subsection{Pure torsion}
\label{sec:verification_torsion}

The second verification case considers pure torsion of the initially
straight beam. In contrast to the bending test, the beam centerline
remains unchanged and the deformation is produced exclusively by the
rotation of the cross-sections about the beam axis. This case therefore
provides a direct verification of the rotational contribution to the mapping independently of centerline deformation.

A linear twist distribution is prescribed along the beam as
\begin{equation}
    \phi(X)
    =
    \frac{\beta X}{L},
    \label{eq:torsion_rotation}
\end{equation}
where $\alpha$ denotes the total twist angle at the right end. Thus, the
cross-section at $X=0$ remains unrotated, whereas the cross-section at
$X=L$ undergoes a rotation $\alpha$ about the $x$-axis. Since the
centerline remains fixed, its analytical displacement is
\begin{equation}
    \boldsymbol{u}^{\mathrm{ex}}_c(X)
    =
    \boldsymbol{0}.
    \label{eq:torsion_centerline_displacement}
\end{equation}

The exact surface motion follows from the rotation of each cross-section
about the undeformed centerline. For a surface point with reference
coordinates $\boldsymbol{X}=(X,Y,Z)^{\mathrm{T}}$, the exact current
position is
\begin{equation}
    \boldsymbol{x}^{\mathrm{ex}}(\boldsymbol{X})
    =
    \begin{bmatrix}
        X\\[1mm]
        Y\cos\phi(X)-Z\sin\phi(X)\\[1mm]
        Y\sin\phi(X)+Z\cos\phi(X)
    \end{bmatrix},
    \label{eq:torsion_exact_position}
\end{equation}
and the corresponding analytical displacement is
\begin{equation}
    \boldsymbol{u}^{\mathrm{ex}}(\boldsymbol{X})
    =
    \begin{bmatrix}
        0\\[1mm]
        Y\left[\cos\phi(X)-1\right]-Z\sin\phi(X)\\[1mm]
        Y\sin\phi(X)+Z\left[\cos\phi(X)-1\right]
    \end{bmatrix}.
    \label{eq:torsion_exact_displacement}
\end{equation}

Three large twist angles,
$\beta=2\pi$, $4\pi$, and $8\pi$, are considered,
corresponding to one, two, and four complete rotations of the
cross-section between the two beam ends, respectively. The same spline
degrees and uniform beam refinement levels as in the bending study are
employed, ranging from $2$ to $64$ knot spans, while the surface
discretization is kept fixed.

The Bernoulli mapper is considered first. Since the centerline remains
straight throughout the deformation, its tangent is unchanged and the
cross-section rotation is governed entirely by the scalar twist field
$\phi(X)$. Accordingly, the centerline displacement is set to zero and
the linear twist field of Eq.~\eqref{eq:torsion_rotation} is prescribed
as the rotational input to the mapper. The test therefore isolates the
ability of the Bernoulli mapping formulation to reconstruct the nonlinear
surface motion generated by twisting the material cross-sections about
an otherwise unchanged centerline.

\begin{figure}[h!]
    \centering
    \begin{subfigure}[t]{0.32\textwidth}
        \centering
        \includegraphics[width=\textwidth]
        {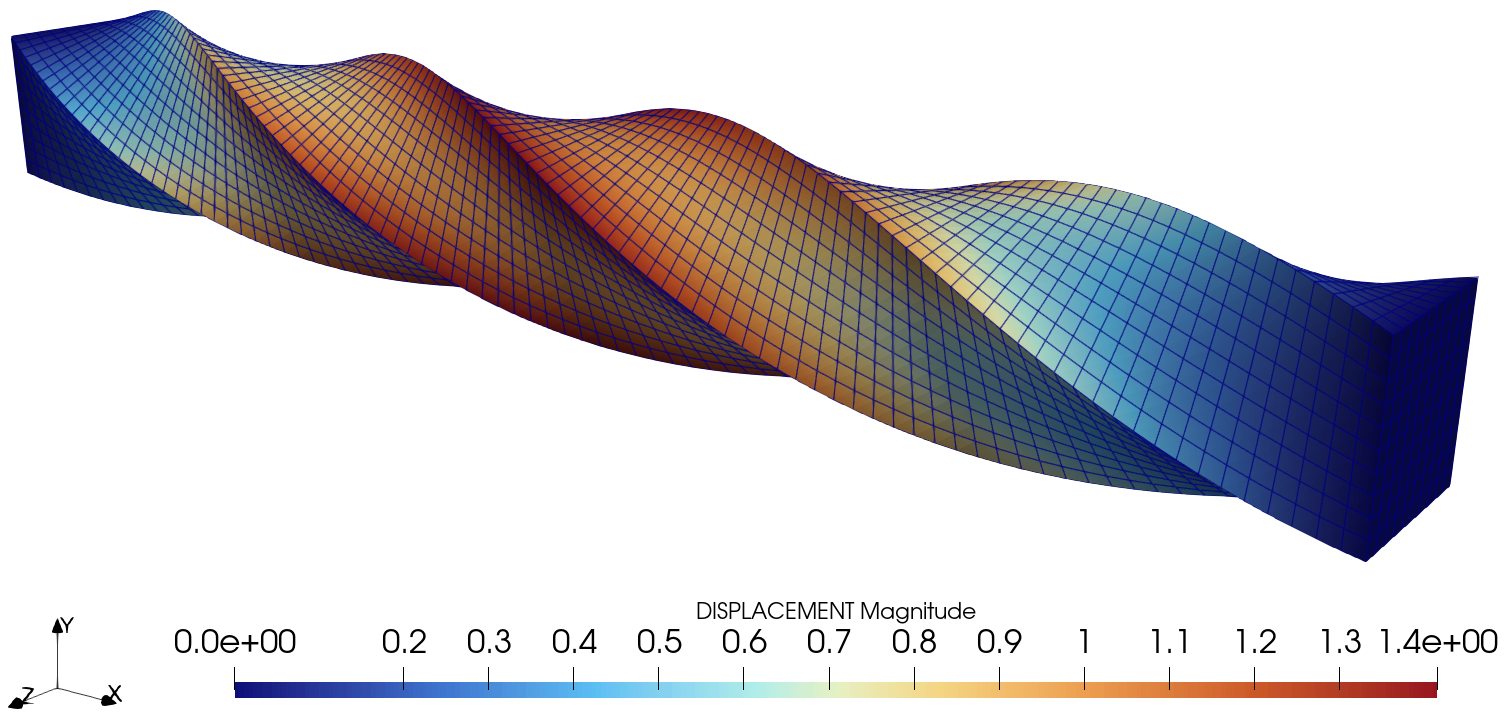}
        \caption{$\alpha=360^\circ$.}
    \end{subfigure}
    \hfill
    \begin{subfigure}[t]{0.32\textwidth}
        \centering
        \includegraphics[width=\textwidth]
        {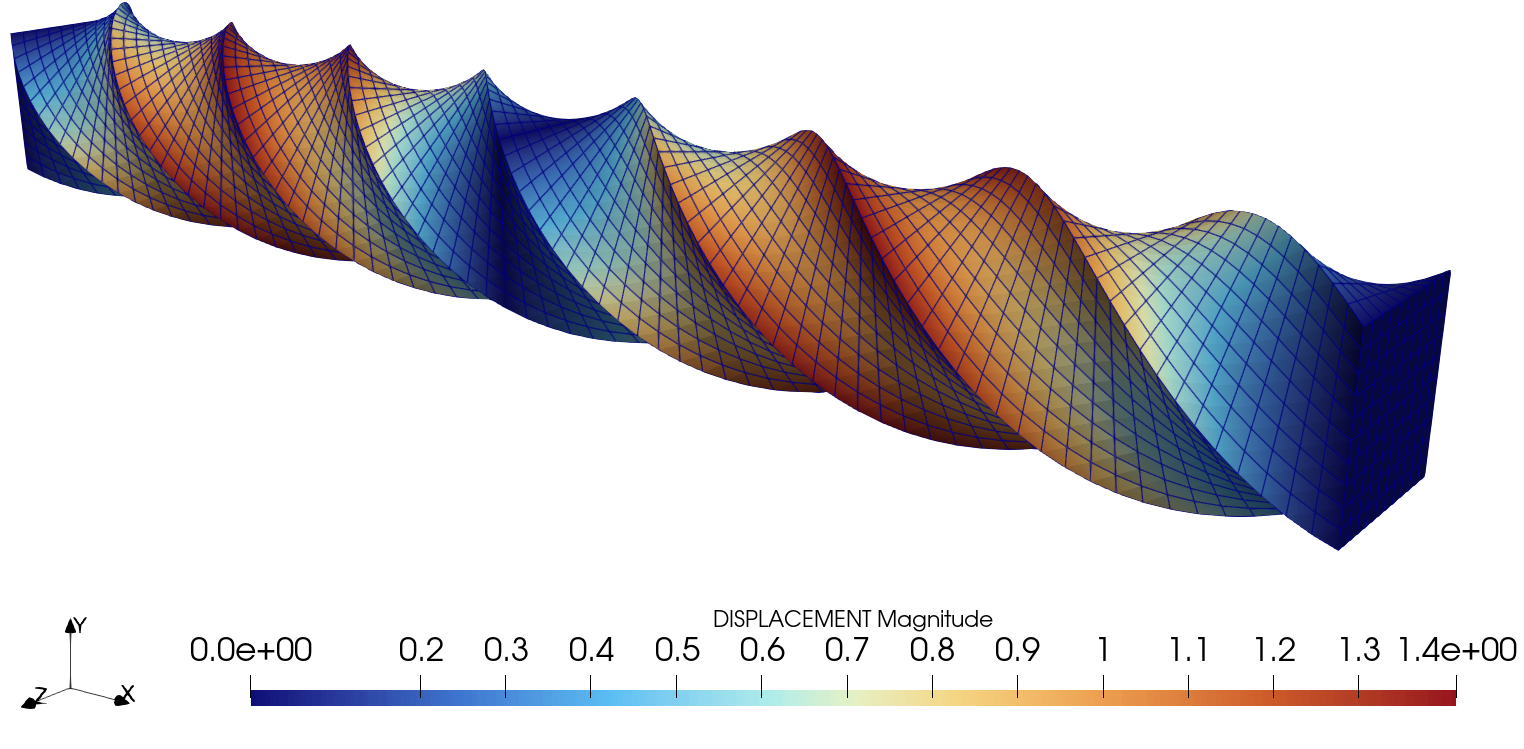}
        \caption{$\alpha=720^\circ$.}
    \end{subfigure}
    \hfill
    \begin{subfigure}[t]{0.32\textwidth}
        \centering
        \includegraphics[width=\textwidth]
        {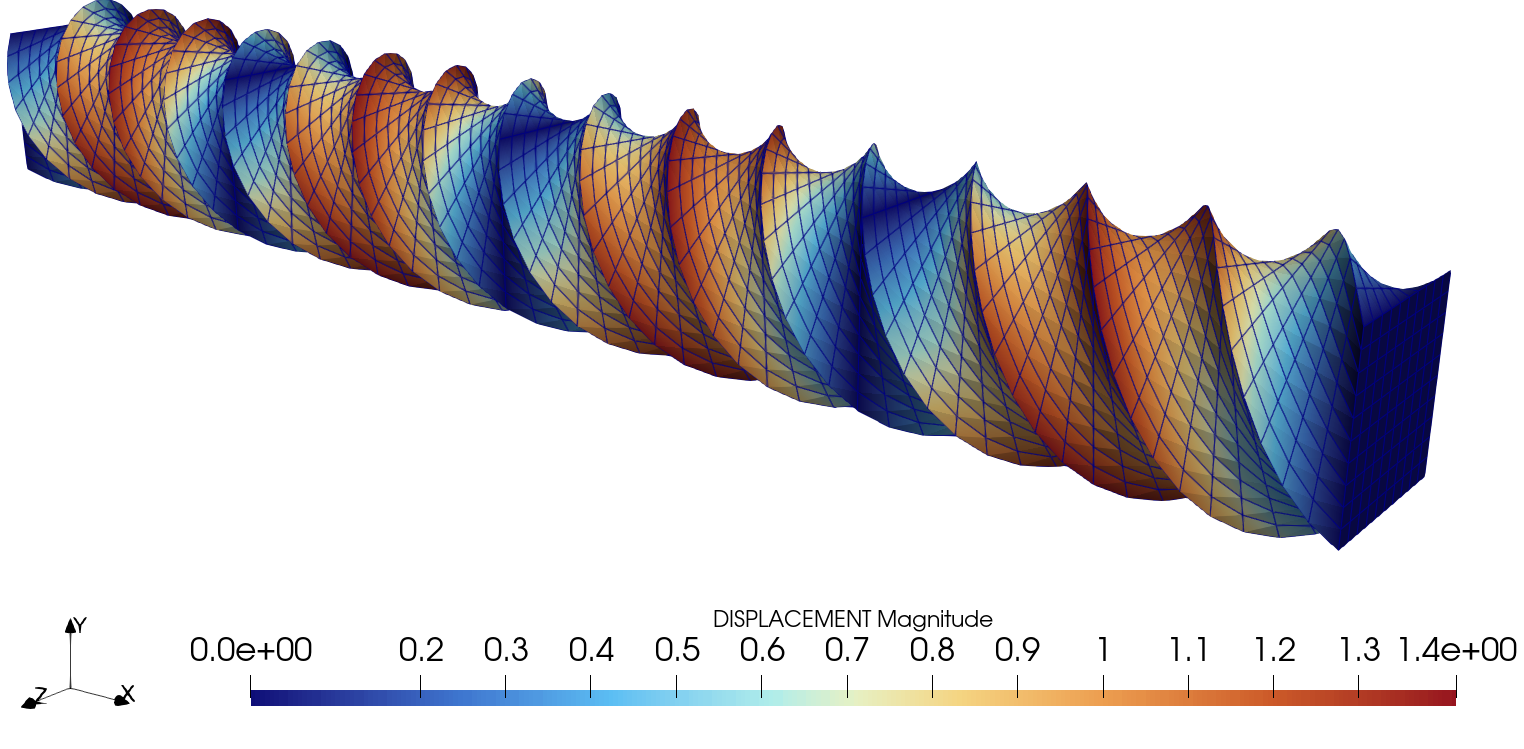}
        \caption{$\alpha=1440^\circ$.}
    \end{subfigure}
    \caption{Surface configurations obtained with the Bernoulli
    beam mapper for the pure-torsion test using the finest beam
    discretization with $64$ knot spans: (a) $\alpha=360^\circ$,
    (b) $\alpha=720^\circ$, and (c) $\alpha=1440^\circ$,
    corresponding to one, two, and four complete twists, respectively.
    The surface is colored by the magnitude of the mapped displacement.}
    \label{fig:bernoulli_torsion_deformed_configurations}
\end{figure}

Figure~\ref{fig:bernoulli_torsion_deformed_configurations} shows the
surface configurations obtained with the Bernoulli mapper for the three
prescribed twist angles using the finest beam discretization with $64$
knot spans. The increasing number of complete twists is recovered while
the beam centerline remains unchanged. The surface is colored by the
magnitude of the mapped displacement.

Unlike the pure-bending case, a conventional convergence rate is not
expected for this test. The prescribed twist field
$\phi(X)=\alpha X/L$ is linear in the beam coordinate and is therefore
represented exactly by the employed spline spaces. Although the exact
surface displacement in Eq.~\eqref{eq:torsion_exact_displacement}
contains nonlinear trigonometric terms, these quantities are not
approximated directly by the beam spline. Instead, the mapper evaluates
the nonlinear cross-section rotation pointwise from the exactly
represented twist field and the fixed material attachment coordinates.
Consequently, the pure-torsion case constitutes an exact-reproduction
test rather than a conventional convergence study.

\begin{figure}[h!]
    \centering
    \includegraphics[width=0.45\textwidth]
    {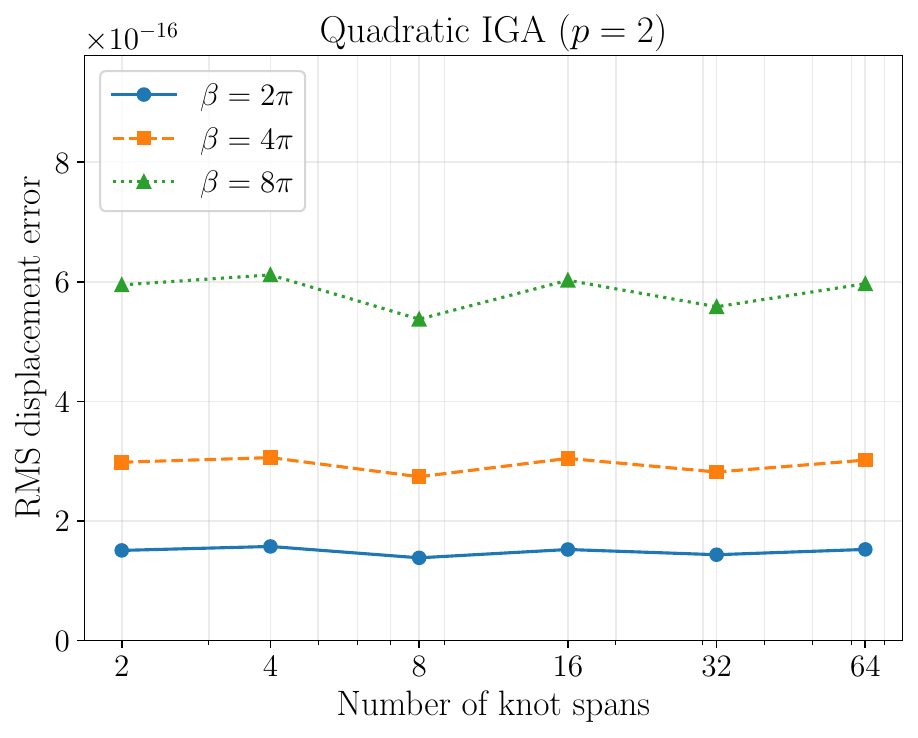}
    \caption{RMS displacement error of the Bernoulli beam
    mapper for the pure-torsion test using quadratic splines ($p=2$)
    and different beam refinement levels. Results are shown for
    $\alpha=360^\circ$, $720^\circ$, and $1440^\circ$.}
    \label{fig:bernoulli_torsion_accuracy}
\end{figure}

Figure~\ref{fig:bernoulli_torsion_accuracy} shows the RMS displacement
error for the quadratic discretization ($p=2$) over all refinement levels
and prescribed twist angles. The error remains of the order of
$10^{-16}$ for $\alpha=2\pi$, $4\pi$, and $8\pi$,
without a systematic dependence on the number of knot spans. The same
machine-precision behavior is obtained for the higher spline degrees and
is therefore not shown. These results demonstrate that the Bernoulli
mapper reproduces the analytical torsional surface motion up to numerical
precision, including the extreme case of four complete twists.

The Timoshenko mapper is verified using the same prescribed twist fields.
In this case, the rotation about the beam axis is represented by the
independent rotational field of the Timoshenko formulation. Since this
field varies linearly along the beam, it is represented exactly by the
employed spline spaces.

\begin{figure}[h!]
    \centering
    \begin{subfigure}[t]{0.32\textwidth}
        \centering
        \includegraphics[width=\textwidth]
        {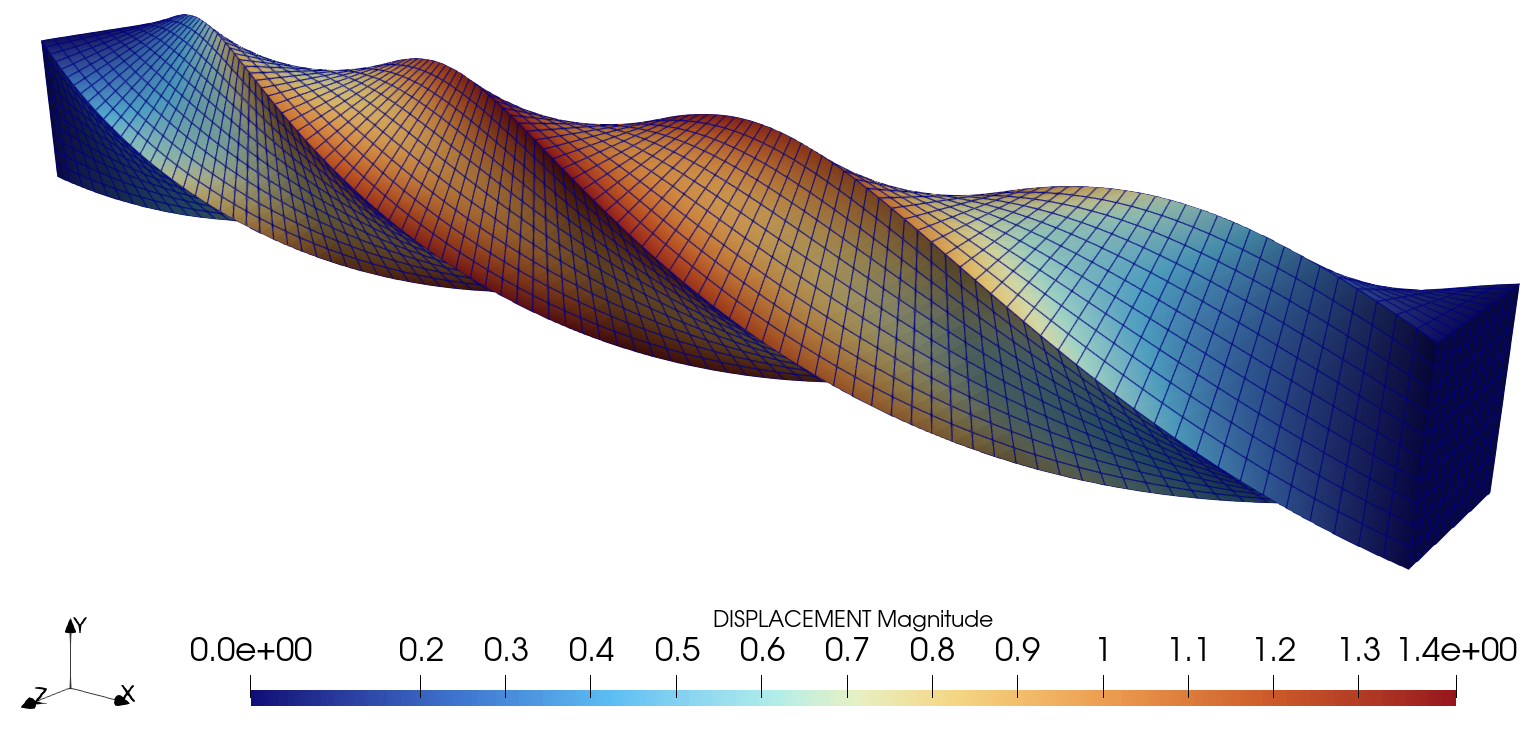}
        \caption{$\alpha=360^\circ$.}
    \end{subfigure}
    \hfill
    \begin{subfigure}[t]{0.32\textwidth}
        \centering
        \includegraphics[width=\textwidth]
        {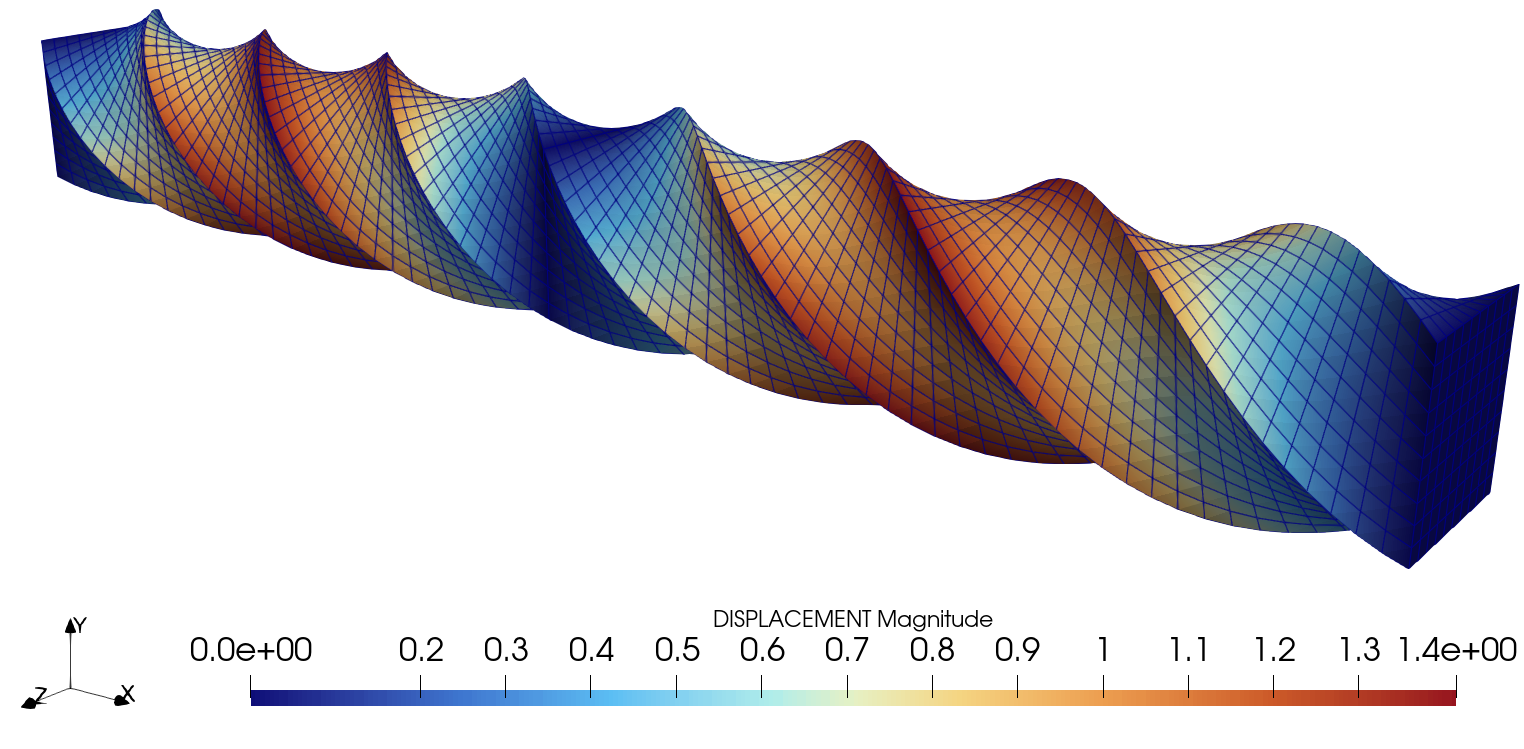}
        \caption{$\alpha=720^\circ$.}
    \end{subfigure}
    \hfill
    \begin{subfigure}[t]{0.32\textwidth}
        \centering
        \includegraphics[width=\textwidth]
        {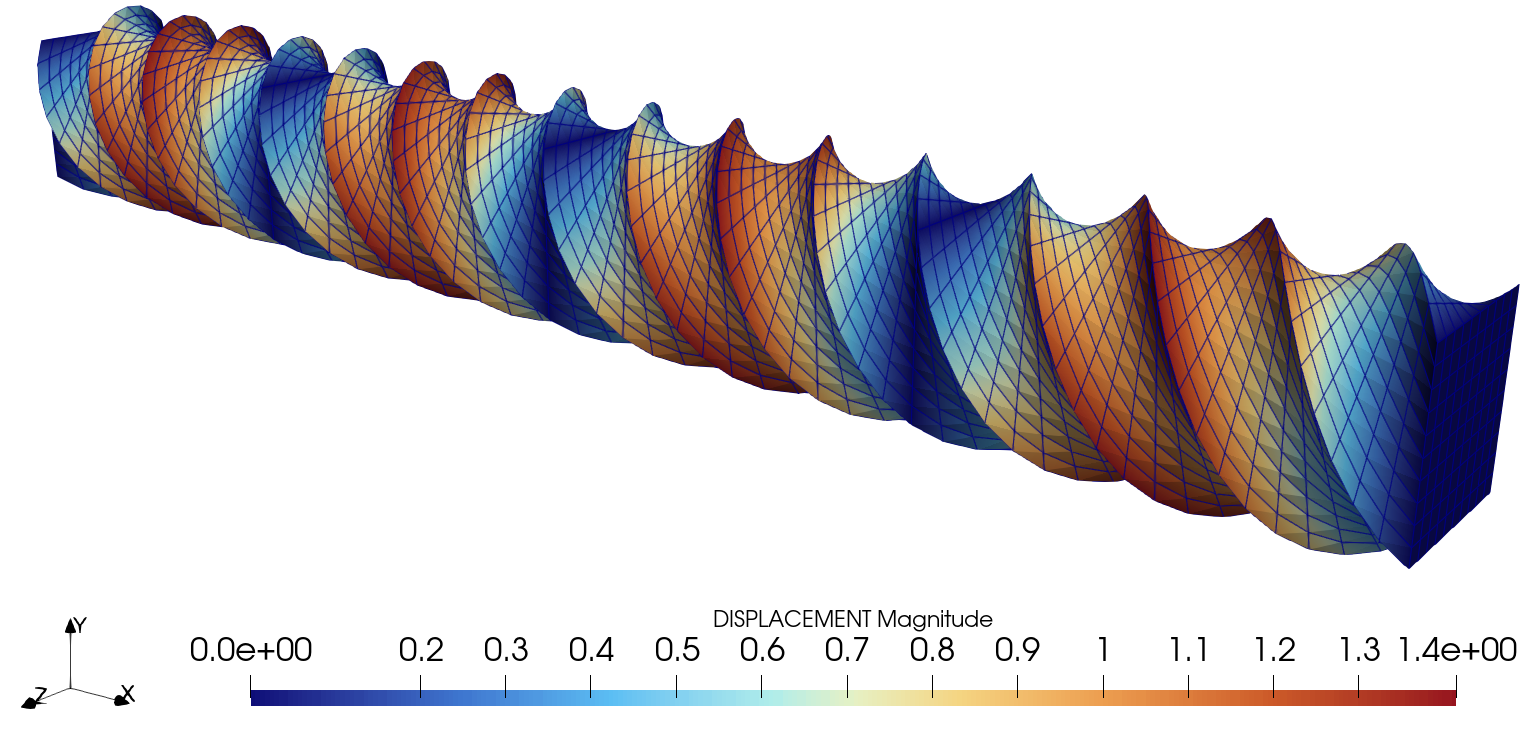}
        \caption{$\alpha=1440^\circ$.}
    \end{subfigure}
    \caption{Surface configurations obtained with the Timoshenko
    beam mapper for the pure-torsion test using $64$ knot
    spans: (a) $\alpha=360^\circ$, (b) $\alpha=720^\circ$, and
    (c) $\alpha=1440^\circ$.}
    \label{fig:timoshenko_torsion_deformed_configurations}
\end{figure}

Figure~\ref{fig:timoshenko_torsion_deformed_configurations} shows the
mapped surface configurations for the three prescribed twist angles using
$64$ knot spans.

\begin{figure}[h!]
    \centering
    \includegraphics[width=0.45\textwidth]
    {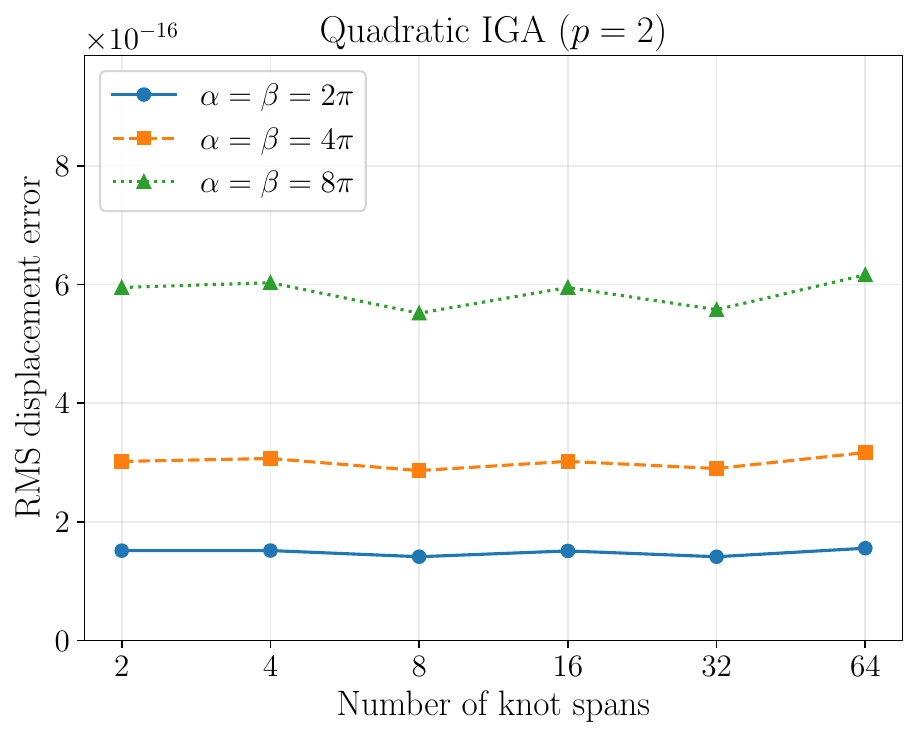}
    \caption{RMS displacement error of the Timoshenko beam
    mapper for the pure-torsion test using quadratic splines ($p=2$).}
    \label{fig:timoshenko_torsion_accuracy}
\end{figure}

As shown in Fig.~\ref{fig:timoshenko_torsion_accuracy}, the RMS
displacement error remains of the order of $10^{-16}$ for all refinement
levels and prescribed twist angles. As for the Bernoulli mapper, the same
behavior is obtained for the higher spline degrees, and only the
quadratic case is therefore reported. The results confirm exact
reproduction of the nonlinear torsional surface motion up to numerical
precision.

\subsection{Combined bending and torsion}
\label{sec:verification_bending_torsion}

The third verification case combines the bending and torsional motions
considered separately in the previous examples, following the combined
bending and twist test of Wang et al.~\cite{Wang2016}. The initially
straight centerline is bent into a circular arc in the $x$--$y$ plane,
while the cross-sections are simultaneously twisted about the beam axis.
Let $\alpha$ and $\beta$ denote the prescribed bending and torsional
angles, respectively, at the beam tip. The corresponding local bending
and twist angles vary linearly along the reference centerline as
\begin{equation}
    \theta(X)
    =
    \frac{\alpha X}{L},
    \qquad
    \phi(X)
    =
    \frac{\beta X}{L}.
    \label{eq:bending_torsion_angles}
\end{equation}
The bending curvature is constant and, as in the pure-bending case, the
radius of the deformed centerline is
\begin{equation}
    r=\frac{L}{\alpha},
    \label{eq:bending_torsion_radius}
\end{equation}
where $\alpha$ is expressed in radians.

For a surface point with reference coordinates
$\boldsymbol{X}=(X,Y,Z)^{\mathrm{T}}$, the cross-section coordinates are
first rotated about the beam axis according to the local twist angle
$\phi(X)$,
\begin{equation}
    \begin{aligned}
        Y^\star
        &=
        Y\cos\phi(X)-Z\sin\phi(X),\\
        Z^\star
        &=
        Y\sin\phi(X)+Z\cos\phi(X).
    \end{aligned}
    \label{eq:bending_torsion_twisted_coordinates}
\end{equation}
The twisted cross-section is subsequently transported along the circular
centerline according to the local bending angle $\theta(X)$. The exact
current position of the surface point is therefore
\begin{equation}
    \boldsymbol{x}^{\mathrm{ex}}(\boldsymbol{X})
    =
    \begin{bmatrix}
        \left(r+Y^\star\right)\sin\theta(X)\\[1mm]
        r\left[\cos\theta(X)-1\right]
        +Y^\star\cos\theta(X)\\[1mm]
        Z^\star
    \end{bmatrix},
    \label{eq:bending_torsion_exact_position}
\end{equation}
and the corresponding exact displacement is
\begin{equation}
    \boldsymbol{u}^{\mathrm{ex}}(\boldsymbol{X})
    =
    \boldsymbol{x}^{\mathrm{ex}}(\boldsymbol{X})
    -
    \boldsymbol{X}.
    \label{eq:bending_torsion_exact_displacement}
\end{equation}
For $\beta=0$, Eq.~\eqref{eq:bending_torsion_exact_position} reduces to
the pure-bending solution, whereas for non-zero $\beta$ it additionally
accounts for the torsional rotation of the cross-sections.

In the present verification, equal tip bending and torsional angles are
prescribed,
\begin{equation}
    \beta=\alpha,
    \label{eq:bending_torsion_equal_angles}
\end{equation}
with $\alpha=20^\circ$, $40^\circ$, and $60^\circ$. The same beam
refinement levels and fixed surface discretization employed in the
previous examples are considered. This test therefore verifies the
ability of the mapping formulations to reconstruct the surface motion
when deformation of the centerline and rotation of the cross-sections
occur simultaneously. The mapped surface displacements are compared with
Eq.~\eqref{eq:bending_torsion_exact_displacement}, and the accuracy is
quantified using the RMS error defined in
Eq.~\eqref{eq:mapping_rms_error}.

\begin{figure}[h!]
    \centering
    \begin{subfigure}[t]{0.32\textwidth}
        \centering
        \includegraphics[width=\linewidth]
        {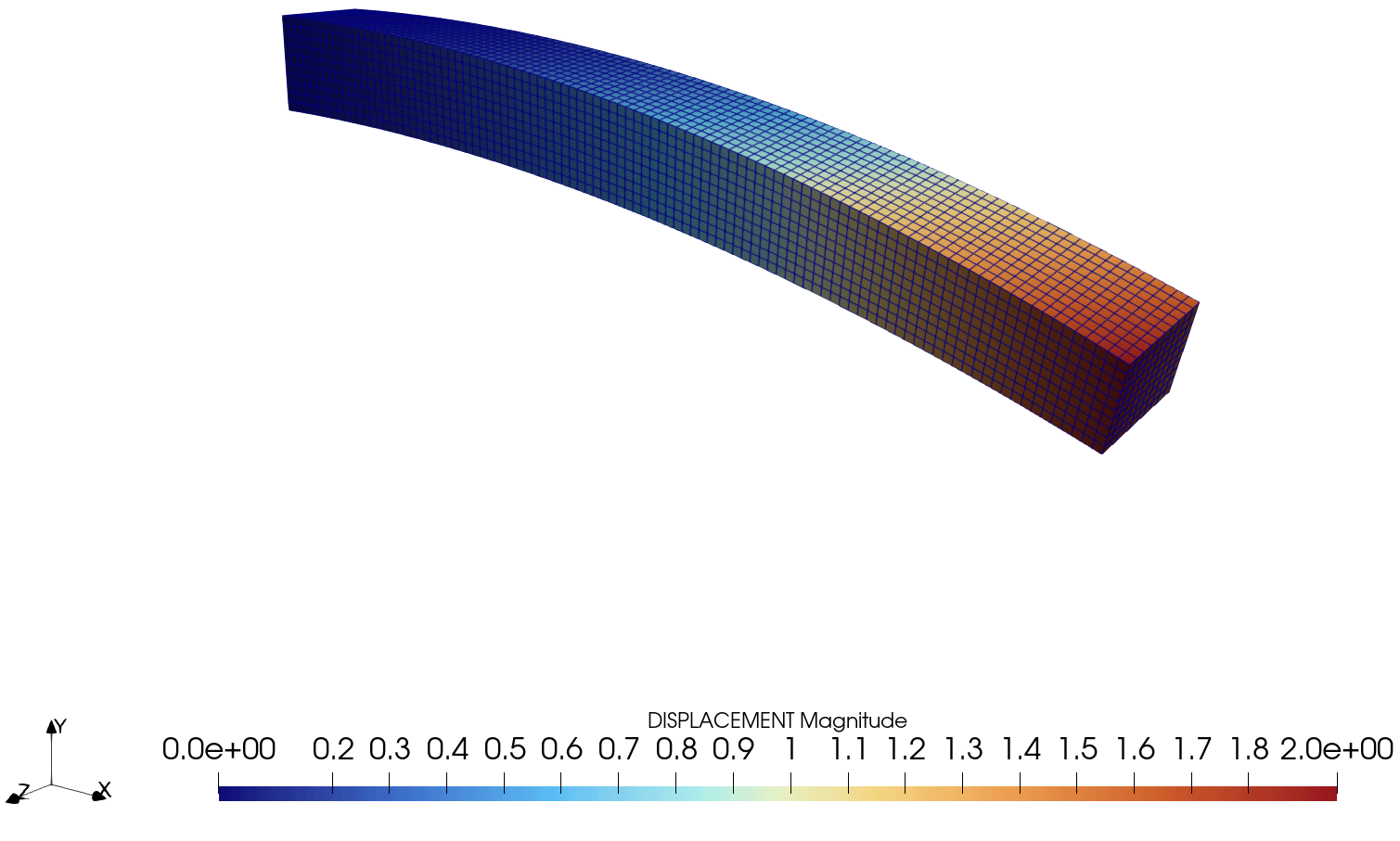}
        \caption{$\alpha=\beta=20^\circ$.}
    \end{subfigure}
    \hfill
    \begin{subfigure}[t]{0.32\textwidth}
        \centering
        \includegraphics[width=\linewidth]
        {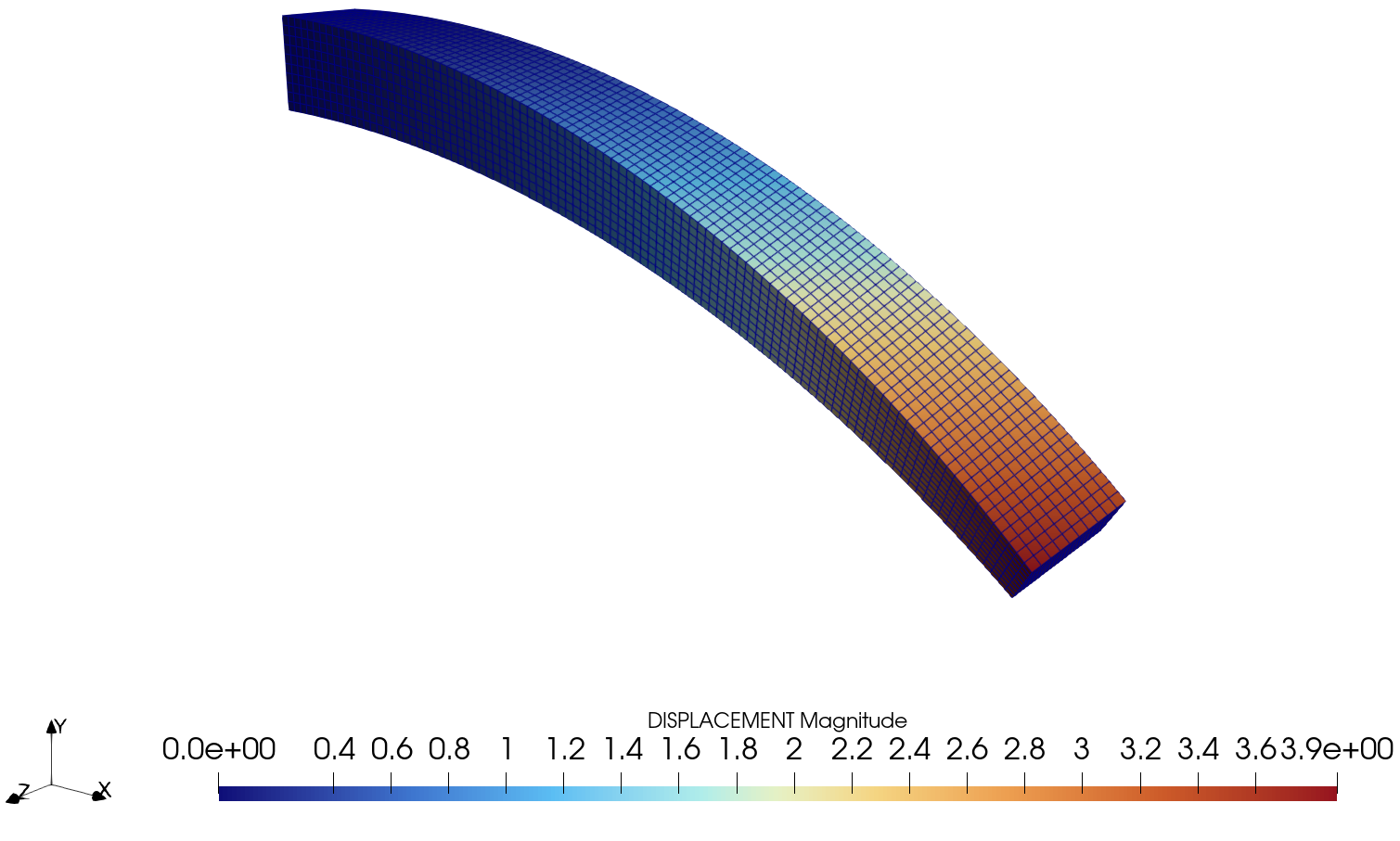}
        \caption{$\alpha=\beta=40^\circ$.}
    \end{subfigure}
    \hfill
    \begin{subfigure}[t]{0.32\textwidth}
        \centering
        \includegraphics[width=\linewidth]
        {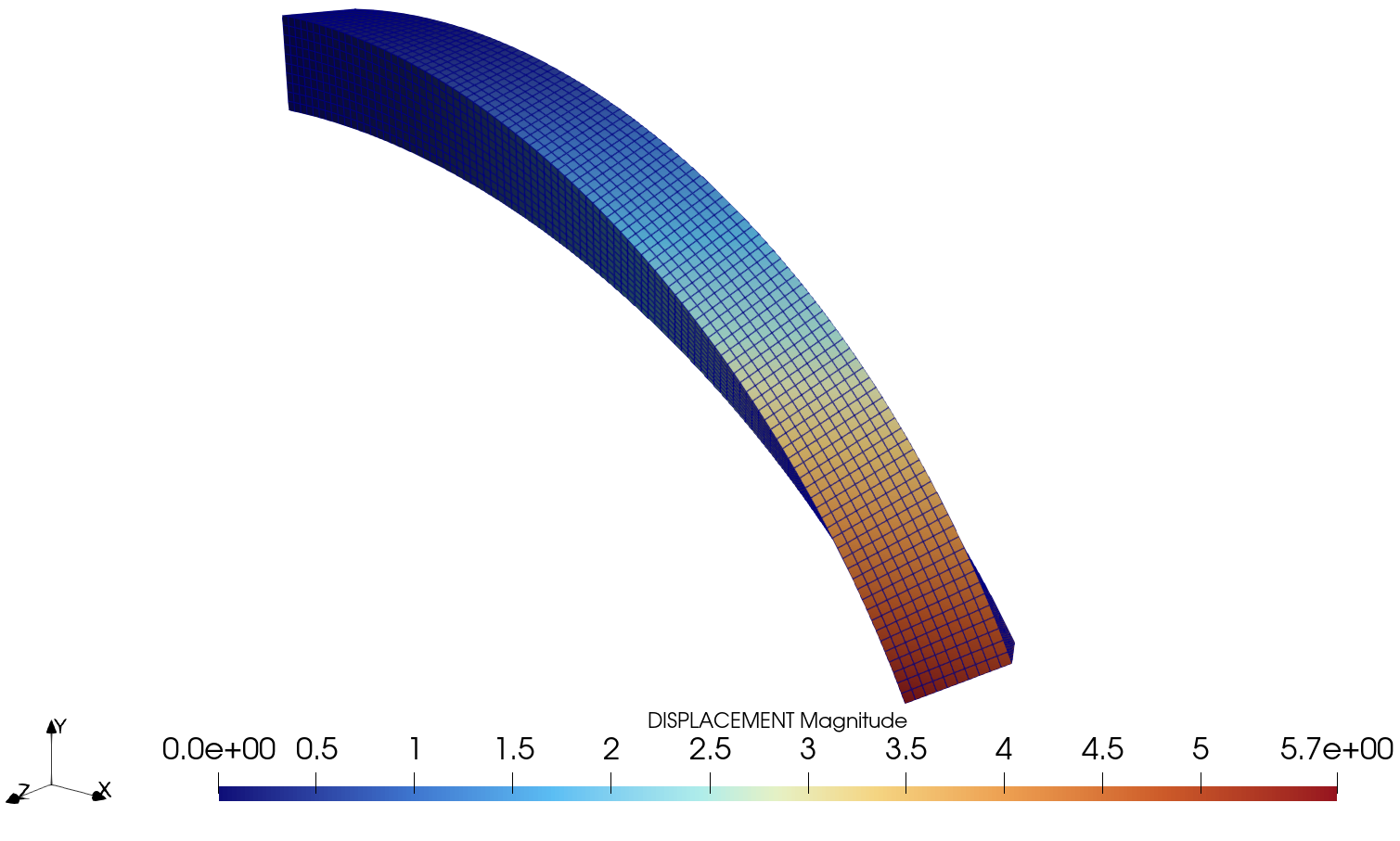}
        \caption{$\alpha=\beta=60^\circ$.}
    \end{subfigure}
    \caption{Surface configurations obtained with the Bernoulli mapper
    for combined bending and torsion using $64$ knot spans. The surface
    is colored by the displacement magnitude.}
    \label{fig:bernoulli_bending_torsion_deformed_configurations}
\end{figure}

The mapped surface configurations obtained with the Bernoulli mapper are
shown in Fig.~\ref{fig:bernoulli_bending_torsion_deformed_configurations}
for the three prescribed deformation amplitudes using $64$ knot spans.

\begin{figure}[h!]
    \centering
    \begin{subfigure}[t]{0.32\textwidth}
        \centering
        \includegraphics[width=\linewidth]
        {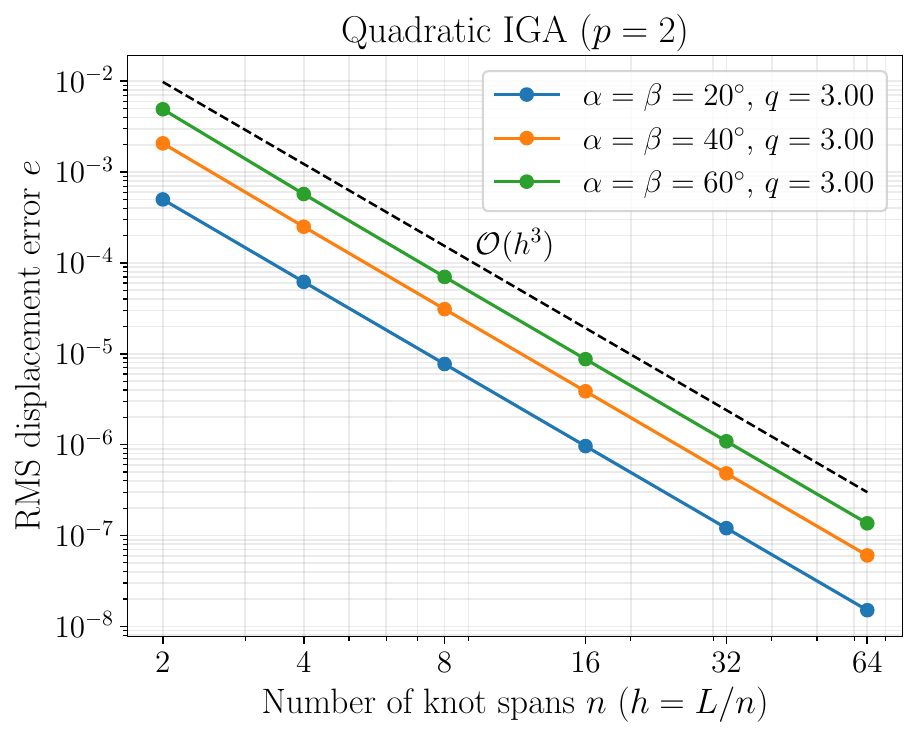}
        \caption{$p=2$.}
    \end{subfigure}
    \hfill
    \begin{subfigure}[t]{0.32\textwidth}
        \centering
        \includegraphics[width=\linewidth]
        {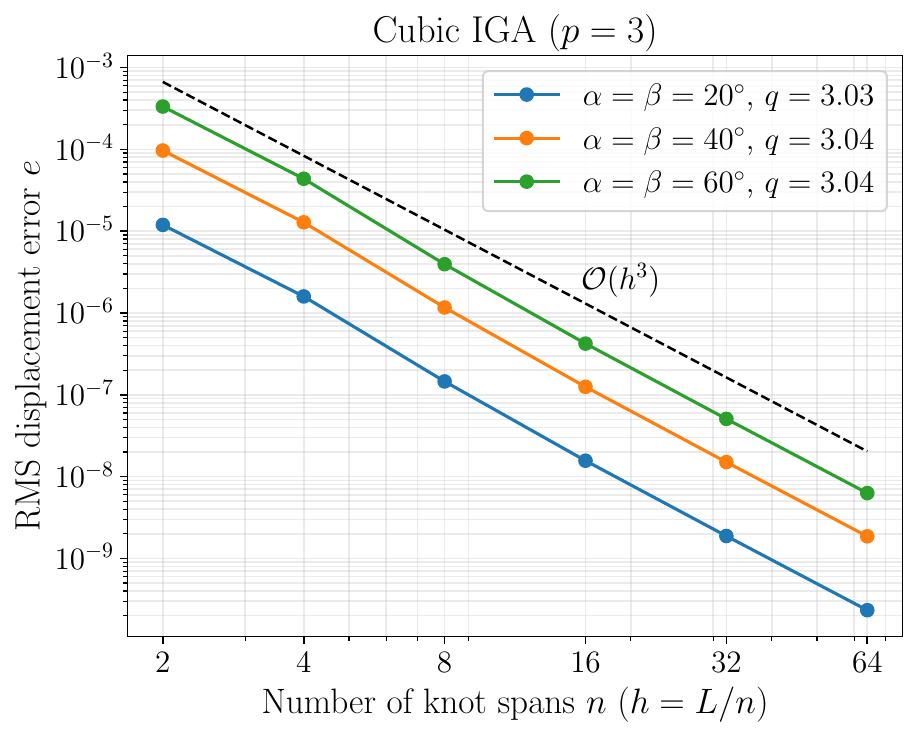}
        \caption{$p=3$.}
    \end{subfigure}
    \hfill
    \begin{subfigure}[t]{0.32\textwidth}
        \centering
        \includegraphics[width=\linewidth]
        {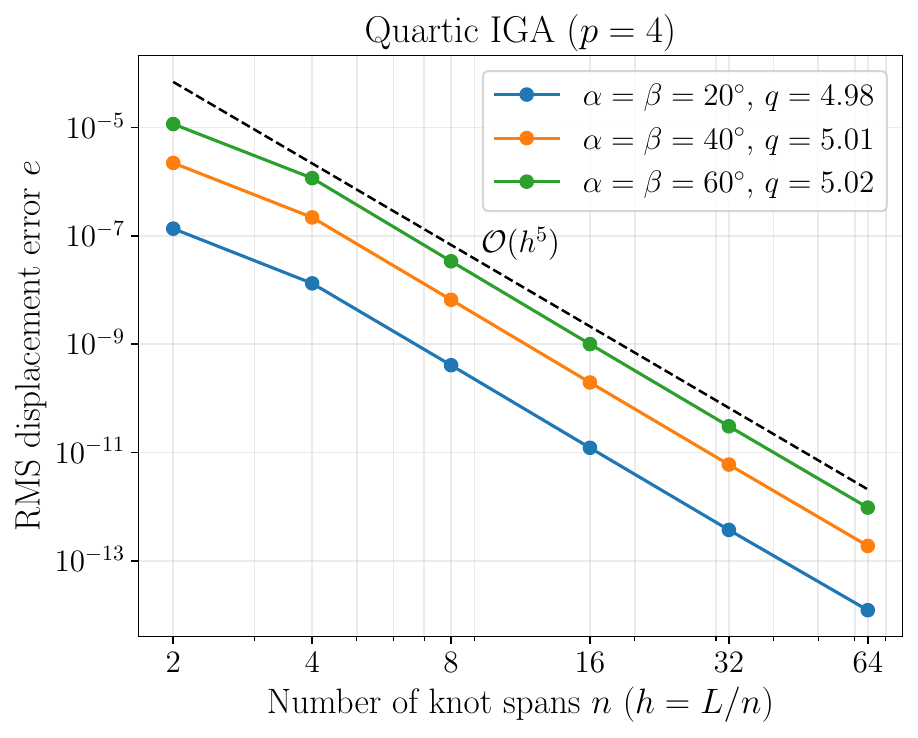}
        \caption{$p=4$.}
    \end{subfigure}
    \caption{RMS displacement error for the Bernoulli mapper under
    combined bending and torsion for different spline degrees and
    refinement levels.}
    \label{fig:bernoulli_bending_torsion_convergence}
\end{figure}

The corresponding RMS displacement errors are shown in Fig.~\ref{fig:bernoulli_bending_torsion_convergence}. The quadratic and cubic B-splines exhibit approximately third-order convergence, whereas the quartic basis reaches approximately fifth-order convergence. Thus, the combined bending and torsion case exhibits the same even--odd pairing observed for pure circular bending. Since the prescribed twist field is linear and can be represented exactly by the spline basis, it does not introduce an additional interpolation error. The convergence behavior is therefore primarily governed by the derivative-based reconstruction of the cross-section orientation from the approximated centerline. The enhanced rates observed for even polynomial degrees are consistent with the geometry-dependent error cancellation discussed for the constant-curvature bending case, rather than being a general property of the Bernoulli formulation.

\begin{figure}[h!]
    \centering
    \begin{subfigure}[t]{0.32\textwidth}
        \centering
        \includegraphics[width=\linewidth]
        {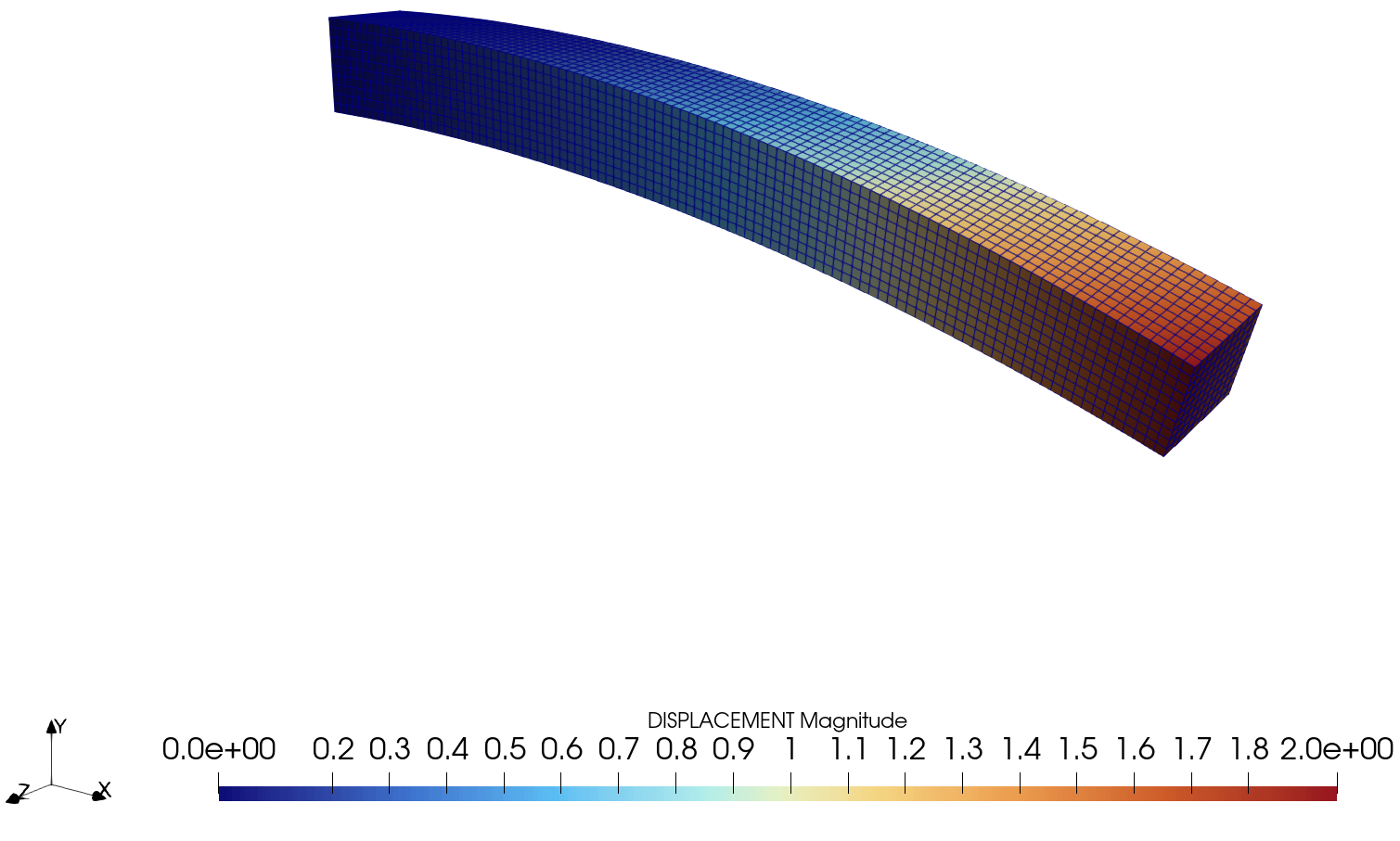}
        \caption{$\alpha=\beta=20^\circ$.}
    \end{subfigure}
    \hfill
    \begin{subfigure}[t]{0.32\textwidth}
        \centering
        \includegraphics[width=\linewidth]
        {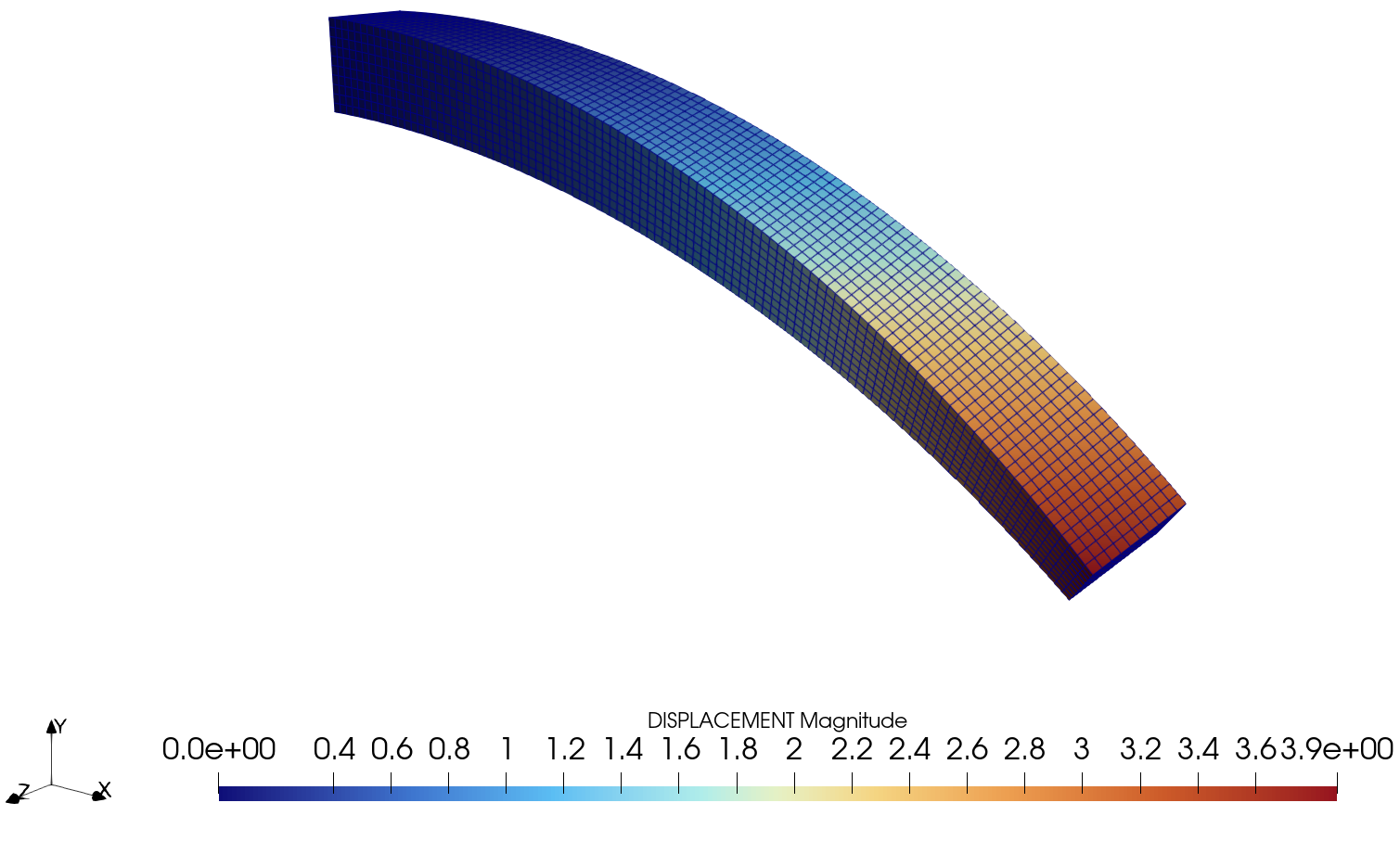}
        \caption{$\alpha=\beta=40^\circ$.}
    \end{subfigure}
    \hfill
    \begin{subfigure}[t]{0.32\textwidth}
        \centering
        \includegraphics[width=\linewidth]
        {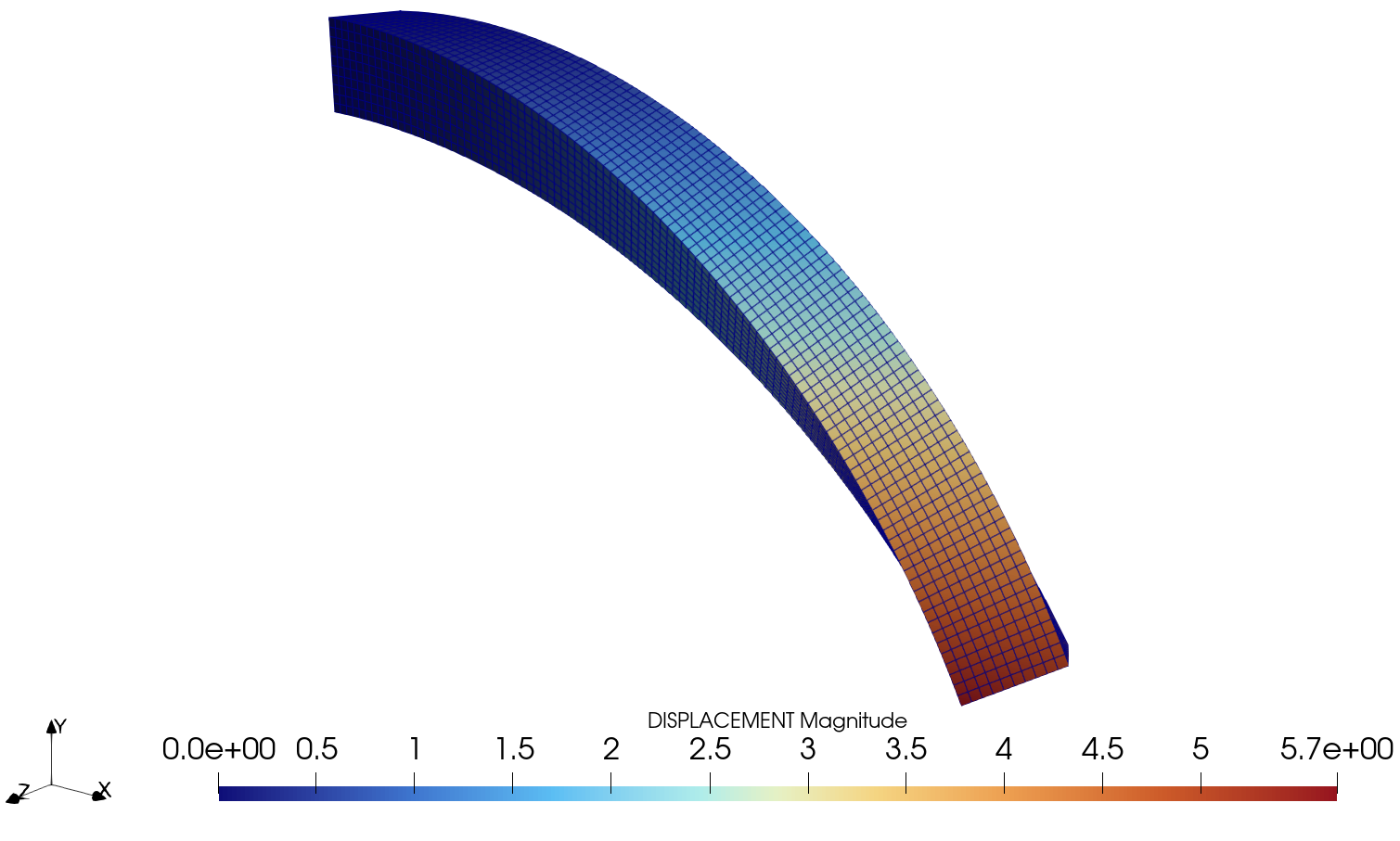}
        \caption{$\alpha=\beta=60^\circ$.}
    \end{subfigure}
    \caption{Surface configurations obtained with the Timoshenko mapper
    for combined bending and torsion using $64$ knot spans. The surface
    is colored by the displacement magnitude.}
    \label{fig:timoshenko_bending_torsion_deformed_configurations}
\end{figure}

The same test is performed with the Timoshenko mapper. The corresponding
surface configurations for the three prescribed deformation amplitudes
are shown in
Fig.~\ref{fig:timoshenko_bending_torsion_deformed_configurations} using
$64$ knot spans.

\begin{figure}[h!]
    \centering
    \begin{subfigure}[t]{0.32\textwidth}
        \centering
        \includegraphics[width=\linewidth]
        {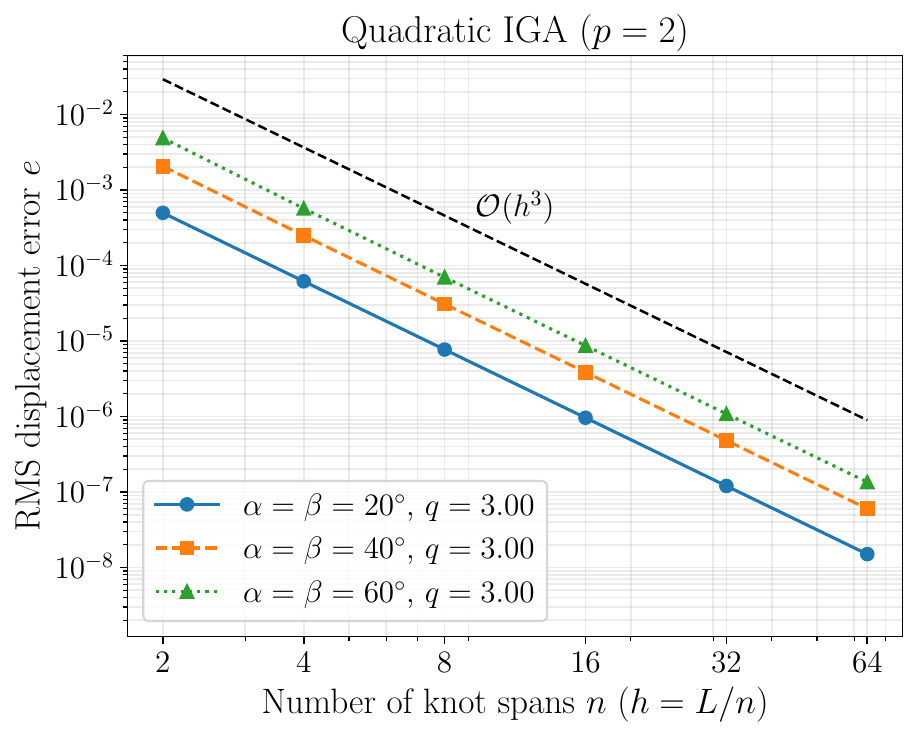}
        \caption{$p=2$.}
    \end{subfigure}
    \hfill
    \begin{subfigure}[t]{0.32\textwidth}
        \centering
        \includegraphics[width=\linewidth]
        {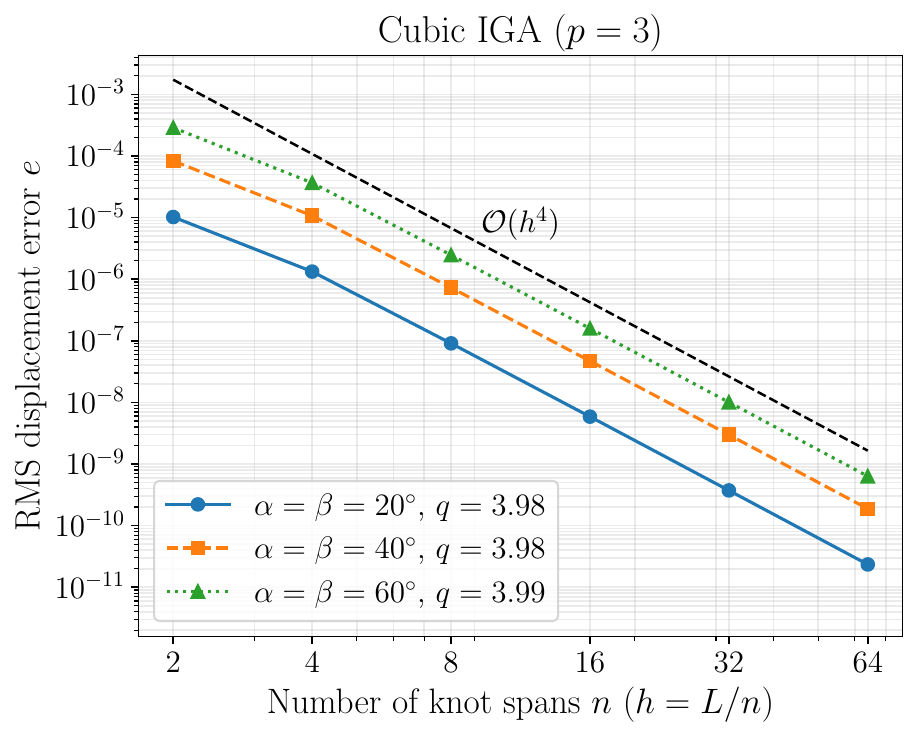}
        \caption{$p=3$.}
    \end{subfigure}
    \hfill
    \begin{subfigure}[t]{0.32\textwidth}
        \centering
        \includegraphics[width=\linewidth]
        {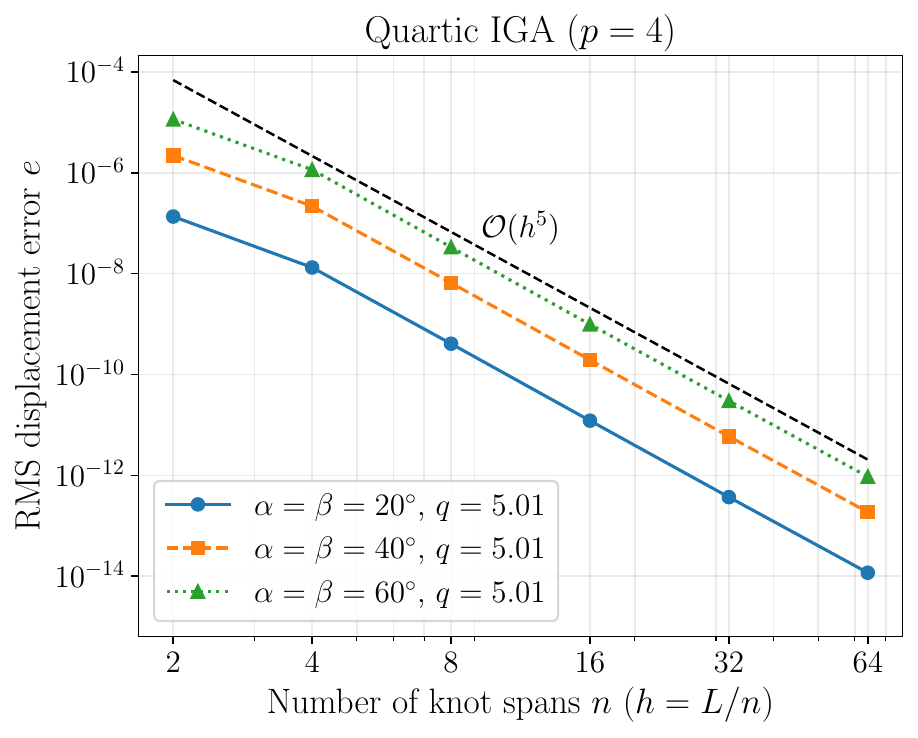}
        \caption{$p=4$.}
    \end{subfigure}
    \caption{RMS displacement error for the Timoshenko mapper under
    combined bending and torsion for different spline degrees and
    refinement levels.}
    \label{fig:timoshenko_bending_torsion_convergence}
\end{figure}

The corresponding RMS displacement errors are shown in
Fig.~\ref{fig:timoshenko_bending_torsion_convergence}. In contrast to the
Bernoulli mapper, the Timoshenko mapper exhibits the expected
$O(h^{p+1})$ convergence for all considered spline degrees. The observed
orders are approximately three, four, and five for $p=2$, $3$, and $4$,
respectively, and are essentially independent of the prescribed
deformation amplitude.

\subsection{Verification of load transfer}
\label{sec:verification_load_transfer}

The reverse transfer of fluid forces to generalized beam loads is
verified through the work-conjugacy condition underlying the proposed
load mapping. For an arbitrary virtual variation of the beam degrees of
freedom $\delta\boldsymbol{q}$ and the corresponding surface displacement
variation $\delta\boldsymbol{u}_S$, consistency of the forward and
inverse mapping operators requires
\begin{equation}
    \delta W_B
    =
    \boldsymbol{f}_B^{\mathrm{T}}\delta\boldsymbol{q}
    =
    \boldsymbol{f}_S^{\mathrm{T}}\delta\boldsymbol{u}_S
    =
    \delta W_S,
    \label{eq:load_transfer_virtual_work}
\end{equation}
where $\boldsymbol{f}_S$ contains the discrete fluid forces on the wet surface and
$\boldsymbol{f}_B$ denotes the generalized beam loads obtained from the
inverse mapping. 

To verify Eq.~\eqref{eq:load_transfer_virtual_work} independently of the
analytical mapping linearization, a finite-difference test is performed
in a strongly deformed configuration. The beam is bent by
$\alpha=60^\circ$ and simultaneously subjected to a linearly varying
twist with tip value $\beta=4\pi$. A nonuniform three-dimensional
force field is prescribed directly at the nodes of the surface mesh
representing the fluid--structure interface, with the nodal force
vector at each surface node $Q$ defined as
\begin{equation}
    \boldsymbol{f}_Q
    =
    \frac{1}{n_Q}
    \begin{bmatrix}
        \sin(0.7X_Q)+0.2Y_Q\\
        \cos(0.3X_Q)+0.1Z_Q\\
        0.5+0.2\sin(X_Q)+Y_Q
    \end{bmatrix},
    \label{eq:load_transfer_surface_force}
\end{equation}
where $(X_Q,Y_Q,Z_Q)$ are the coordinates of surface node $Q$
in the reference configuration and $n_Q$ is the total number
of surface nodes. The prescribed force field is thus defined
independently of the current deformation and is transferred
to the generalized beam loads $\boldsymbol{f}_B$ using the
inverse mapper explained in Eq.~\eqref{eq:work_conjugate_load_transfer}.

Three independent virtual directions
$\delta\boldsymbol{q}^{(k)}$, $k=1,2,3$, are prescribed
as perturbations of the beam degrees of freedom about
the current deformed configuration. Each direction
combines translational and rotational variations and
is used to evaluate the consistency of the load-transfer
operator, rather than representing an actual structural
displacement. For each
direction, the beam-side virtual work is computed from
the generalized loads obtained through the inverse mapper as
\begin{equation}
    \delta W_B
    =
    \boldsymbol{f}_B^{\mathrm{T}}\delta\boldsymbol{q}.
    \label{eq:beam_virtual_work_verification}
\end{equation}

To evaluate the corresponding surface-side virtual work
independently of the analytical mapping linearization,
the nonlinear forward mapper is evaluated at two perturbed
beam configurations. The resulting surface displacement
variation is approximated by the centered finite difference
\begin{equation}
    \delta\boldsymbol{u}_S^{\mathrm{FD}}
    =
    \frac{
        \boldsymbol{u}_S(\boldsymbol{q}
        +\varepsilon\;\delta\boldsymbol{q})
        -
        \boldsymbol{u}_S(\boldsymbol{q}
        -\varepsilon\;\delta\boldsymbol{q})
    }{2\varepsilon},
    \label{eq:surface_variation_finite_difference}
\end{equation}
where $\varepsilon$ controls the perturbation magnitude.
For the Bernoulli formulation, the perturbations are
applied directly to the centerline displacement and scalar
twist degrees of freedom. For the Timoshenko formulation,
rotational perturbations are interpreted as spatial
incremental rotations and applied multiplicatively to
the current cross-sectional orientation, consistently
with the rotation update introduced in
Section~\ref{sec:timoshenko_beam}.

The surface-side virtual work is then evaluated using
the prescribed surface forces and the finite-difference
displacement variation,
\begin{equation}
    \delta W_S^{\mathrm{FD}}
    =
    \boldsymbol{f}_S^{\mathrm{T}}
    \delta\boldsymbol{u}_S^{\mathrm{FD}}.
    \label{eq:surface_virtual_work_verification}
\end{equation}
The discrepancy between the beam- and surface-side
virtual work is quantified by the normalized error
\begin{equation}
    e_W
    =
    \frac{
        \left|
        \delta W_S^{\mathrm{FD}}-\delta W_B
        \right|
    }{
        \displaystyle
        \sum_i
        \left|f_{B,i}\,\delta q_i\right|
    }.
    \label{eq:normalized_virtual_work_error}
\end{equation}

\begin{figure}[h!]
    \centering
    \begin{subfigure}[t]{0.7\textwidth}
        \centering
        \includegraphics[width=\linewidth]{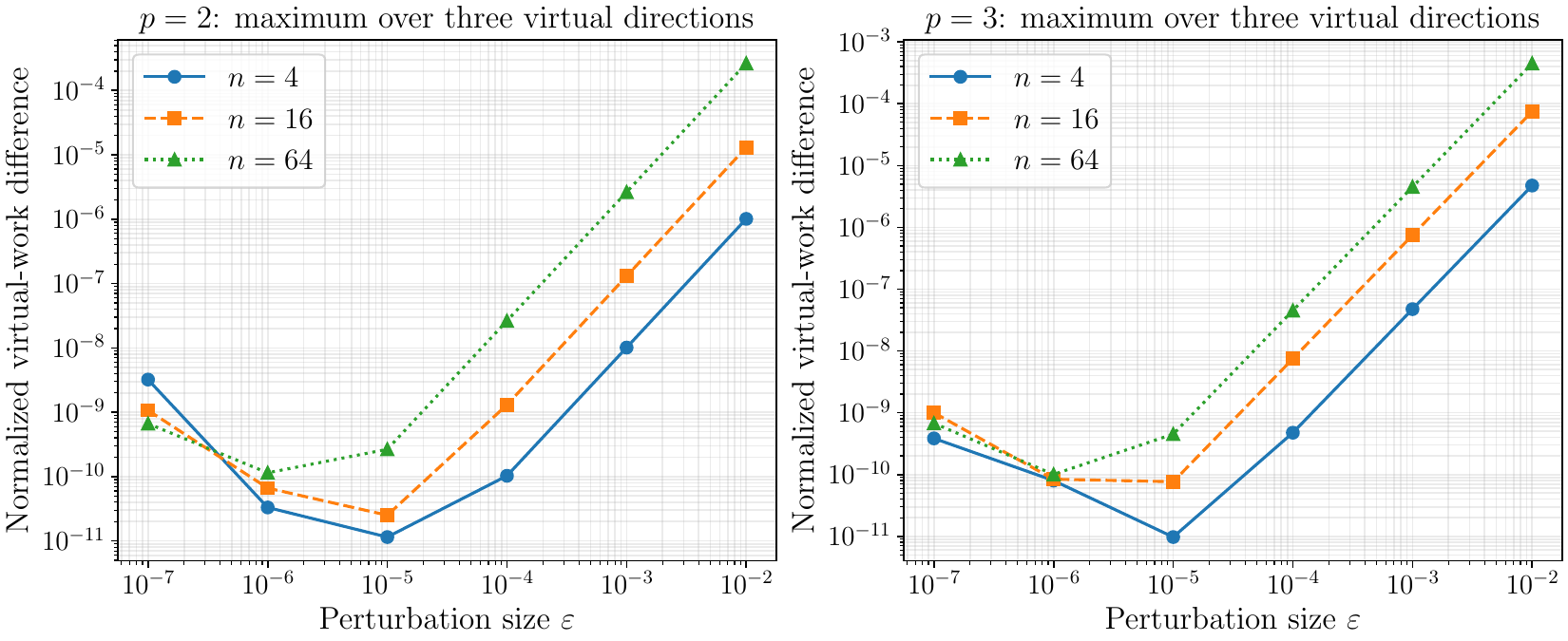}
        \caption{Bernoulli mapper.}
        \label{fig:virtual_work_bernoulli}
    \end{subfigure}

    \vspace{0.4cm}

    \begin{subfigure}[t]{0.7\textwidth}
        \centering
        \includegraphics[width=\linewidth]
        {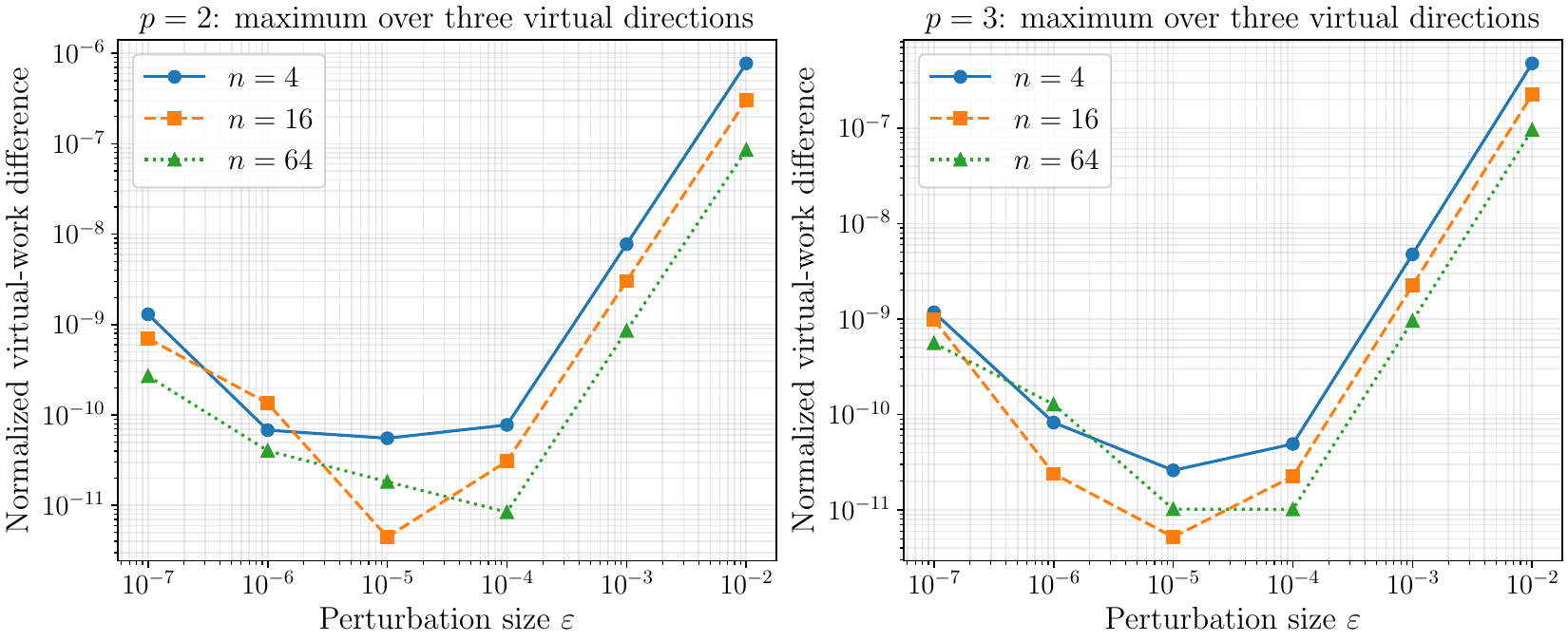}
        \caption{Timoshenko mapper.}
        \label{fig:virtual_work_timoshenko}
    \end{subfigure}

    \caption{Normalized virtual-work difference obtained from the
    independent finite-difference verification of the load-transfer
    operator: (a) Bernoulli mapper and (b) Timoshenko mapper. For each
    perturbation size, the maximum error over three virtual directions
    is reported.}
    \label{fig:load_transfer_virtual_work}
\end{figure}

Figure~\ref{fig:load_transfer_virtual_work} presents the maximum
normalized virtual-work error over the three prescribed virtual
directions for the Bernoulli and Timoshenko mappers. Results are
reported for spline degrees $p=2$ and $p=3$, beam discretizations
with $n=4$, $16$, and $64$ knot spans, and perturbation sizes
ranging from $\varepsilon=10^{-2}$ to $10^{-7}$.

For both formulations, the error initially decreases with
decreasing perturbation size, following the expected
$\mathcal{O}(\varepsilon^2)$ truncation error of the centered
finite-difference approximation. Minimum errors of approximately
$10^{-11}$--$10^{-10}$ are reached for
$\varepsilon\approx10^{-5}$--$10^{-6}$, beyond which round-off
errors associated with subtracting nearly identical surface
displacement fields become dominant. This behavior is consistent
across the considered spline degrees and discretizations.
The close agreement between the beam- and surface-side virtual
work confirms the numerical consistency of the load-transfer
operators with the corresponding nonlinear displacement mappings
for the tested configurations and virtual directions.

\section{Fluid--structure interaction numerical examples}
\label{sec:fsi_numerical_examples}

Having verified the kinematic and load-transfer operators independently,
the proposed beam mapping formulations are now assessed in
partitioned fluid--structure interaction (FSI) simulations. In contrast
to the previous verification examples, the interface quantities are here
obtained as part of the coupled solution and are exchanged between
non-matching fluid and beam discretizations throughout the simulation.

For all examples, the fluid and structural subproblems are coupled using
a partitioned Dirichlet--Neumann scheme. The structural interface motion
is transferred to the fluid side as a Dirichlet condition, while the
resulting fluid loads are mapped back to the structure as Neumann data.
Strong coupling within each time step is achieved by a
Gauss--Seidel iteration, accelerated using the Multi-Vector Quasi-Newton
(MVQN) method \cite{Bogaers2014}. The proposed beam mapping operators are employed in both
coupling directions to exchange the interface quantities between the beam
and fluid discretizations.

Three numerical examples of increasing complexity are considered. The first two are based on the Mok \cite{mok2001accelerated} and Turek--Hron \cite{TurekHron2006} FSI benchmarks and are used to assess the mapping approach for flexible beam-like structures interacting with an incompressible flow. The final example considers the Goland wing flutter benchmark \cite{10.1115/1.4009489}, employing an irrotational potential-flow formulation with a Lagrangian wake model to investigate a three-dimensional aeroelastic problem involving coupled bending and torsion. Together, these examples assess the robustness and applicability of the proposed beam-to-surface mapping strategy in coupled FSI problems.

\subsection{Mok benchmark}
\label{sec:fsi_mok}

The first FSI example is based on the benchmark proposed by
Mok et al.~\cite{mok2001accelerated}. It consists of a flexible wall attached to
the lower boundary of a two-dimensional channel and subjected to an
incompressible flow. The structure undergoes large displacements during
the simulation, making this example suitable for assessing the proposed
beam mapping in a coupled setting. Both geometrically nonlinear isogeometric beam formulations, Bernoulli and Timoshenko, are considered, together with their corresponding mapping operators.

\begin{figure}[h!]
    \centering
    \includegraphics[width=0.95\textwidth]{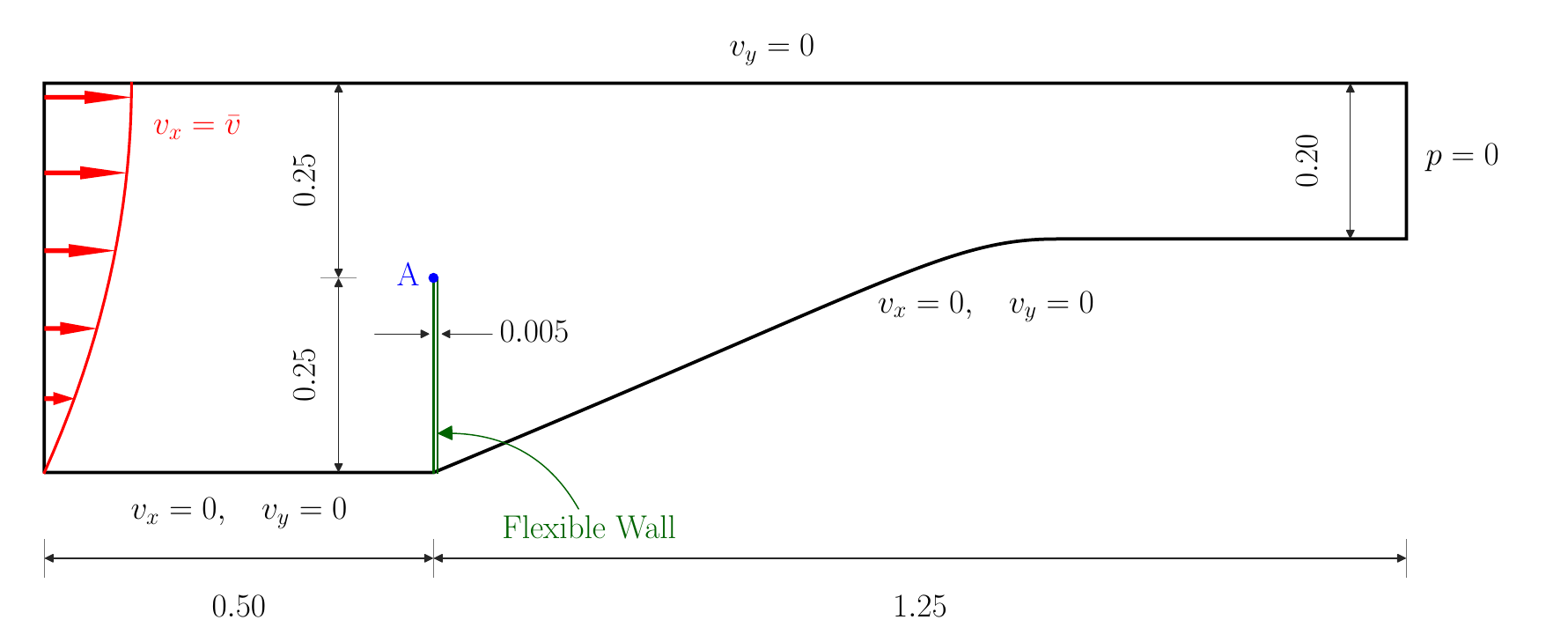}
    \caption{Geometry and boundary conditions of the Mok FSI benchmark
    (adapted from Vald\'es~\cite{valdes2005}).}
    \label{fig:mok_geometry}
\end{figure}

The geometry and boundary conditions are shown in Fig.~\ref{fig:mok_geometry}. The channel has an inlet height of $0.5\,\mathrm{m}$ and contains a vertical flexible wall located $0.5\,\mathrm{m}$ downstream of the inlet. The wall is clamped at the bottom, extends $0.25\,\mathrm{m}$ into the flow, and has a thickness of $0.005\,\mathrm{m}$. Downstream of the flexible wall, the lower channel boundary gradually rises, reducing the channel height to $0.20\,\mathrm{m}$ at the outlet, located $1.25\,\mathrm{m}$ from the wall.

A parabolic velocity profile is prescribed at the inlet according to
\begin{equation}
    v_x(y,t)
    =
    4\bar{v}(t)\,y(1-y),
    \qquad
    v_y(y,t)=0,
    \label{eq:mok_inlet_profile}
\end{equation}
where the reference velocity is gradually increased during the first
$10\,\mathrm{s}$ as
\begin{equation}
    \bar{v}(t)
    =
    \begin{cases}
        \displaystyle
        \frac{0.06067}{2}
        \left[
            1-\cos\left(\frac{\pi t}{10}\right)
        \right],
        & t\leq10\,\mathrm{s},
        \\[3mm]
        0.06067,
        & t>10\,\mathrm{s}.
    \end{cases}
    \label{eq:mok_reference_velocity}
\end{equation}
A zero-pressure condition is imposed at the outlet. A slip condition is
applied at the upper wall, whereas no-slip conditions are imposed on the
remaining rigid walls and on the fluid--structure interface.

The fluid has a density of
$\rho_f=956\,\mathrm{kg/m^3}$ and a dynamic viscosity of
$\mu_f=0.145\,\mathrm{Pa\,s}$. The fluid domain is discretized with
6042 linear triangular elements using a stabilized variational
multiscale formulation based on algebraic subgrid scales
(ASGS)~\cite{CODINA20012681}.

The flexible structure is clamped at its root and is modeled as a
geometrically nonlinear isogeometric beam. The material parameters are
$E=2.3\times10^{6}\,\mathrm{Pa}$, $\nu=0.45$, and
$\rho_s=1500\,\mathrm{kg/m^3}$. A rectangular cross-section with width
$b=1.0\,\mathrm{m}$ and height $h=0.005\,\mathrm{m}$ is considered,
giving a cross-sectional area of $A=0.005\,\mathrm{m^2}$. The beam
centerline is represented by a cubic B-spline curve with eight knot
spans and 11 control points. The same centerline discretization and
material properties are used for the Bernoulli and Timoshenko models.

\begin{figure}[h!]
    \centering
    \includegraphics[width=0.28\textwidth]
        {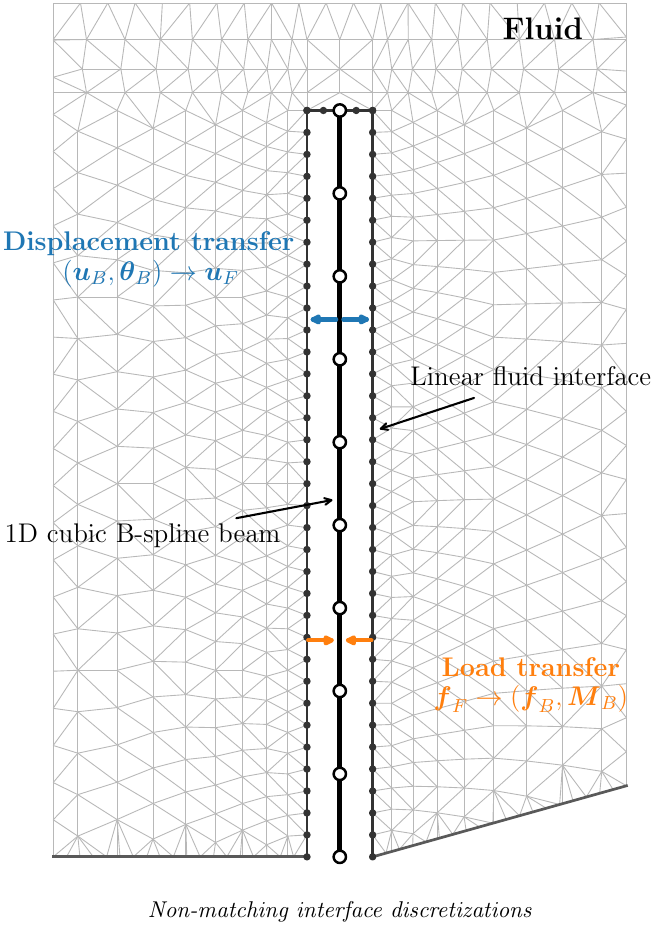}
    \caption{Schematic representation of the non-matching FSI interface
    in the Mok example. The structural model is represented by a
    one-dimensional cubic B-spline beam, while the wet surface is
    discretized independently by linear fluid-interface elements.
    Beam kinematics are transferred to the fluid interface, whereas
    fluid-interface forces are mapped back to the generalized beam
    degrees of freedom. The interface thickness is exaggerated for
    visualization.}
    \label{fig:mok_mapping_scheme}
\end{figure}

On the fluid side, the wet surface of the flexible structure is
represented by linear interface elements. The fluid interface and the
one-dimensional beam discretization are therefore non-matching and have
different geometric representations, as illustrated schematically in
Fig.~\ref{fig:mok_mapping_scheme}. At each coupling iteration, the beam
kinematics are transferred to the wet surface using the corresponding
Bernoulli or Timoshenko beam mapper. The fluid forces are
transferred back to the generalized beam degrees of freedom using the
associated work-conjugate load mapping.  The simulation is performed using the Newmark time integration scheme
with a constant time step of $\Delta t=0.025\,\mathrm{s}$
up to a final time of $t=25\,\mathrm{s}$.

\begin{figure}[h!]
    \centering

    \includegraphics[width=0.95\textwidth]{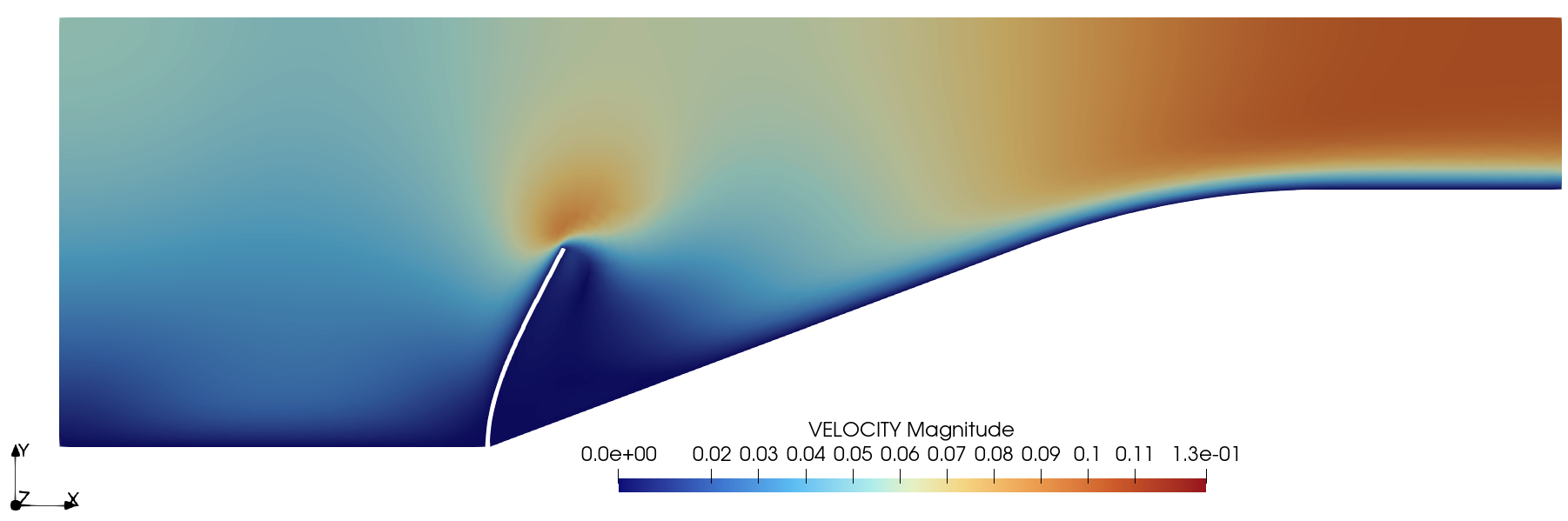}
    \centerline{(a) $t=8\,\mathrm{s}$}

    \vspace{3mm}

    \includegraphics[width=0.95\textwidth]{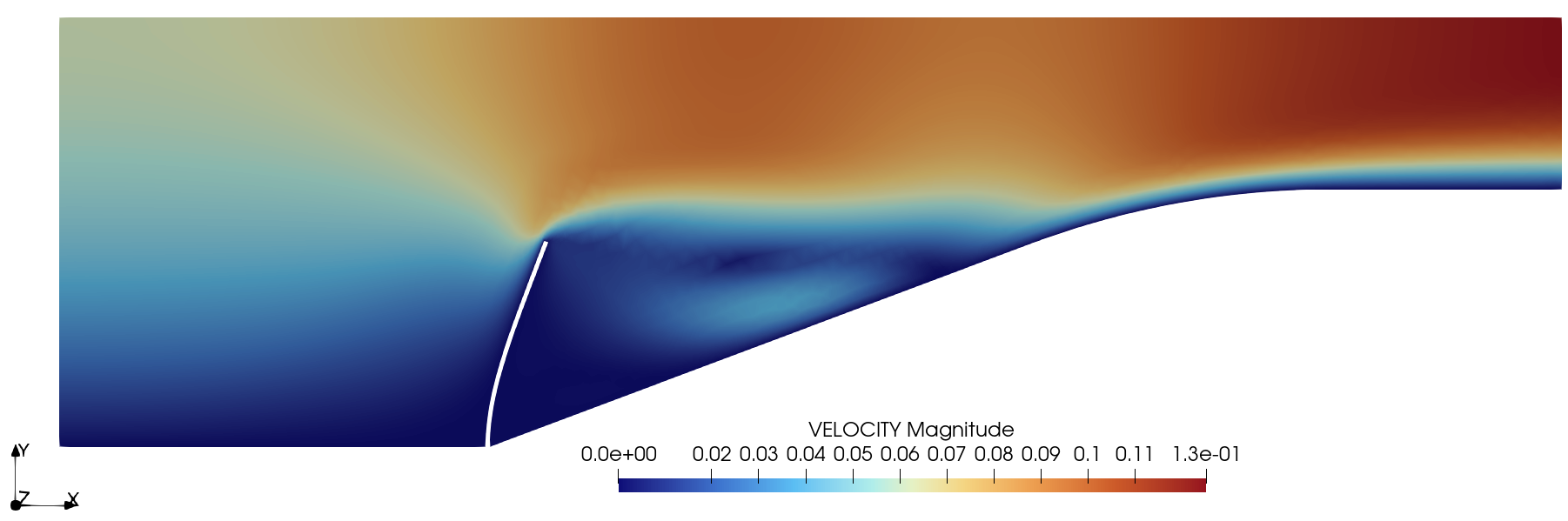}
    \centerline{(b) $t=20\,\mathrm{s}$}

    \caption{Velocity field magnitude and deformed fluid--structure interface
    for the Mok benchmark at (a) $t=8\,\mathrm{s}$ and
    (b) $t=20\,\mathrm{s}$.}
    \label{fig:mok_velocity}
\end{figure}

Figure~\ref{fig:mok_velocity} shows the fluid velocity magnitude and
the deformed fluid--structure interface at $t=8\,\mathrm{s}$ and
$t=20\,\mathrm{s}$. The snapshots illustrate the large deformation of
the flexible wall during the coupled simulation.

For the quantitative assessment, the results obtained with the two beam formulations are compared against two reference solutions. The first corresponds to the body-fitted results reported by Zorrilla et al.~\cite{ZORRILLA2020113179}, considering their finest discretization (Mesh 2), consisting of approximately 65,500 $Q_1/P_1$ triangular fluid elements and 6,400 linear quadrilateral structural elements using a total-Lagrangian formulation. The second reference is computed using the same fluid mesh and coupling setup as in the present beam simulations, but replacing the one-dimensional beam with a two-dimensional isogeometric Kirchhoff--Love shell~\cite{Kiendl2009IGA_KL} acting in membrane action. This second reference enables a more direct assessment of the beam-based representations, since the fluid discretization and coupling procedure remain unchanged, while only the structural model and its associated interface treatment differ.

The main discretization characteristics of the four simulations are
summarized in Table~\ref{tab:mok_discretizations}. The isogeometric shell
reference employs a quadratic isogeometric surface with
$4\times88$ knot spans and 540 control points. In contrast, both beam
models use a cubic centerline discretization with eight knot spans and
only 11 control points. The three Kratos simulations use the same fluid mesh
consisting of 6042 linear triangular elements.

\begin{table}[h!]
    \centering
    \caption{Discretizations used for the numerical comparison of the
    Mok benchmark.}
    \label{tab:mok_discretizations}
    \begin{tabular}{lcccc}
        \hline
        & Zorrilla et al.~\cite{ZORRILLA2020113179}
        & IGA KL shell reference
        & Bernoulli
        & Timoshenko \\
        \hline

        Structural formulation
        & Total Lagrangian
        & KL shell
        & Bernoulli beam
        & Timoshenko beam \\

        Structural discretization
        & FEM
        & IGA
        & IGA
        & IGA \\

        Polynomial degree
        & 1
        & 2
        & 3
        & 3 \\

        FEM elements
        & $\approx 6\,400$
        & --
        & --
        & -- \\

        IGA knot spans
        & --
        & $4\times88$
        & 8
        & 8 \\

        IGA control points
        & --
        & 540
        & 11
        & 11 \\

        Fluid discretization
        & FEM
        & FEM
        & FEM
        & FEM \\

        Fluid elements
        & $\approx 65\,500$
        & 6042
        & 6042
        & 6042 \\

        Fluid element type
        & $Q_1/P_1$ triangles
        & $P_1/P_1$ triangles
        & $P_1/P_1$ triangles
        & $P_1/P_1$ triangles \\

        \hline
    \end{tabular}
\end{table}

\begin{figure}[h!]
    \centering

    \begin{subfigure}[t]{0.49\textwidth}
        \centering
        \includegraphics[width=\textwidth]{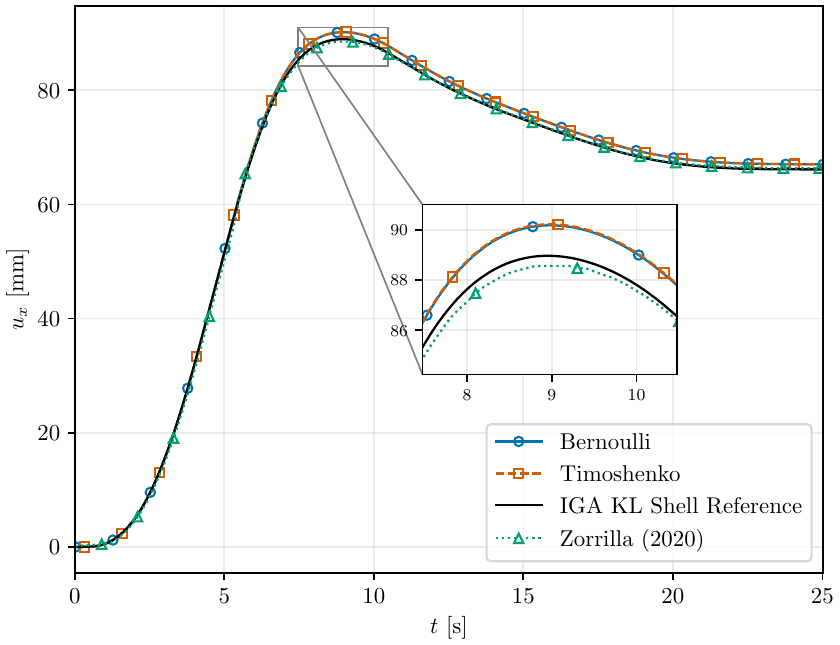}
        \caption{Horizontal tip displacement.}
        \label{fig:mok_tip_displacement}
    \end{subfigure}
    \hfill
    \begin{subfigure}[t]{0.49\textwidth}
        \centering
        \includegraphics[width=\textwidth]{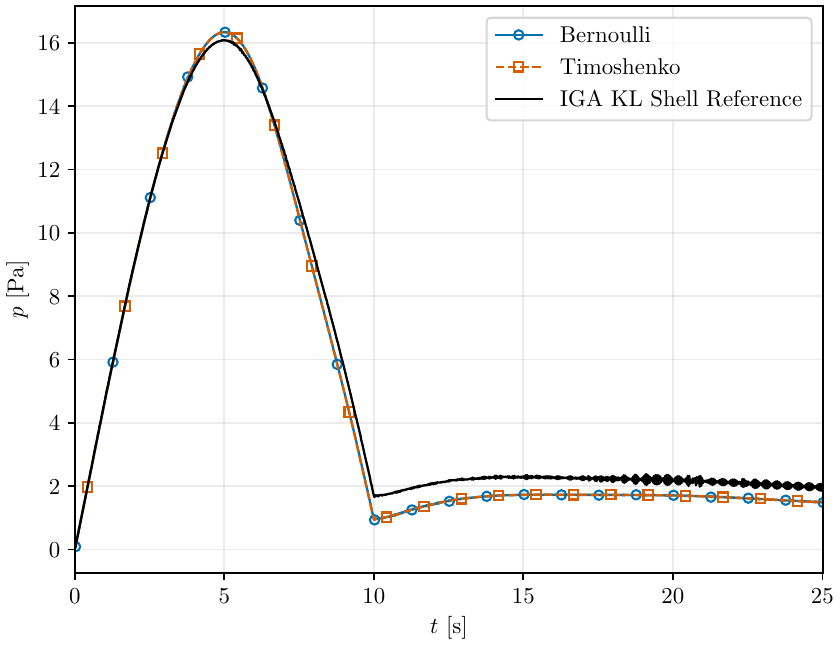}
        \caption{Pressure evolution at the beam tip.}
        \label{fig:mok_tip_pressure}
    \end{subfigure}

    \caption{Transient response of the Mok benchmark:
(a) horizontal tip displacement obtained with the Bernoulli and
Timoshenko beam formulations, the isogeometric Kirchhoff--Love
shell reference, and the body-fitted results of Zorrilla
et al.~\cite{ZORRILLA2020113179};
(b) pressure evolution at the beam tip for the two beam
formulations and the isogeometric Kirchhoff--Love shell reference.}
    \label{fig:mok_response}
\end{figure}

The structural and fluid responses of the Mok benchmark are compared in Fig.~\ref{fig:mok_response}. The horizontal tip displacement, shown in Fig.~\ref{fig:mok_tip_displacement}, increases during the initial loading phase, reaches a maximum at approximately $t=9\,\mathrm{s}$, and subsequently decreases towards a quasi-steady value. The Bernoulli and Timoshenko solutions are nearly indistinguishable, predicting maximum displacements of $90.18\,\mathrm{mm}$ and $90.22\,\mathrm{mm}$, respectively, compared with $88.97\,\mathrm{mm}$ for the isogeometric Kirchhoff--Love shell reference. These values correspond to relative peak-displacement errors of $1.36\%$ and $1.41\%$, while the relative $L^2$ errors over the complete time history are $1.36\%$ and $1.39\%$, respectively. The close agreement is consistent with the predominantly bending-driven response and the similar kinematic assumptions of the beam and Kirchhoff--Love shell formulations. The body-fitted results of Zorrilla et al.~\cite{ZORRILLA2020113179} follow the same overall evolution, with a slightly lower maximum displacement and a comparable quasi-steady response. The remaining differences may be attributed to the different structural formulations, fluid discretizations, and coupling procedures.

The pressure response at the tip region, shown in Fig.~\ref{fig:mok_tip_pressure}, reaches its maximum at approximately $t=5$--$6\,\mathrm{s}$, followed by a rapid decrease until approximately $t=10\,\mathrm{s}$ and a subsequent transition towards an almost constant value. The Bernoulli and Timoshenko formulations closely reproduce the pressure response of the isogeometric Kirchhoff--Love shell reference, showing good agreement in both the transient evolution and the quasi-steady pressure.

\begin{figure}[h!]
    \centering

    \begin{subfigure}[t]{0.49\textwidth}
        \centering
        \includegraphics[width=\textwidth]{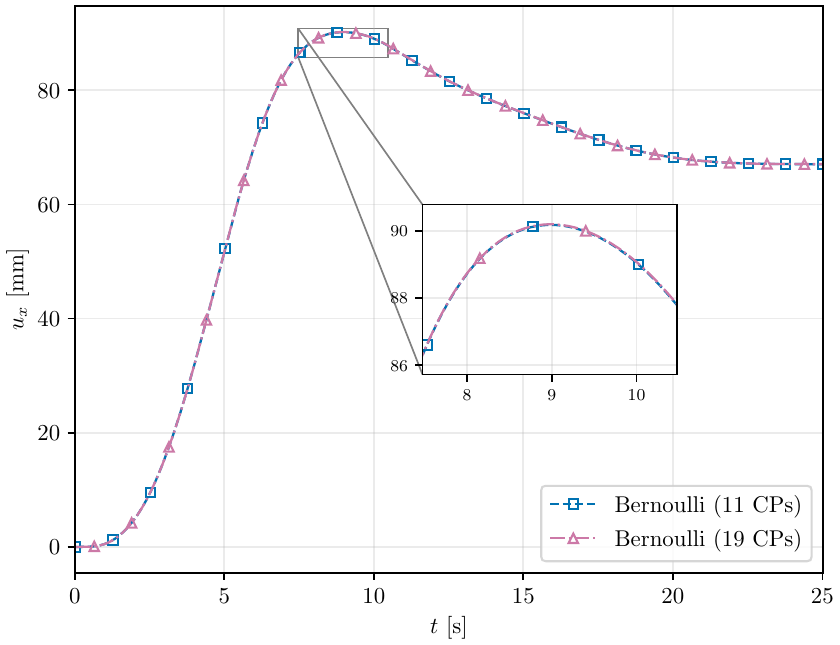}
        \caption{Bernoulli beam.}
        \label{fig:mok_cps_refinement_bernoulli}
    \end{subfigure}
    \hfill
    \begin{subfigure}[t]{0.49\textwidth}
        \centering
        \includegraphics[width=\textwidth]{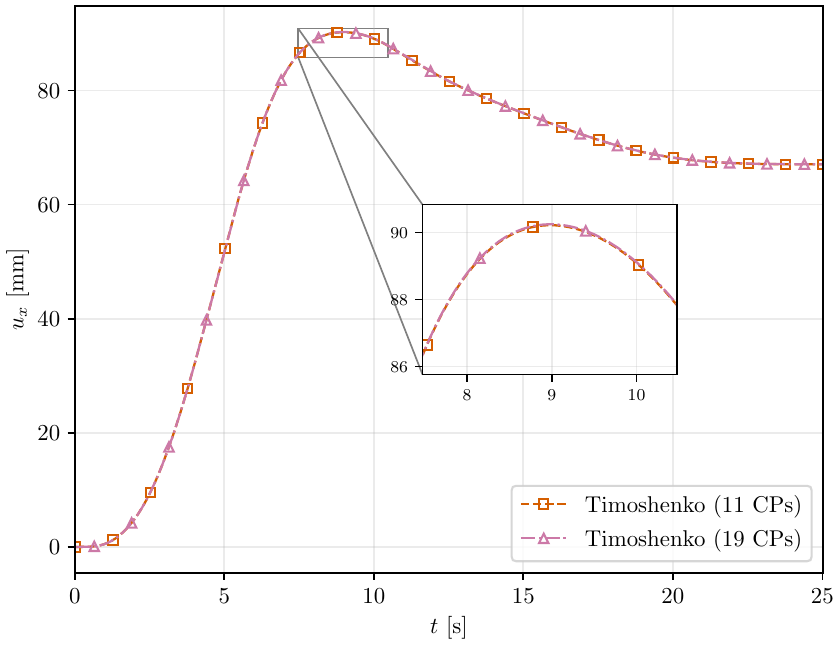}
        \caption{Timoshenko beam.}
        \label{fig:mok_cps_refinement_timoshenko}
    \end{subfigure}

    \caption{Horizontal tip displacement of the Mok benchmark for two
    beam centerline discretizations (11 and 19 control points): (a)
    Bernoulli and (b) Timoshenko formulations.}
    \label{fig:mok_cps_refinement}
\end{figure}

To assess the sensitivity of the coupled response to the beam
centerline discretization, the tip displacement obtained with 11
control points is compared against a refined discretization of 19
control points, for both beam formulations, keeping the cubic
polynomial degree and all remaining modeling choices unchanged.
As shown in Fig.~\ref{fig:mok_cps_refinement}, the two discretizations
yield virtually indistinguishable tip-displacement histories for both
the Bernoulli and Timoshenko beams, confirming that 11 control points
already provide a converged centerline representation for this
example.

Overall, both beam mapping formulations show good agreement with the coupled response of the isogeometric Kirchhoff--Love shell reference, while requiring substantially fewer structural degrees of freedom. The agreement in both tip displacement and fluid pressure supports the applicability of the proposed mapping approach to geometrically nonlinear beam-based FSI simulations.

\subsection{Turek--Hron benchmark}
\label{sec:turek_hron}

As a second fluid--structure interaction example, the proposed
beam mapping operators are assessed using a beam-based adaptation of the
Turek--Hron FSI2 benchmark~\cite{TurekHron2006}. The original benchmark
consists of an incompressible laminar flow through a two-dimensional
channel containing a rigid circular cylinder with a flexible structure
attached to its downstream side. The interaction between the flow and the
flexible structure gives rise to a strongly coupled oscillatory response,
making this configuration a well-established reference problem for the
verification of fluid--structure interaction methods.

In the present work, the flexible structure is replaced by a
one-dimensional geometrically nonlinear isogeometric beam, while its
finite-thickness boundary is retained as the fluid--structure interface,
similarly to the configuration shown in
Fig.~\ref{fig:mok_mapping_scheme}. 

\begin{figure}[h!]
    \centering
    \begin{subfigure}[t]{0.95\textwidth}
        \centering
        \includegraphics[width=\textwidth]{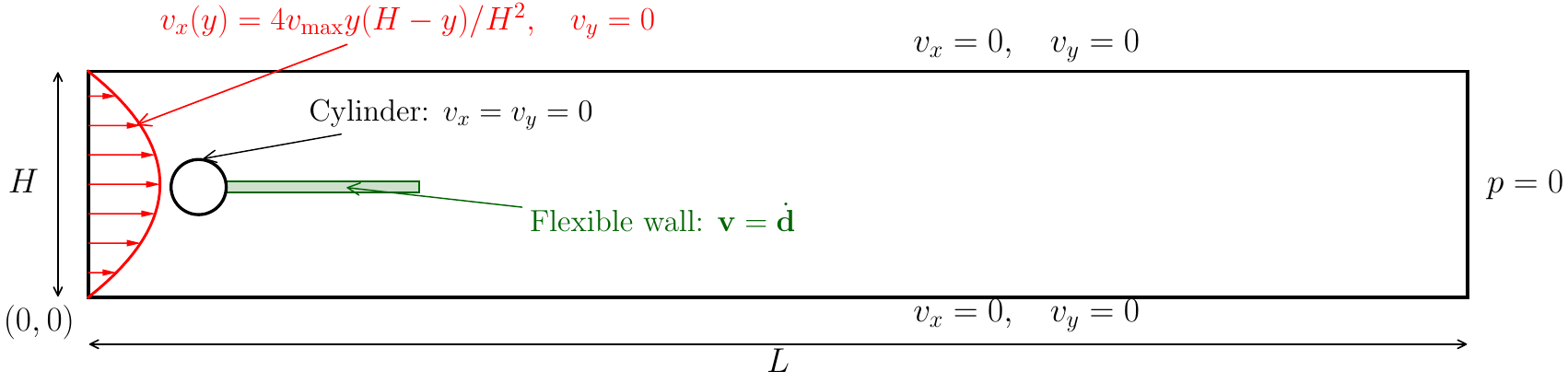}
        \caption{Fluid domain and boundary conditions for the FSI Turek benchmark.}
        \label{fig:turek_fluid_domain}
    \end{subfigure}

    \vspace{0.5em}

    \begin{subfigure}[t]{0.55\textwidth}
        \centering
        \includegraphics[width=\textwidth]{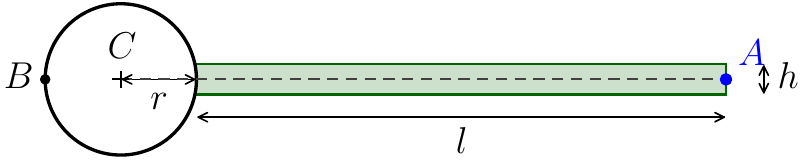}
        \caption{Cylinder–beam geometry and characteristic dimensions. The dashed line denotes the beam centerline; A and B identify reference points.}
        \label{fig:turek_structural_domain}
    \end{subfigure}

    \caption{Geometry and boundary conditions of the beam-based Turek--Hron FSI2 configuration,
    adapted from~\cite{TurekHron2006}.}
    \label{fig:turek_geometry}
\end{figure}

The geometry is based on the FSI2 configuration proposed by Turek and
Hron~\cite{TurekHron2006} and is illustrated in
Fig.~\ref{fig:turek_geometry}. The channel has length
$L=2.5\,\mathrm{m}$ and height $H=0.41\,\mathrm{m}$. A rigid circular
cylinder of radius $r=0.05\,\mathrm{m}$ is centered at
$C=(0.2,0.2)\,\mathrm{m}$. The flexible structure is attached to the
downstream side of the cylinder and has length $l=0.35\,\mathrm{m}$ and
height $h=0.02\,\mathrm{m}$. Its centerline, indicated by the dashed line
in Fig.~\ref{fig:turek_structural_domain}, defines the beam geometry.

The geometric, fluid, and structural parameters are summarized in Table~\ref{tab:turek_parameters}. The fluid is modeled as an incompressible Newtonian fluid using the same element formulation as in the Mok benchmark, namely linear triangular elements with a stabilized variational multiscale formulation based on algebraic subgrid scales (ASGS)~\cite{CODINA20012681}. The original two-dimensional elastic structure is replaced by a one-dimensional beam model.

\begin{table}[h!]
    \centering
    \caption{Parameters of the beam-based Turek--Hron FSI2 configuration.}
    \label{tab:turek_parameters}
    \begin{tabular}{lll@{\qquad}lll}
        \hline
        Parameter & Symbol & Value
        & Parameter & Symbol & Value \\
        \hline
        \noalign{\vskip 0.1cm}
        Channel length
        & $L$ & $2.5\,\mathrm{m}$
        & Fluid density
        & $\rho_f$ & $1000\,\mathrm{kg/m^3}$ \\

        Channel height
        & $H$ & $0.41\,\mathrm{m}$
        & Dynamic viscosity
        & $\mu_f$ & $1\,\mathrm{Pa\,s}$ \\

        Cylinder radius
        & $r$ & $0.05\,\mathrm{m}$
        & Mean inlet velocity
        & $\bar{v}$ & $1\,\mathrm{m/s}$ \\

        Cylinder center
        & $C$ & $(0.2,0.2)\,\mathrm{m}$
        & Structural density
        & $\rho_s$ & $10^4\,\mathrm{kg/m^3}$ \\

        Structure length
        & $l$ & $0.35\,\mathrm{m}$
        & Young's modulus
        & $E$ & $1.4\times10^6\,\mathrm{Pa}$ \\

        Structure height
        & $h$ & $0.02\,\mathrm{m}$
        & Poisson's ratio
        & $\nu$ & $0.4$ \\
        \hline
    \end{tabular}
\end{table}

The fluid boundary conditions follow the FSI2 configuration of
Turek and Hron~\cite{TurekHron2006}. A no-slip condition is imposed on
the upper and lower channel walls and on the surface of the rigid
cylinder, while zero pressure is prescribed at the outlet. At the inlet,
a parabolic velocity profile is imposed,
\begin{equation}
    v_x(y,t)
    =
    1.5\,\bar{v}\,
    \frac{y(H-y)}{(H/2)^2}\,
    \gamma(t),
    \qquad
    v_y(y,t)=0,
    \label{eq:turek_inlet}
\end{equation}
where the startup function
\begin{equation}
    \gamma(t)
    =
    \begin{cases}
        \dfrac{1-\cos(\pi t/2)}{2},
        & 0 \leq t < 2\,\mathrm{s}, \\[2mm]
        1,
        & t \geq 2\,\mathrm{s}
    \end{cases}
    \label{eq:turek_ramp}
\end{equation}
smoothly increases the inlet velocity from rest during the first
$2\,\mathrm{s}$. On the moving fluid--structure interface, the no-slip
condition is imposed using the velocity obtained from the beam
mapping. The beam is clamped at its attachment to the rigid cylinder and
is otherwise free.

The numerical discretizations employed for the comparison are summarized in Table~\ref{tab:turek_discretizations}. The Bernoulli and Timoshenko models employ cubic isogeometric beam discretizations with eight knot spans, while the isogeometric Kirchhoff--Love shell reference uses quadratic basis functions with $8\times8$ knot spans. For all three models considered in the present work, the same fluid mesh consisting of 29\,926 linear triangular finite elements is employed. The structural dynamics are integrated using the Newmark scheme, and the simulations are performed up to $t=20\,\mathrm{s}$ with a constant time step of $\Delta t=0.002\,\mathrm{s}$. For comparison, the results reported by Zorrilla et al.~\cite{ZORRILLA2020113179}, obtained using approximately 28\,000 linear triangular fluid elements (Q1P1) and 2\,600 linear quadrilateral Total Lagrangian structural elements with the same time-step size, are also considered.

\begin{table}[h!]
    \centering
    \caption{Discretizations used for the numerical comparison of the
    beam-based Turek--Hron FSI2 benchmark.}
    \label{tab:turek_discretizations}
    \begin{tabular}{lcccc}
        \hline
        & Zorrilla et al.~\cite{ZORRILLA2020113179}
        & IGA KL shell reference
        & Bernoulli
        & Timoshenko \\
        \hline

        Structural model
        & Total Lagrangian solid
        & KL shell
        & Bernoulli beam
        & Timoshenko beam \\

        Structural discretization
        & FEM
        & IGA
        & IGA
        & IGA \\

        Polynomial degree
        & 1
        & 2
        & 3
        & 3 \\

        FEM elements
        & $\approx 2\,600$
        & --
        & --
        & -- \\

        IGA knot spans
        & --
        & $8\times8$
        & 8
        & 8 \\

        Fluid discretization
        & FEM
        & FEM
        & FEM
        & FEM \\

        Fluid elements
        & $\approx 28\,000$
        & 29\,926
        & 29\,926
        & 29\,926 \\

       Fluid element type
        & $Q_1/P_1$ triangles
        & $P_1/P_1$ triangles
        & $P_1/P_1$ triangles
        & $P_1/P_1$ triangles \\
        
        Time step
        & $0.002\,\mathrm{s}$
        & $0.002\,\mathrm{s}$
        & $0.002\,\mathrm{s}$
        & $0.002\,\mathrm{s}$ \\

        \hline
    \end{tabular}
\end{table}

\begin{figure}[h!]
    \centering
    \begin{subfigure}[t]{0.95\textwidth}
        \centering
        \includegraphics[width=\textwidth]{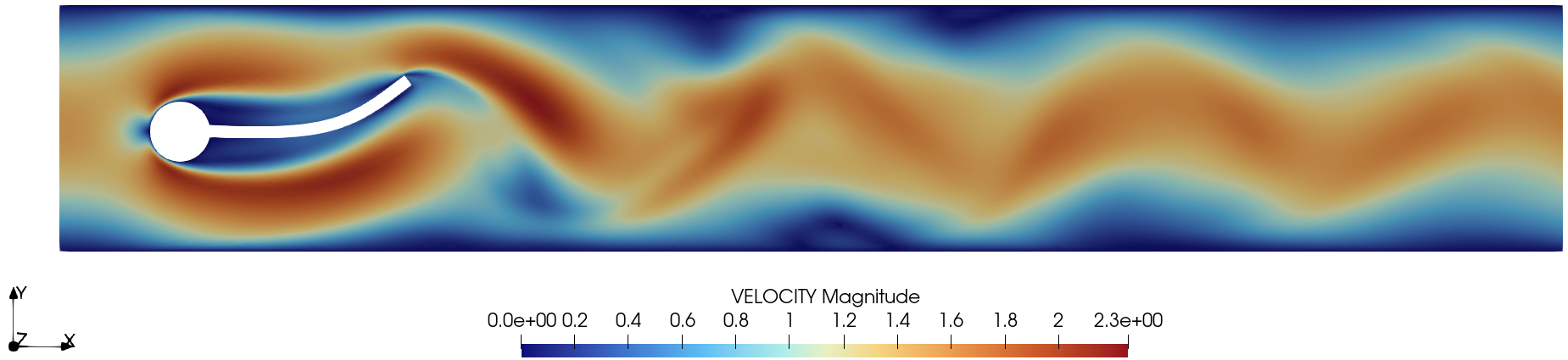}
        \caption{$t=10\,\mathrm{s}$.}
        \label{fig:turek_velocity_10}
    \end{subfigure}

    \vspace{0.5em}

    \begin{subfigure}[t]{0.95\textwidth}
        \centering
        \includegraphics[width=\textwidth]{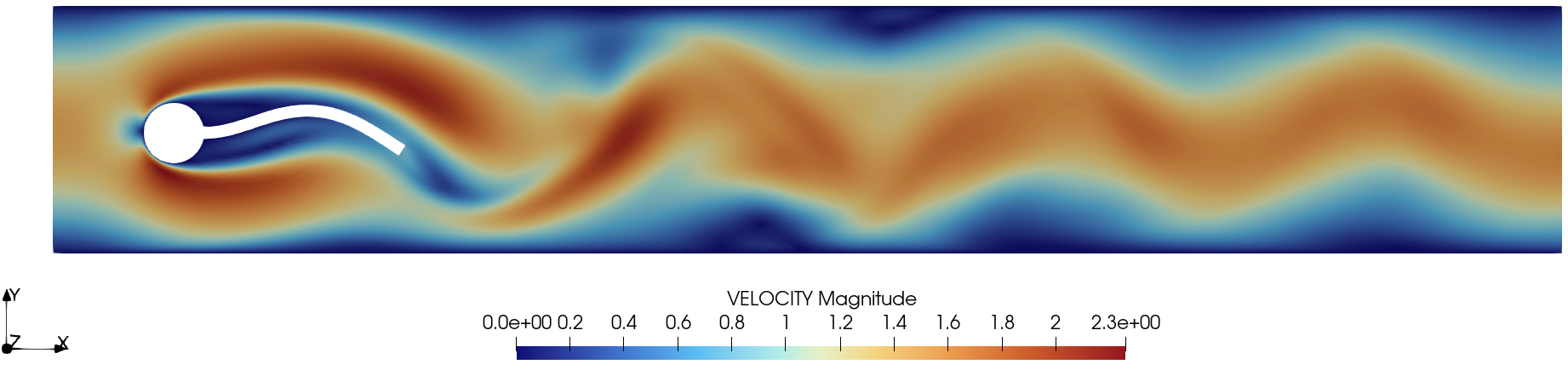}
        \caption{$t=12\,\mathrm{s}$.}
        \label{fig:turek_velocity_12}
    \end{subfigure}

    \caption{Instantaneous fluid velocity-magnitude fields for the
    beam-based Turek--Hron FSI2 configuration at
    (a) $t=10\,\mathrm{s}$ and (b) $t=12\,\mathrm{s}$.}
    \label{fig:turek_velocity}
\end{figure}

Figure~\ref{fig:turek_velocity} shows instantaneous velocity-magnitude
fields at $t=10\,\mathrm{s}$ and $t=12\,\mathrm{s}$. At these instants,
the initial transient has developed into the characteristic oscillatory
fluid--structure response. The deformation of the flexible structure
induces an alternating wake downstream of the cylinder, with the
position of the structure and the associated velocity field changing
between the two instants. The snapshots also illustrate the large
deformation undergone by the structure and the corresponding motion of
the reconstructed finite-thickness fluid--structure interface.

\begin{figure}[h!]
    \centering
    \begin{subfigure}[t]{0.49\textwidth}
        \centering
        \includegraphics[width=\textwidth]{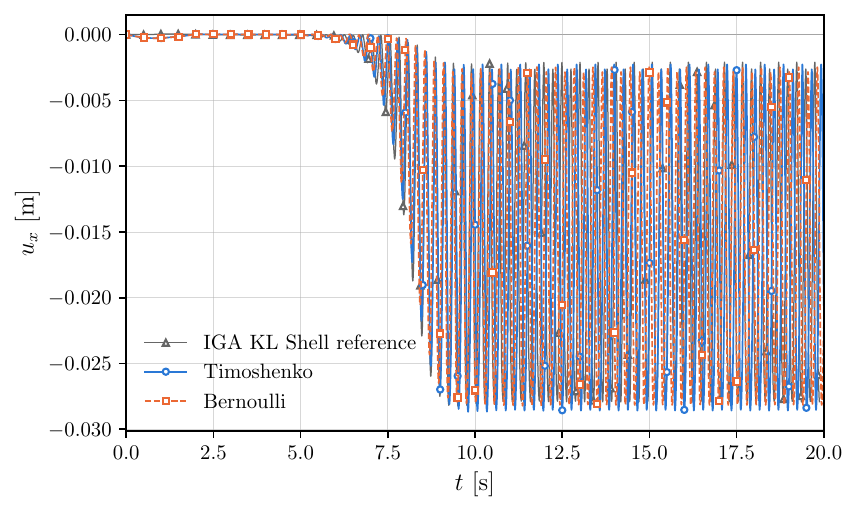}
        \caption{Horizontal displacement $u_x$.}
        \label{fig:turek_ux}
    \end{subfigure}
    \hfill
    \begin{subfigure}[t]{0.49\textwidth}
        \centering
        \includegraphics[width=\textwidth]{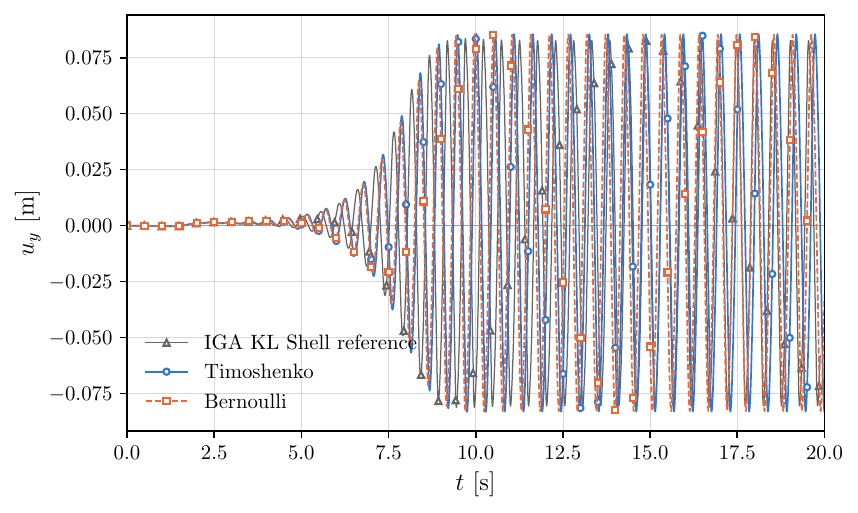}
        \caption{Vertical displacement $u_y$.}
        \label{fig:turek_uy}
    \end{subfigure}

    \caption{Tip-displacement histories for the Bernoulli and Timoshenko
    beam formulations compared with the isogeometric
    Kirchhoff--Love shell reference solution.}
    \label{fig:turek_tip_history}
\end{figure}

The coupled response is evaluated through the horizontal and vertical
displacement histories of the beam tip $A$, indicated in
Fig.~\ref{fig:turek_structural_domain}. Figure~\ref{fig:turek_tip_history}
compares the results obtained with the Bernoulli and Timoshenko beam
models with an isogeometric Kirchhoff--Love shell reference \cite{Kiendl2009IGA_KL} of the same
configuration. The latter retains the finite-thickness structural
representation and therefore provides a reference for assessing the
effect of replacing the structural domain by a one-dimensional beam.

Over the complete interval $0\leq t\leq20\,\mathrm{s}$, the three
structural descriptions reproduce the same characteristic evolution of
the coupled response. Following the initial startup phase, the
oscillation amplitude increases rapidly and subsequently approaches an
approximately periodic regime. This behavior is observed for both the
horizontal and vertical tip displacements. The two beam formulations
closely reproduce the displacement amplitudes and oscillation frequency
obtained with the isogeometric shell reference. In particular, the
Timoshenko solution remains very close to the shell response, whereas
the Bernoulli solution exhibits a phase difference while maintaining
comparable amplitudes and frequency.

\begin{figure}[h!]
    \centering
    \begin{subfigure}[t]{0.49\textwidth}
        \centering
        \includegraphics[width=\textwidth]{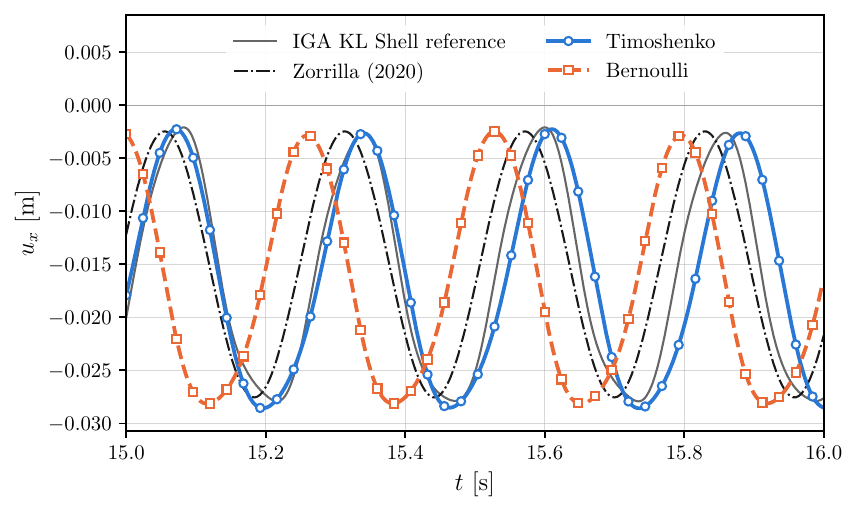}
        \caption{Horizontal displacement $u_x$.}
        \label{fig:turek_ux_zoom}
    \end{subfigure}
    \hfill
    \begin{subfigure}[t]{0.49\textwidth}
        \centering
        \includegraphics[width=\textwidth]{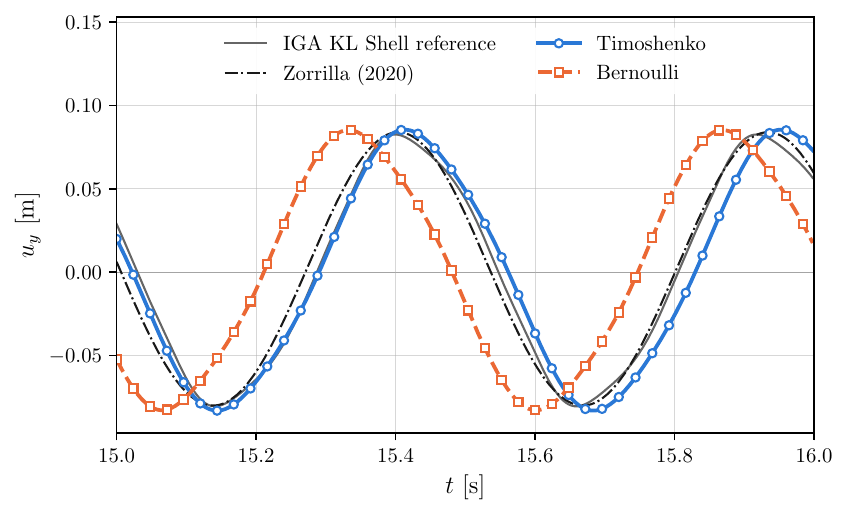}
        \caption{Vertical displacement $u_y$.}
        \label{fig:turek_uy_zoom}
    \end{subfigure}

    \caption{Detailed comparison of the periodic tip response over
    $15\leq t\leq16\,\mathrm{s}$. The Bernoulli and Timoshenko beam
    solutions are compared with the isogeometric Kirchhoff--Love shell
    model and the reference results of Zorrilla et
    al.~\cite{ZORRILLA2020113179}.}
    \label{fig:turek_tip_zoom}
\end{figure}

To examine the established oscillatory response in greater detail,
Fig.~\ref{fig:turek_tip_zoom} shows the interval
$15\leq t\leq16\,\mathrm{s}$. In addition to the isogeometric shell
solution, the reference body-fitted response reported by Zorrilla et
al.~\cite{ZORRILLA2020113179} is included. This interval allows the amplitude,
frequency, and relative phase of the different solutions to be compared
more clearly.

The detailed comparison shows very good agreement in the dominant characteristics of the response. For the horizontal displacement, the different models predict nearly identical oscillation amplitudes and frequencies. The Timoshenko response follows the isogeometric shell solution particularly closely, whereas the Bernoulli solution exhibits a phase lead of approximately $0.1\,\mathrm{s}$ while preserving similar amplitudes and frequencies. Phase differences between numerical solutions were also reported by Zorrilla et al.~\cite{ZORRILLA2020113179} when comparing body-fitted and embedded FSI formulations for the same benchmark. A similar behavior is observed for the vertical displacement. In the established oscillatory regime ($15 \leq t \leq 20\,\mathrm{s}$), the Bernoulli and Timoshenko formulations predict vertical tip-displacement amplitudes of $84.06\,\mathrm{mm}$ and $84.29\,\mathrm{mm}$, respectively, with corresponding oscillation frequencies of $1.865\,\mathrm{Hz}$ and $1.854\,\mathrm{Hz}$. The differences between the two beam formulations remain below $0.6\%$ for both quantities. The peak positive and negative displacements predicted by both beam models also agree closely with the shell and literature reference solutions.

\subsection{Goland wing flutter benchmark}
\label{sec:goland_wing}

The third example considers the uniform cantilever wing introduced
by Goland~\cite{10.1115/1.4009489}, a classical benchmark for the
prediction of aeroelastic flutter. Flutter is a dynamic instability
arising from the interaction between structural deformation and
unsteady aerodynamic loads, leading to self-excited oscillations
above a critical freestream velocity. For the Goland wing, the
instability is characterized by coupled flapwise bending and torsion,
making this configuration particularly relevant for assessing the
transfer of rotational motion and eccentric aerodynamic loads between
the structural and aerodynamic discretizations.

Both the Bernoulli and Timoshenko beam formulations are considered
together with their corresponding mapping operators.
The DUST--MBDyn results reported by Savino et al.~\cite{Savino2021}
are used as an additional reference based on three-dimensional
aerodynamics.

\begin{figure}[h!]
    \centering
    \includegraphics[width=\linewidth]{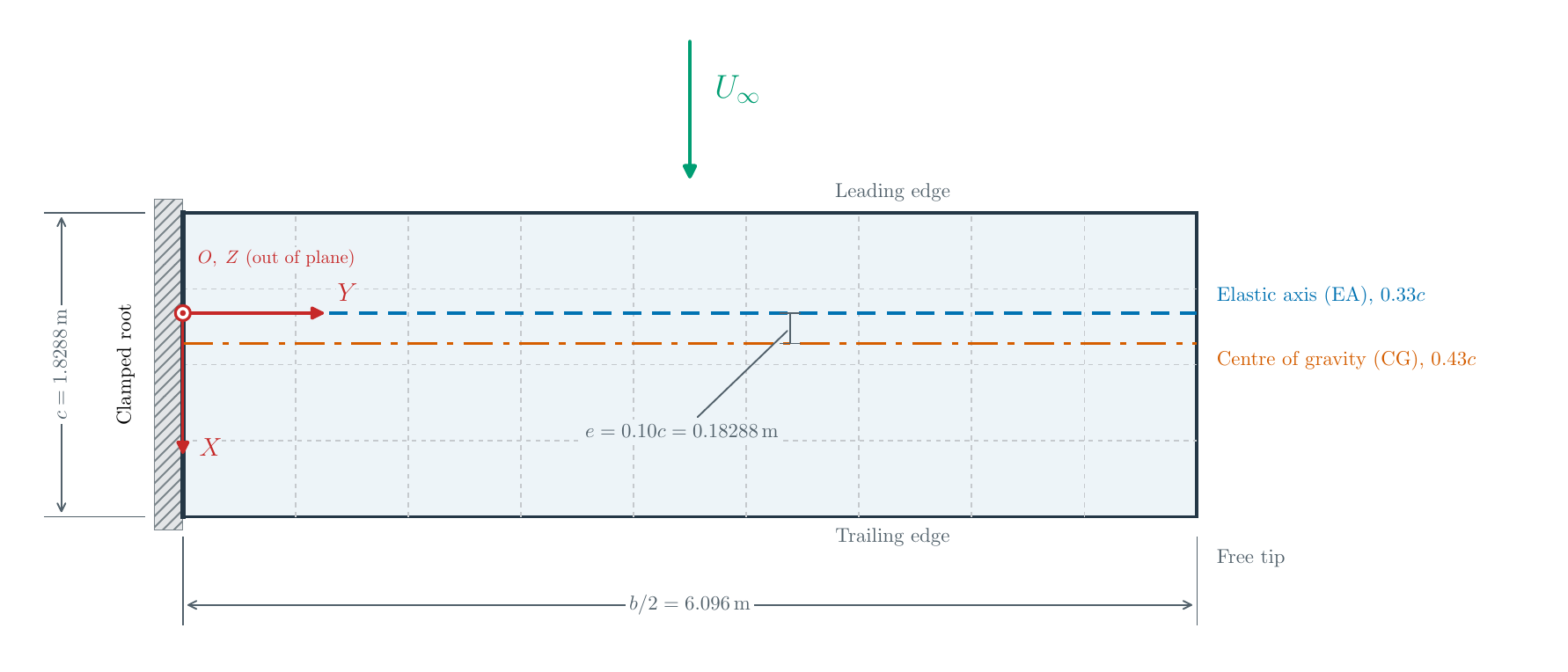}
    \caption{Goland wing geometry, boundary condition, coordinate
    system, and schematic aerodynamic surface panels. EA and CG
    denote the elastic axis and sectional center of gravity,
    respectively. The illustrated panels do not represent the
    actual computational discretization.}
    \label{fig:goland_geometry}
\end{figure}

The modeled wing has a rectangular planform with a cantilever
semi-span of $L=6.096\,\mathrm{m}$ and a chord of
$c=1.8288\,\mathrm{m}$. The structural reference line coincides
with the elastic axis, located at $33\%$ of the chord measured
from the leading edge. The sectional center of gravity is located
at $43\%$ of the chord, resulting in a chordwise eccentricity of
$e=0.18288\,\mathrm{m}$. The global $X$, $Y$, and $Z$ axes denote
the streamwise, spanwise, and flapwise directions, respectively.
The beam is clamped at the root and free at the tip.
Figure~\ref{fig:goland_geometry} illustrates the geometry,
coordinate system, and a schematic representation of the
aerodynamic surface discretization.

The principal parameters are summarized in
Table~\ref{tab:goland_parameters}, following the configuration
reported by Savino et al.~\cite{Savino2021}. The sectional
properties are prescribed to reproduce the target line mass,
flapwise bending rigidity, torsional rigidity, and mass moment
of inertia about the elastic axis. The center-of-gravity
eccentricity is accounted for through the translation--rotation
coupling terms of the sectional mass matrix.

\begin{table}[h!]
    \centering
    \caption{Principal parameters of the Goland wing configuration.}
    \label{tab:goland_parameters}
    \begin{tabular}{lll}
        \hline
        Parameter & Symbol & Value \\
        \hline
        Cantilever semi-span & $L$ & $6.096\,\mathrm{m}$ \\
        Nominal chord & $c$ & $1.8288\,\mathrm{m}$ \\
        Elastic-axis position & $x_{\mathrm{EA}}/c$ & $0.33$ \\
        Center-of-gravity position & $x_{\mathrm{CG}}/c$ & $0.43$ \\
        Sectional eccentricity & $e$ & $0.18288\,\mathrm{m}$ \\
        Mass per unit length & $m$ & $35.71\,\mathrm{kg/m}$ \\
        Flapwise bending rigidity & $EI$ &
        $9.77\times10^{6}\,\mathrm{N\,m^{2}}$ \\
        Torsional rigidity & $GJ$ &
        $0.99\times10^{6}\,\mathrm{N\,m^{2}}$ \\
        Sectional inertia about the elastic axis
        & $I_{\theta}$ & $8.64\,\mathrm{kg\,m}$ \\
        Freestream density & $\rho_f$ & $1.02\,\mathrm{kg/m^{3}}$ \\
        Initial angle of attack & $\alpha_0$ & $0.05^{\circ}$ \\
        \hline
    \end{tabular}
\end{table}

The beam centerline is represented by a single cubic B-spline
curve with an open knot vector and uniformly spaced internal
knots. The baseline structural discretization comprises 16 knot
spans and 19 control points. At the root, the Bernoulli formulation
constrains the centerline displacement, slope, and scalar twist,
whereas the Timoshenko formulation constrains the displacement
and independent cross-sectional rotations. In the equivalent-section
model employed here, the in-plane bending rigidity is set equal
to the flapwise bending rigidity.

Prior to aerodynamic coupling, the structural discretizations are
verified through an eigenvalue analysis. The first natural
frequencies obtained with the Bernoulli and Timoshenko formulations
are $7.6645\,\mathrm{Hz}$ and $7.6534\,\mathrm{Hz}$, respectively,
in close agreement with the first bending frequency of
$7.66\,\mathrm{Hz}$ reported by Savino et al.~\cite{Savino2021}.
This comparison provides an independent verification of the
structural model before introducing the aerodynamic coupling.

\begin{figure}[h!]
    \centering
    \includegraphics[width=0.60\linewidth]{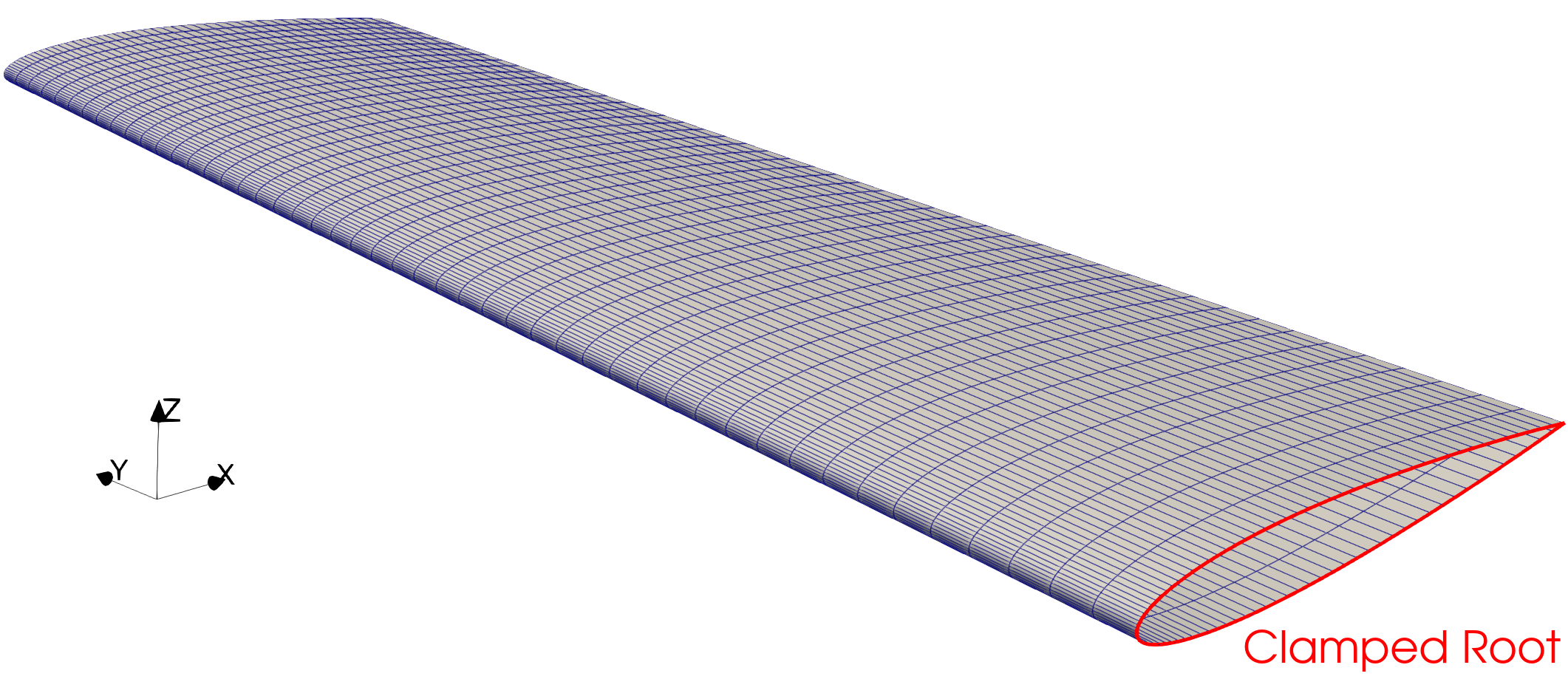} 
\caption{Aerodynamic surface discretization of the Goland wing
employed in DUST, comprising $48\times48$ surface panels (SP).
The panels are uniformly distributed in the spanwise direction
and concentrated towards the leading edge in the chordwise
direction. The clamped wing root is highlighted in red.}
    \label{fig:goland_aero_mesh}
\end{figure}

The aerodynamic problem is solved using DUST~\cite{TUGNOLI2021106804},
which combines potential-flow boundary elements with a Lagrangian
representation of the wake. Further details on the aerodynamic
formulation and its mathematical foundations are provided in the
cited reference. The aerodynamic surface is discretized using
$48\times48$ panels in the chordwise and spanwise directions,
respectively, as shown in Fig.~\ref{fig:goland_aero_mesh}. The panels
are uniformly distributed along the span, whereas the chordwise
discretization is refined towards the leading edge to better resolve
the larger pressure gradients occurring in this region. Up to
$200\,000$ particles are employed for the wake representation, and a
time-step size of $\Delta t=0.001\,\mathrm{s}$ is used for the coupled
simulations.

DUST and the isogeometric structural solver in Kratos
Multiphysics~\cite{dadvand2010,dadvand2013,KratosMultiphysics2025}
are coupled through a partitioned co-simulation approach~\cite{bucher2024}.
At each coupling step, the proposed mapping operators transfer
the structural motion to the aerodynamic surface and the resulting
aerodynamic loads back to the beam. In addition to surface
displacements, DUST requires the corresponding surface velocities
to impose the slip boundary condition on the moving aerodynamic
surface.

The surface velocities are obtained by differentiating the
reconstructed surface positions with respect to time.
For the Timoshenko formulation, the surface reconstruction
$\mathbf{x}_Q=\mathbf{x}_c+\mathbf{R}\boldsymbol{\rho}_{0,Q}$
gives
\begin{equation}
\begin{aligned}
    \mathbf{v}_Q
    &=
    \dot{\mathbf{x}}_c
    +
    \dot{\mathbf{R}}\boldsymbol{\rho}_{0,Q}
    \\
    &=
    \dot{\mathbf{x}}_c
    +
    \boldsymbol{\omega}\times\boldsymbol{\rho}_Q
    \\
    &=
    \mathbf{H}_Q^{\mathrm{Tim}}\mathbf{v}_b,
\end{aligned}
\end{equation}
where $\boldsymbol{\omega}$ is the spatial angular velocity
of the cross-section, defined by
$\dot{\mathbf{R}}=[\boldsymbol{\omega}]_\times\mathbf{R}$.
The resulting expression has the same structure as the
linearized surface displacement, with translational and
angular velocities replacing the corresponding virtual
translations and rotations. Here, $\mathbf{v}_b$ collects
the control-point translational and spatial angular velocities.

For the Bernoulli formulation, the surface velocity follows
from the time derivative of the centerline- and director-based
reconstruction. Applying the chain rule gives
\begin{equation}
\begin{aligned}
    \mathbf{v}_Q
    &=
    \dot{\mathbf{x}}_c
    +
    \rho_{2,Q}\dot{\mathbf{a}}_2
    +
    \rho_{3,Q}\dot{\mathbf{a}}_3
    \\
    &=
    \left(
    \frac{\partial\mathbf{x}_c}{\partial\mathbf{q}_b}
    +
    \rho_{2,Q}\frac{\partial\mathbf{a}_2}{\partial\mathbf{q}_b}
    +
    \rho_{3,Q}\frac{\partial\mathbf{a}_3}{\partial\mathbf{q}_b}
    \right)\dot{\mathbf{q}}_b
    \\
    &=
    \mathbf{H}_Q^{\mathrm{Ber}}\dot{\mathbf{q}}_b.
\end{aligned}
\end{equation}
The director derivatives account for both the change in
cross-section orientation induced by the evolving centerline
tangent and the rotation associated with the twist rate.
These contributions are already included in the linearization
of the surface reconstruction, allowing the same Bernoulli
mapping operator to transfer the generalized beam velocities,
consisting of control-point translational velocities and
twist rates, to the aerodynamic surface.

Thus, for both beam formulations, the surface velocities are
evaluated using the same configuration-dependent kinematic
mapping operators previously derived for the displacement
linearization, ensuring consistency between the displacement
and velocity transfers.

A small initial incidence of $\alpha_0=0.05^{\circ}$ is prescribed
to excite the coupled bending--torsion response, following
Savino et al.~\cite{Savino2021}. The aeroelastic stability is
assessed from the transient evolution of the flapwise tip
displacement and tip twist as the freestream velocity is varied.
Below the flutter boundary, the oscillation amplitude decays,
whereas above the critical velocity it grows. The modal frequency
and growth rate are extracted from the tip-displacement histories
using the matrix-pencil method~\cite{Hua1990}, and the flutter
boundary is determined from the zero crossing of the growth rate
by linear interpolation between neighboring freestream velocities.

\begin{figure}[h!]
    \centering
    \begin{minipage}[t]{0.49\linewidth}
        \centering
        \includegraphics[width=\linewidth]
        {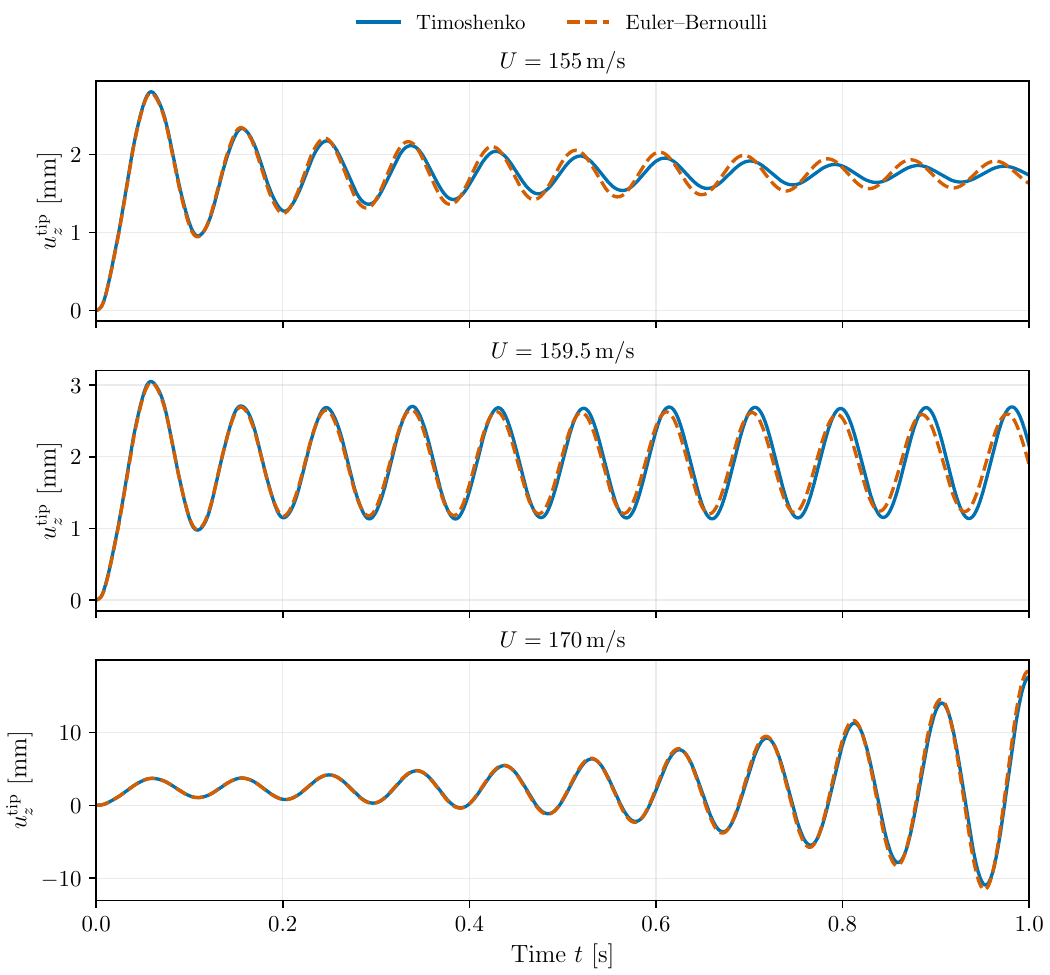}
        \caption*{(a) Flapwise tip displacement}
    \end{minipage}
    \hfill
    \begin{minipage}[t]{0.49\linewidth}
        \centering
        \includegraphics[width=\linewidth]
        {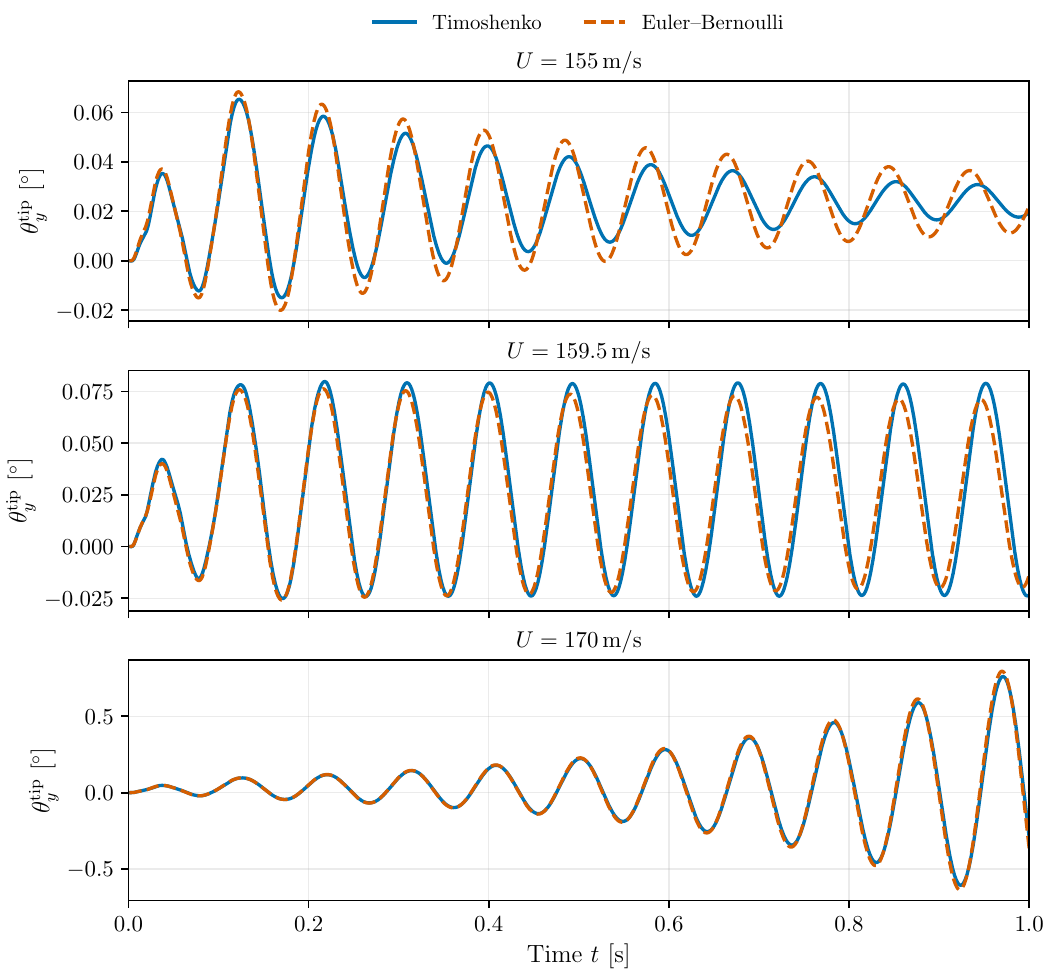}
        \caption*{(b) Tip twist}
    \end{minipage}

    \caption{Transient response of the Goland wing obtained with the
    Bernoulli and Timoshenko beam formulations at
    $U_\infty=155$, $159.5$, and $170\,\mathrm{m/s}$:
    (a) flapwise tip displacement and (b) tip twist.}
    \label{fig:goland_tip_response}
\end{figure}

Figure~\ref{fig:goland_tip_response} shows the flapwise tip
displacement and tip twist for three representative freestream
velocities. At
$U_\infty=155\,\mathrm{m/s}$, the oscillations decay, indicating a
stable response. At $U_\infty=159.5\,\mathrm{m/s}$, the response is
only weakly damped and lies close to the flutter boundary, whereas
at $U_\infty=170\,\mathrm{m/s}$ rapidly growing oscillations indicate
an unstable bending--torsion response. The Bernoulli and Timoshenko
formulations show very similar behavior in all three regimes.

The matrix-pencil analysis yields a flutter velocity and frequency of
$U_f=159.97\,\mathrm{m/s}$ and $f_f=10.94\,\mathrm{Hz}$ for the
Bernoulli formulation, and $U_f=159.53\,\mathrm{m/s}$ and
$f_f=10.89\,\mathrm{Hz}$ for the Timoshenko formulation. The flutter
velocities differ by only $0.28\%$, confirming the close agreement
between the two mapping formulations.

Table~\ref{tab:goland_flutter_comparison} compares the present flutter predictions with representative results from the literature. The reported flutter velocities can be broadly divided into two groups according to the aerodynamic formulation. Two-dimensional strip-theory approaches predict flutter velocities of approximately $135$--$137\,\mathrm{m/s}$, whereas three-dimensional aerodynamic models yield values of approximately $164$--$174\,\mathrm{m/s}$. As discussed by Savino et al.~\cite{Savino2021}, this difference highlights the importance of three-dimensional aerodynamic effects in the flutter analysis of low-aspect-ratio wings, which are not captured by strip theory. Since the present simulations employ the three-dimensional surface-panel formulation of DUST, comparisons with the latter group are more relevant.

Among these studies, Savino et al.~\cite{Savino2021} provide a particularly relevant reference, as they also employ DUST coupled with the multibody solver MBDyn. Their reported flutter velocity of $174.2\,\mathrm{m/s}$ is higher than the present predictions of $159.97\,\mathrm{m/s}$ and $159.53\,\mathrm{m/s}$ for the Bernoulli and Timoshenko formulations, respectively. However, Savino et al. use a distance-based interpolation between structural and aerodynamic nodes, whereas the present work employs geometrically nonlinear isogeometric beams with kinematically consistent mapping operators. Differences in structural modeling and coupling strategies therefore need to be considered when interpreting the comparison. Under identical aerodynamic and coupling conditions, the two beam formulations predict flutter velocities differing by only $0.28\%$, with corresponding flutter frequencies close to the reference values of approximately $11\,\mathrm{Hz}$.

\begin{table}[h!]
    \centering
    \caption{Comparison of flutter speed and frequency for the
    Goland wing.}
    \label{tab:goland_flutter_comparison}
    \begin{tabular}{l l c c}
        \hline
        Reference & Model & $U_f$ [m/s] & $f_f$ [Hz] \\
        \hline
        Goland~\cite{10.1115/1.4009489}
        & Analytical
        & 137.2 & 11.25 \\
        Patil et al.~\cite{Patil2000}
        & Intrinsic beam + strip theory
        & 135.6 & 11.17 \\
        Wang et al.~\cite{Wang2006}
        & ZAERO
        & 174.3 & -- \\
        Wang et al.~\cite{Wang2006}
        & Intrinsic beam + UVLM
        & 163.8 & -- \\
        Murua et al.~\cite{Murua2012}
        & SHARP, displacement beam + UVLM
        & 165.0 & 10.98 \\
        Savino et al.~\cite{Savino2021}
        & MBDyn + strip theory
        & 135.1 & 11.07 \\
        Savino et al.~\cite{Savino2021}
        & DUST (VL)--MBDyn
        & 168.2 & 10.84 \\
        Savino et al.~\cite{Savino2021}
        & DUST (SP)--MBDyn
        & 174.2 & 11.06 \\
        \hline
        \textbf{Present work}
        & \textbf{DUST (SP)--IGA Bernoulli}
        & \textbf{159.97} & \textbf{10.94} \\
        \textbf{Present work}
        & \textbf{DUST(SP)--IGA Timoshenko}
        & \textbf{159.53} & \textbf{10.89} \\
        \hline
    \end{tabular}
\end{table}

\begin{figure}[h!]
    \centering
    \begin{minipage}[t]{0.49\linewidth}
        \centering
        \includegraphics[width=\linewidth]
        {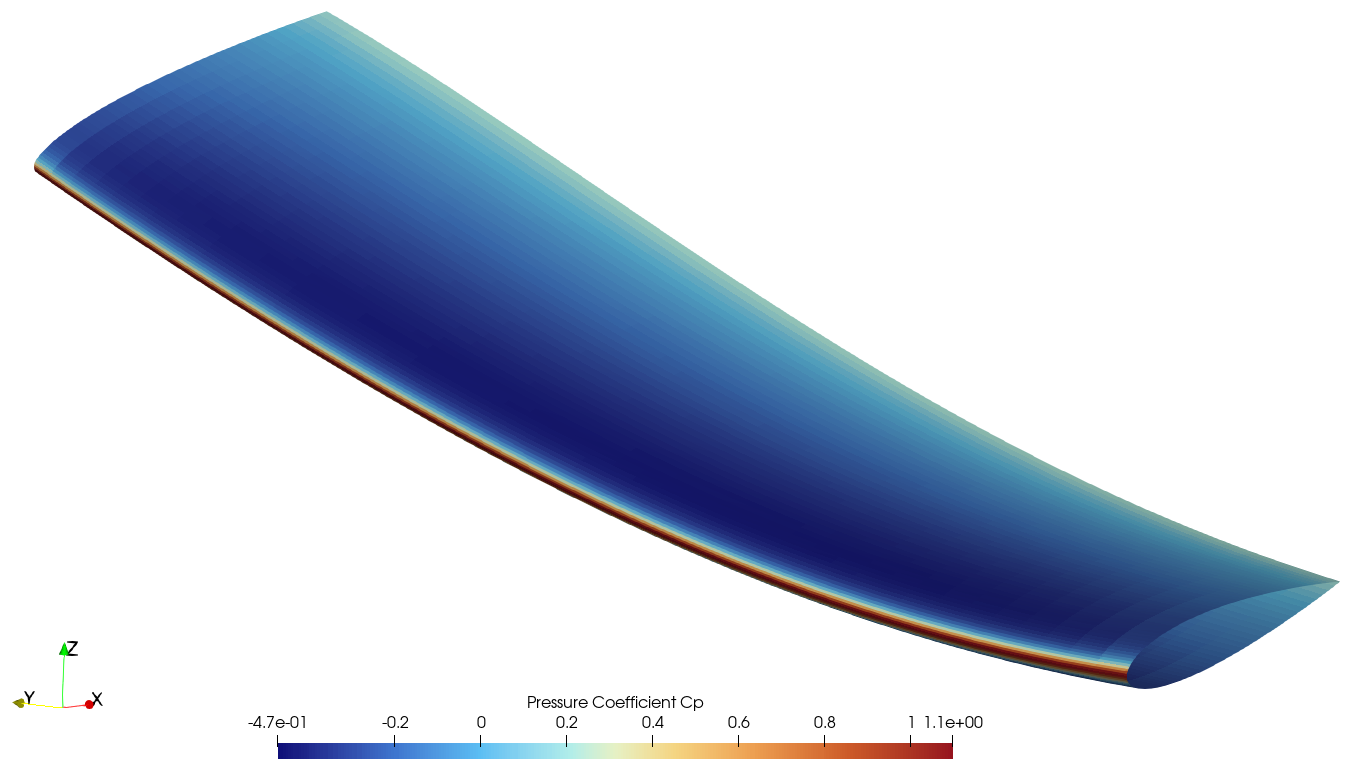}
        \caption*{(a) Pressure coefficient}
    \end{minipage}
    \hfill
    \begin{minipage}[t]{0.49\linewidth}
        \centering
        \includegraphics[width=\linewidth]
        {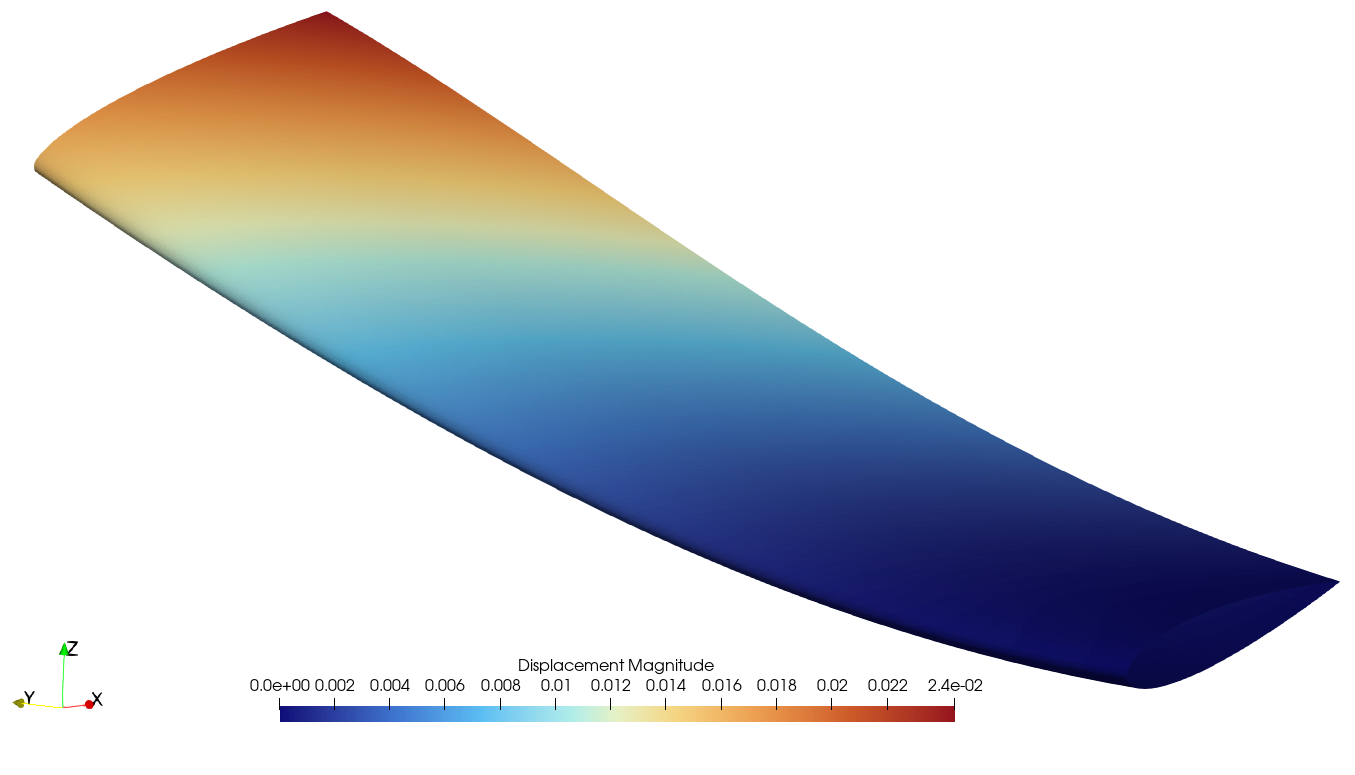}
        \caption*{(b) Displacement magnitude}
    \end{minipage}

    \caption{Unstable aeroelastic response of the Goland wing at
    $U_\infty=170\,\mathrm{m/s}$ and $t=1\,\mathrm{s}$:
    (a) pressure-coefficient distribution and
    (b) displacement magnitude. The deformation is magnified by a
    factor of 100 for visualization.}
    \label{fig:goland_deformed_fields}
\end{figure}

Fig.~\ref{fig:goland_deformed_fields} shows the unstable
response at $U_\infty=170\,\mathrm{m/s}$ and $t=1\,\mathrm{s}$.
The deformed configuration, magnified by a factor of 100 for
visualization, illustrates the coupled bending--torsion motion, while
the pressure-coefficient distribution shows the corresponding
aerodynamic loading, with pronounced gradients near the leading edge.

Finally, the computational cost of the two mappers is assessed by
varying the fluid-interface size and beam size separately. To keep the
comparison independent of changes in coupling convergence, the mapper
calls are replayed using data from 30 saved time steps. The plots report
the median execution times across these states, with error bars spanning
the 25th to 75th percentiles. Measurements are performed on an Intel
Core i9-14900K processor using one mapper thread and a Release build.

The forward cost is defined as the average execution time of one
displacement-transfer call and one velocity-transfer call. The reverse
cost is the execution time of one load-transfer call. A complete
mapping cycle comprises one call to each of these three operations.

\begin{figure}[h!]
    \centering
    \includegraphics[width=0.85\linewidth]
        {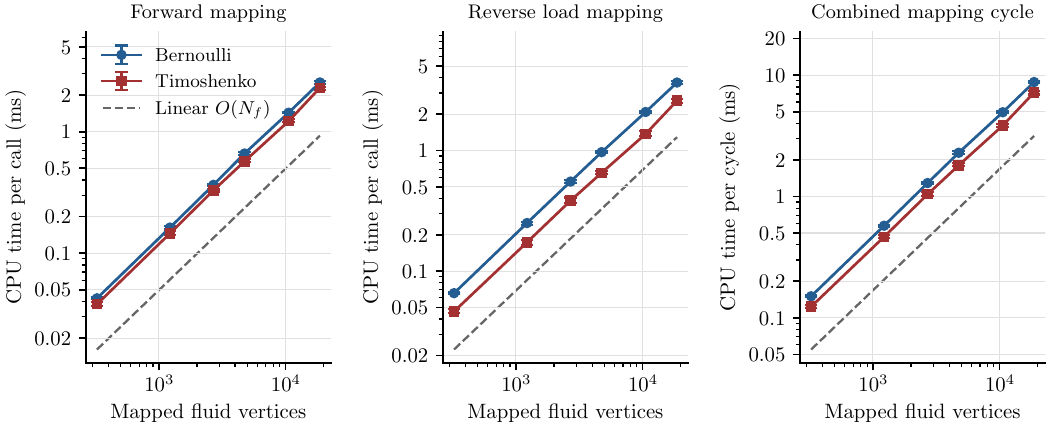}
    \caption{CPU cost of the Bernoulli and Timoshenko mappers versus
    the number of mapped fluid-interface vertices, with a fixed
    cubic beam containing 16 knot spans and 19 control points.
    Left: average cost of one displacement-transfer call and one
    velocity-transfer call. Middle: cost of one reverse load-transfer
    call. Right: combined cost of one displacement, one velocity, and
    one load-transfer call. Points show medians over data from 30 saved
    time steps; error bars span the 25th to 75th percentiles. Dashed
    lines indicate linear reference scaling.}
    \label{fig:goland_mapping_cost_fluid}
\end{figure}

Figure~\ref{fig:goland_mapping_cost_fluid} shows the results for a fixed
cubic beam with 16 knot spans and 19 control points. The fluid-interface
size ranges from 325 to 18721 mapped vertices. For both formulations,
the measured costs increase approximately linearly with the number of
mapped vertices over the tested range, as indicated by the linear
reference slopes. The median costs are consistently lower for the
Timoshenko mapper, with the largest difference observed in the reverse
load transfer.

\begin{figure}[h!]
    \centering
    \includegraphics[width=0.85\linewidth]
        {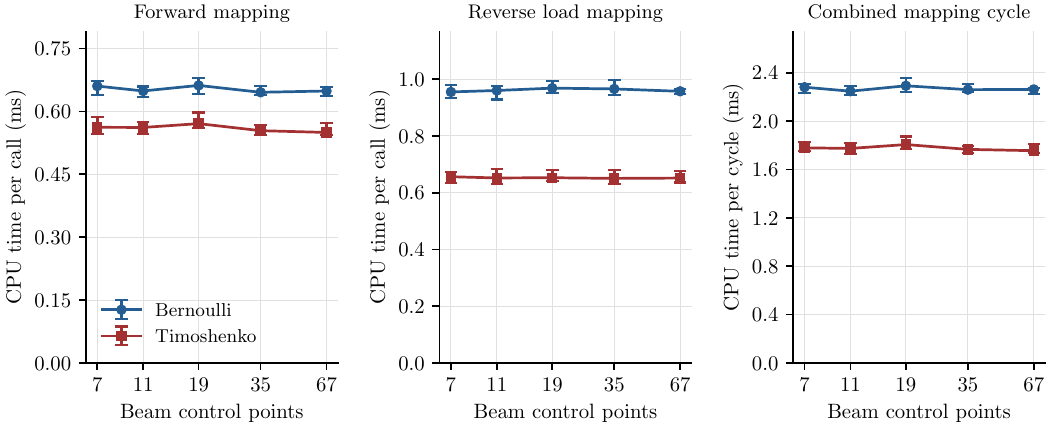}
    \caption{CPU cost of the Bernoulli and Timoshenko mappers versus
    beam control-point count, with cubic degree and a fixed fluid
    interface containing 4753 mapped vertices.
    Left: average cost of one displacement-transfer call and one
    velocity-transfer call. Middle: cost of one reverse load-transfer
    call. Right: combined cost of one displacement, one velocity, and
    one load-transfer call. Points show medians over data from 30 saved
    time steps; error bars span the 25th to 75th percentiles.}
    \label{fig:goland_mapping_cost_beam}
\end{figure}

Figure~\ref{fig:goland_mapping_cost_beam} shows the results for a fixed
fluid interface containing 4753 mapped vertices, while the cubic beam
discretization varies from 7 to 67 control points. The measured costs
show little variation with beam size over the tested range. For all
three operations, the median cost of the Timoshenko mapper remains
lower than that of the Bernoulli mapper.

\section{Conclusions}
\label{sec:conclusions}

This work introduced consistent mapping operators for coupling
geometrically nonlinear isogeometric Bernoulli and Timoshenko beams
with fluid discretizations in partitioned FSI. The proposed
framework addresses the differences in discretization and
kinematic representation between the structural and fluid domains,
enabling the consistent transfer of displacements and loads
across nonmatching coupling interfaces. To this end, fluid-side
interface points are associated with the beam in the reference
configuration and reconstructed from the current centerline
and cross-sectional kinematics. The consistent linearization
of this reconstruction defines the displacement mapping operator,
whose transpose transfers fluid forces to generalized beam
forces and moments in a work-conjugate manner.

Analytical bending, torsion, and combined-deformation tests verified accurate surface reconstruction under large rotations. The Timoshenko mapper achieved the expected $\mathcal{O}(h^{p+1})$ convergence, while the Bernoulli mapper showed convergence governed by the derivative-based reconstruction of the cross-section orientation. Virtual-work tests in strongly deformed configurations confirmed the work-conjugacy of both mappings, with relative discrepancies of approximately $10^{-11}$--$10^{-10}$ at the optimal finite-difference step sizes.

In the Mok benchmark, both beam formulations closely reproduced
the coupled response of the isogeometric Kirchhoff--Love shell
reference, demonstrating the ability of the proposed mappings
to accurately represent the fluid--structure interaction.
The beam-based Turek--Hron example further confirmed their
applicability to strongly coupled oscillatory FSI.

For the Goland wing, the predicted flutter velocities were 159.97 m/s for the Bernoulli model and 159.53 m/s for the Timoshenko model, with corresponding frequencies of 10.94 Hz and 10.89 Hz. The predicted frequencies agree closely with published values around 11 Hz, while the flutter velocities fall within the range of approximately 135--174 m/s reported using different structural and aerodynamic models.

Taken together, these results demonstrate that the proposed
mappings enable consistent coupling between reduced-dimensional
isogeometric beam models and independently discretized fluid
interfaces in partitioned FSI, provided that rigid cross-sectional
kinematics adequately represent the structural deformation.
Since the beam models do not capture warping or local
cross-sectional deformation, problems in which these effects
are significant require a richer structural model.

\appendix
\section{Detailed linearization of the interface reconstruction}
\label{app:derivatives}

This appendix provides the explicit linearizations used to construct the
configuration-dependent mapping tangents introduced in Section~\ref{sec:mapping_linearization}.
The corresponding work-conjugate load transfer follows directly from the
transpose of these tangents, as shown in Section~\ref{sec:load_transfer}, and is therefore not
repeated here.

\subsection{Bernoulli interface reconstruction}

Consider a point $Q$ belonging to the fluid--structure interaction (FSI)
coupling interface, attached to the beam at the fixed parameter $\xi_Q$.
All quantities below are evaluated at this parameter, and the subscript
$Q$ is omitted where no ambiguity arises.

The current centerline position, unit tangent, and twist angle are
obtained from the isogeometric interpolation as
\begin{equation}
    \mathbf{x}_c
    =
    \sum_i R_i(\mathbf{P}_i+\mathbf{u}_i),
    \qquad
    \mathbf{t}
    =
    \frac{\mathbf{x}_{c,\xi}}{\|\mathbf{x}_{c,\xi}\|},
    \qquad
    \phi
    =
    \sum_i R_i\phi_i.
\end{equation}
Here, $\mathbf{P}_i$, $\mathbf{u}_i$, and $\phi_i$ denote the
reference control-point positions, translational displacements,
and twist degrees of freedom, respectively.

The interface point is attached to the reference cross-section
through the fixed transverse offset $\boldsymbol{\rho}_0$,
defined in Section~\ref{sec:material_attachment}.
Its current position is reconstructed by first transporting
this offset from the reference tangent $\mathbf{T}$ to the
current tangent $\mathbf{t}$ and then applying the twist $\phi$.

The first operation uses the smallest rotation
$\boldsymbol{\Lambda}(\mathbf{T},\mathbf{t})$, which aligns
the two tangents without introducing additional twist.
The resulting vector
\begin{equation}
    \mathbf{b}
    =
    \boldsymbol{\Lambda}(\mathbf{T},\mathbf{t})
    \boldsymbol{\rho}_0
\end{equation}
represents the reference offset after the bending-induced
rotation of the cross-section, but before torsion is applied.
Using Rodrigues' rotation formula, it can be evaluated as
\begin{equation}
    \mathbf{b}
    =
    (\mathbf{T}\cdot\mathbf{t})\boldsymbol{\rho}_0
    +
    (\mathbf{T}\times\mathbf{t})\times\boldsymbol{\rho}_0
    +
    \frac{
        (\mathbf{T}\times\mathbf{t})
        [(\mathbf{T}\times\mathbf{t})\cdot\boldsymbol{\rho}_0]
    }{
        1+\mathbf{T}\cdot\mathbf{t}
    }.
    \label{eq:app_transport}
\end{equation}

Since $\mathbf{b}$ is perpendicular to the current tangent,
the additional rotation through the twist angle $\phi$
gives the current interface offset
\begin{equation}
    \boldsymbol{\rho}_Q
    =
    \cos\phi\,\mathbf{b}
    +
    \sin\phi\,(\mathbf{t}\times\mathbf{b}).
    \label{eq:app_bern_offset}
\end{equation}
The current position and displacement of the interface point
are therefore
\begin{equation}
    \mathbf{x}_Q
    =
    \mathbf{x}_c+\boldsymbol{\rho}_Q,
    \qquad
    \mathbf{u}_Q
    =
    \mathbf{x}_Q-\mathbf{X}_Q,
    \label{eq:app_bern_position}
\end{equation}
where $\mathbf{X}_Q$ is its fixed reference position.

\paragraph{Consistent linearization}

The objective is to determine the pointwise mapping operator
$\mathbf{H}_Q^{\mathrm{Ber}}$, which relates variations of the beam
degrees of freedom to variations of the reconstructed interface
displacement:
\begin{equation}
    \delta\mathbf{u}_Q
    =
    \underbrace{
        \frac{\partial\mathbf{u}_Q}{\partial\mathbf{q}}
    }_{\mathbf{H}_Q^{\mathrm{Ber}}}
    \delta\mathbf{q}.
    \label{eq:app_bern_mapping_definition}
\end{equation}
Here, $\mathbf{q}$ collects the translational and twist
degrees of freedom of the active beam control points.
Because the reference interface position $\mathbf{X}_Q$
is fixed, $\delta\mathbf{u}_Q=\delta\mathbf{x}_Q$.
Consequently, linearizing Eq.~\eqref{eq:app_bern_position}
gives
\begin{equation}
    \delta\mathbf{u}_Q
    =
    \delta\mathbf{x}_c
    +
    \delta\boldsymbol{\rho}_Q.
    \label{eq:app_bern_du}
\end{equation}

The first contribution follows directly from the centerline
interpolation:
\begin{equation}
    \delta\mathbf{x}_c
    =
    \sum_i R_i\delta\mathbf{u}_i.
    \label{eq:app_bern_dxc}
\end{equation}

The second contribution accounts for changes in the
cross-sectional orientation. Since the current interface
offset depends on both the unit tangent and the twist angle,
its variation can be expressed using the chain rule as
\begin{equation}
    \delta\boldsymbol{\rho}_Q
    =
    \frac{\partial\boldsymbol{\rho}_Q}
         {\partial\mathbf{t}}
    \delta\mathbf{t}
    +
    \frac{\partial\boldsymbol{\rho}_Q}
         {\partial\phi}
    \delta\phi.
    \label{eq:app_bern_chain_rule}
\end{equation}

We first consider the variation of the unit tangent.
Differentiating its normalized definition yields
\begin{equation}
    \delta\mathbf{t}
    =
    \frac{
        \mathbf{I}-\mathbf{t}\otimes\mathbf{t}
    }{
        \|\mathbf{x}_{c,\xi}\|
    }
    \delta\mathbf{x}_{c,\xi}.
\end{equation}
Using the derivative of the centerline interpolation,
this becomes
\begin{equation}
    \delta\mathbf{t}
    =
    \frac{
        \mathbf{I}-\mathbf{t}\otimes\mathbf{t}
    }{
        \|\mathbf{x}_{c,\xi}\|
    }
    \sum_i R_{i,\xi}\delta\mathbf{u}_i.
    \label{eq:app_dt}
\end{equation}
The projection
$\mathbf{I}-\mathbf{t}\otimes\mathbf{t}$
removes the component parallel to the tangent, since
changes in its magnitude do not affect its direction.

Next, we determine how a change in the tangent modifies
the reconstructed interface offset. Differentiating
Eq.~\eqref{eq:app_bern_offset} gives
\begin{align}
    \delta\boldsymbol{\rho}_Q
    ={}&
    \cos\phi\,\delta\mathbf{b}
    +
    \sin\phi
    \left(
        \delta\mathbf{t}\times\mathbf{b}
        +
        \mathbf{t}\times\delta\mathbf{b}
    \right)
    \nonumber\\
    &+
    \left[
        -\sin\phi\,\mathbf{b}
        +
        \cos\phi\,(\mathbf{t}\times\mathbf{b})
    \right]\delta\phi.
    \label{eq:app_drho_bern}
\end{align}

The variation $\delta\mathbf{b}$ follows from the
smallest-rotation transport. Since the reference tangent
$\mathbf{T}$ and offset $\boldsymbol{\rho}_0$ are fixed,
differentiation of Eq.~\eqref{eq:app_transport} gives
\begin{align}
    \delta\mathbf{b}
    ={}&
    (\mathbf{T}\cdot\delta\mathbf{t})
    \boldsymbol{\rho}_0
    +
    (\mathbf{T}\times\delta\mathbf{t})
    \times\boldsymbol{\rho}_0
    \nonumber\\
    &+
    \frac{
        (\mathbf{T}\times\delta\mathbf{t})
        [(\mathbf{T}\times\mathbf{t})\cdot\boldsymbol{\rho}_0]
    }{
        1+\mathbf{T}\cdot\mathbf{t}
    }
    \nonumber\\
    &+
    \frac{
        (\mathbf{T}\times\mathbf{t})
        [(\mathbf{T}\times\delta\mathbf{t})
        \cdot\boldsymbol{\rho}_0]
    }{
        1+\mathbf{T}\cdot\mathbf{t}
    }
    \nonumber\\
    &-
    \frac{
        (\mathbf{T}\times\mathbf{t})
        [(\mathbf{T}\times\mathbf{t})\cdot\boldsymbol{\rho}_0]
        (\mathbf{T}\cdot\delta\mathbf{t})
    }{
        (1+\mathbf{T}\cdot\mathbf{t})^2
    }.
    \label{eq:app_db}
\end{align}

Equations~\eqref{eq:app_drho_bern} and~\eqref{eq:app_db}
provide the derivative of the reconstructed interface
offset with respect to the current tangent.
For compactness, this derivative is denoted by
\begin{equation}
    \mathbf{K}
    =
    \left.
    \frac{\partial\boldsymbol{\rho}_Q}
         {\partial\mathbf{t}}
    \right|_{\phi},
    \label{eq:app_bern_K}
\end{equation}
where the subscript indicates that the twist angle is
held fixed. In particular, $\mathbf{K}$ is the linear
operator defined by
\begin{equation}
    \mathbf{K}\,\delta\mathbf{t}
    =
    \cos\phi\,\delta\mathbf{b}
    +
    \sin\phi
    \left(
        \delta\mathbf{t}\times\mathbf{b}
        +
        \mathbf{t}\times\delta\mathbf{b}
    \right),
    \label{eq:app_bern_K_action}
\end{equation}
with $\delta\mathbf{b}$ evaluated from
Eq.~\eqref{eq:app_db}.

The derivative with respect to the twist angle follows
directly from Eq.~\eqref{eq:app_bern_offset}:
\begin{equation}
    \frac{\partial\boldsymbol{\rho}_Q}{\partial\phi}
    =
    -\sin\phi\,\mathbf{b}
    +
    \cos\phi\,(\mathbf{t}\times\mathbf{b}).
    \label{eq:app_bern_drho_dphi}
\end{equation}

\paragraph{Pointwise mapping operator}

Having obtained the derivatives of the interface offset,
we now express the interface displacement variation in
terms of the beam degrees of freedom.

Substituting the previous results into
Eq.~\eqref{eq:app_bern_du} gives
\begin{equation}
    \delta\mathbf{u}_Q
    =
    \delta\mathbf{x}_c
    +
    \mathbf{K}\,\delta\mathbf{t}
    +
    \frac{\partial\boldsymbol{\rho}_Q}{\partial\phi}
    \delta\phi.
    \label{eq:app_bern_du_expanded}
\end{equation}

Using Eqs.~\eqref{eq:app_bern_dxc} and~\eqref{eq:app_dt},
together with
\begin{equation}
    \delta\phi
    =
    \sum_i R_i\delta\phi_i,
\end{equation}
we obtain
\begin{align}
    \delta\mathbf{u}_Q
    ={}&
    \sum_i
    \left[
        R_i\mathbf{I}
        +
        \frac{R_{i,\xi}}{\|\mathbf{x}_{c,\xi}\|}
        \mathbf{K}
        \left(
            \mathbf{I}-\mathbf{t}\otimes\mathbf{t}
        \right)
    \right]\delta\mathbf{u}_i
    \nonumber\\
    &+
    \sum_i
    R_i
    \left[
        -\sin\phi\,\mathbf{b}
        +
        \cos\phi\,(\mathbf{t}\times\mathbf{b})
    \right]\delta\phi_i.
    \label{eq:app_bern_du_final}
\end{align}

Comparing this expression with the definition of the
mapping operator in Eq.~\eqref{eq:app_bern_mapping_definition},
the translational block associated with control point $i$
is identified as
\begin{equation}
    \mathbf{H}_{Qi}^{u}
    =
    R_i\mathbf{I}
    +
    \frac{R_{i,\xi}}{\|\mathbf{x}_{c,\xi}\|}
    \mathbf{K}
    \left(
        \mathbf{I}-\mathbf{t}\otimes\mathbf{t}
    \right)
    \label{eq:app_bern_translation_column}
\end{equation}
and the corresponding twist block as
\begin{equation}
    \mathbf{H}_{Qi}^{\phi}
    =
    R_i
    \left[
        -\sin\phi\,\mathbf{b}
        +
        \cos\phi\,(\mathbf{t}\times\mathbf{b})
    \right].
    \label{eq:app_bern_twist_column}
\end{equation}

The pointwise Bernoulli mapping operator associated
with control point $i$ is therefore
\begin{equation}
    \mathbf{H}_{Qi}^{\mathrm{Ber}}
    =
    \begin{bmatrix}
        \mathbf{H}_{Qi}^{u}
        &
        \mathbf{H}_{Qi}^{\phi}
    \end{bmatrix},
    \label{eq:app_H_bern}
\end{equation}
where the translational and twist blocks have dimensions
$3\times3$ and $3\times1$, respectively.
Assembly over the active control points yields
$\mathbf{H}_Q^{\mathrm{Ber}}$.

The smallest-rotation transport requires
$1+\mathbf{T}\cdot\mathbf{t}>0$.
The antiparallel configuration
$\mathbf{T}\cdot\mathbf{t}=-1$, for which the smallest
rotation is not uniquely defined, is excluded.

\subsection{Timoshenko interface reconstruction}

For the Timoshenko formulation, the cross-sectional orientation
is an independent kinematic field. The current offset of an
interface point $Q$ is reconstructed directly from the current
material directors as
\begin{equation}
    \boldsymbol{\rho}_Q
    =
    \rho_{2,Q}\boldsymbol{a}_2
    +
    \rho_{3,Q}\boldsymbol{a}_3.
\end{equation}

Consistently with the multiplicative rotation update in
Eq.~\eqref{eq:timoshenko_rotation_update}, an infinitesimal
spatial rotational variation $\delta\boldsymbol{\varphi}$
produces
\begin{equation}
    \delta\boldsymbol{\rho}_Q
    =
    \delta\boldsymbol{\varphi}
    \times\boldsymbol{\rho}_Q
    =
    -[\boldsymbol{\rho}_Q]_\times
    \delta\boldsymbol{\varphi}.
\end{equation}
Here, the variation is evaluated at vanishing incremental
rotation.

The centerline and rotational variations are interpolated
using the same NURBS basis functions as
\begin{equation}
    \delta\boldsymbol{x}_c
    =
    \sum_i R_{i,p}(\xi_Q)\delta\boldsymbol{u}_i,
    \qquad
    \delta\boldsymbol{\varphi}
    =
    \sum_i R_{i,p}(\xi_Q)\delta\boldsymbol{\varphi}_i.
\end{equation}

Consequently, the interface displacement variation becomes
\begin{equation}
    \delta\boldsymbol{u}_Q
    =
    \sum_i R_{i,p}(\xi_Q)
    \left(
        \delta\boldsymbol{u}_i
        -
        [\boldsymbol{\rho}_Q]_\times
        \delta\boldsymbol{\varphi}_i
    \right).
\end{equation}

The corresponding pointwise mapping operator associated
with control point $i$ is therefore
\begin{equation}
    \boldsymbol{H}_{Qi}^{\mathrm{Tim}}
    =
    R_{i,p}(\xi_Q)
    \begin{bmatrix}
        \boldsymbol{I}
        &
        -[\boldsymbol{\rho}_Q]_\times
    \end{bmatrix}.
\end{equation}
The resulting operator is used for the work-conjugate
load transfer described in
Section~\ref{sec:load_transfer}.

\section*{Declarations}
\paragraph{\textbf{Conflict of interest}} 
The authors have no competing interests to declare that are relevant to the content of this article.

\section*{Data availability}
\noindent Data will be made available on request.

\section*{Acknowledgments}
\noindent The authors gratefully acknowledge the \emph{Design for IGA-type discretization workflows (GECKO)} project. The Design for IGA-type discretization workflows has received funding from the European Union’s Horizon Europe research and Innovation programme under grant agreement No.~101073106, Call: HORIZON-MSCA-2021-DN-01.

\bibliographystyle{elsarticle-num}
\bibliography{A_camarotti_paper-refs}

\end{document}